\documentclass[aps,preprintnumbers,floatfix,nofootinbib]{revtex4}
  \pdfoutput=1

  \usepackage{epsfig}
  \usepackage{amsmath}
  \usepackage{graphicx}
  \usepackage{amssymb}
  \usepackage{axodraw}
  \usepackage{color}
  \usepackage{url}

\newcommand{\diag}{{\rm diag}}

\newcommand{\eq}{Eq.}
\newcommand{\eqs}{Eqs.}
\newcommand{\Fig}{Fig.}

\newcommand{\Sec}{Sec.}

\newcommand{\be}{\begin{equation}}
\newcommand{\ee}{\end{equation}}
\def\aprle{\buildrel < \over {_{\sim}}}
\def\aprge{\buildrel > \over {_{\sim}}}
\newcommand{\bea}{\begin{eqnarray}}
\newcommand{\eea}{\end{eqnarray}}

\begin{document} 

 \title{Neutrino Propagation in Matter}
 
 \author{Mattias Blennow}
 \affiliation{Max-Planck-Institut f\"ur Kernphysik, Saupfercheckweg 1, 69117 Heidelberg, Germany \\ and \\ KTH Royal Institute of Technology, AlbaNova University Center,
Roslagstullsbacken 21, 106 91 Stockholm, Sweden}

 \author{Alexei Yu.\ Smirnov}
 \affiliation{The Abdus Salam International Centre for Theoretical Physics, 
                I-34100 Trieste, Italy}
                
 \label{ch-05:smirnov}

\begin{abstract}

We describe the effects of neutrino propagation in the matter
of the Earth relevant for experiments with atmospheric
and accelerator neutrinos and aimed at the determination of the
neutrino mass hierarchy and CP-violation.
These include (i) the resonance enhancement of
neutrino oscillations in matter with constant
or nearly constant density,
(ii) adiabatic conversion in matter with slowly changing
density, (iii) parametric enhancement of oscillations
in a multi-layer medium, (iv) oscillations in thin layers of matter.
We present the results of semi-analytic descriptions of
flavor transitions for the cases of small density perturbations,
in the limit of large densities and  for small density widths.
Neutrino oscillograms of the Earth and their structure after
determination of the 1-3 mixing are described.
A possibility to identify the  neutrino mass hierarchy with the
atmospheric neutrinos and multi-megaton scale detectors having low 
energy thresholds is explored. The potential of future 
accelerator experiments to establish the hierarchy is outlined.

\end{abstract}

\maketitle

\section{Introduction}
\label{sec-05:intro}

Neutrinos are eternal travelers: once produced (especially at low energies) they have 
little chance to interact and be absorbed. Properties of neutrino fluxes:  
flavor compositions, lepton charge asymmetries, energy spectra of encode 
information. Detection the neutrinos brings unique knowledge  
about their sources, properties of medium,  space-time they propagated 
as well as on neutrinos themselves. 

Neutrino propagation in matter is vast area of research which 
covers variety of different aspects: from conceptual ones to applications.  
This includes propagation in matter (media) with (i) different properties
(unpolarized, polarized, moving, turbulent, fluctuating, with neutrino components,  etc),
(ii) different density profiles, and (iii) in different energy regions. 
The applications cover neutrino propagation in matter of the Earth and the Sun, 
supernova and relativistic jets as well as neutrinos in the Early Universe. 

The impact of matter on neutrino oscillations was first studied by
Wolfenstein in 1978~\cite{05-Wolfenstein:1977ue}.
He marked that matter suppresses oscillations of the solar neutrinos
propagating in the Sun and supernova neutrinos inside a star.
He considered a hypothetical experiments with neutrinos propagating
through 1000~km of rock,
 something that today is no longer only a thought but actual experimental
reality.
Later Barger et~al~\cite{05-Barger:1980tf} have observed that matter can also  enhance
oscillations
at certain energies. The work of Wolfenstein was expanded upon in papers
by Mikheev and
Smirnov~\cite{05-Mikheev:1986gs,05-Mikheev:1986wj,05-Mikheev:1986if}, in
particular, in the context of the solar neutrino problem.
Essentially two new effects have been proposed:
the resonant enhancement of neutrino oscillations in matter with constant
and  nearly constant density and
the adiabatic flavor conversion in matter with slowly changing density. It
was marked that the first effect
can be realized for neutrinos crossing the matter of the Earth.
The second one can take place in propagation  of solar neutrinos from the
dense solar core via the resonance region inside the Sun to
the surface with negligible density. This adiabatic flavor transformation,
called later the MSW effect,  was proposed as a solution
of the solar neutrino problem.

Since the appearance of these seminal papers, neutrino flavor evolution in
background matter
were studied extensively including the treatment of propagation in media
which are not consisting simply of matter at rest,
but also backgrounds that take on a more general form.
For instance, in a thermal field theory approach~\cite{05-Notzold:1987ik},
effects of finite temperature and density can be taken readily into
account. If neutrinos are dense enough,
new  type of effects can arise due to the neutrino background itself,
causing a collective behavior
in the flavor evolution. This type of effect could have a significant
impact on neutrinos in the early Universe
and in central parts of collapsing stars.

There has been a great progress in treatments of neutrino conversion in
matter,
both from an analytical and a pure computational points of view.
From the analytical side, the description  of three-flavor neutrino
oscillations
in matter is given by a plethora of formulas containing information
that may be hard to get a proper grasp of without introducing approximations.
 Luckily, given the parameter values inferred from experiments, various
perturbation theories
 and series expansions in small parameters can be developed.
 In this review we will explain the basic physical effects important  for
the current
 and next generation neutrino oscillation experiments and provide  the
relevant formalism.
 We present an  updated picture of oscillations and conversion given the
current knowledge
 on the neutrino oscillation parameters.

In this paper we focus mainly on aspects related to future experiments 
with atmospheric and accelerator neutrinos. The main goals of these experiments 
are to (i) establish the neutrino mass hierarchy, (ii) discover CP-violation 
in the lepton sector and determination of  the CP-violating phase, 
(iii) precisely measure the neutrino parameters, in particular, 
the deviation of 2-3 mixing from maximal, and (iv) search for sterile neutrinos and new neutrino interactions. 

Accelerator and atmospheric neutrinos propagate in the matter of the Earth. 
Therefore we mainly concentrate on effects of neutrino propagation in the Earth, 
{\it i.e.}, in usual electrically neutral and non-relativistic matter. 
We update existing results on effects of neutrino propagation 
in view of the recent determination of the 1-3 mixing.

The review is organized as follows: 
In \Sec~\ref{sec-05:properties} we consider properties of neutrinos in matter, in particular, 
mixing in matter and effective masses (eigenvalues of the Hamiltonian);  
we derive equations which describe the propagation. 
\Sec~\ref{sec-05:neutrinopropagationindifferentmedia} is devoted to various effects  relevant for neutrino propagating in the Earth.  
We consider the properties of the oscillation/conversion probabilities 
in different channels. 
In \Sec~\ref{sec-05:masshierarchy} we explore the effects of 
the neutrino mass hierarchy and CP-violating phase on the 
atmospheric neutrino fluxes and neutrino beams from accelerators.  
Conclusions and outlook are presented in \Sec~\ref{sec-05:conclusions}. 


\section{Neutrino properties in matter}
\label{sec-05:properties}

We will consider the system of 3 flavor neutrinos, 
$\nu_f^T \equiv (\nu_e,\nu_{\mu},\nu_{\tau})$,  mixed in vacuum: 
\be
\nu_f = U_{PMNS} \nu_m .  
\label{eq-05:pmns}
\ee
Here $U_{PMNS}$ is the Pontecorvo-Maki-Nakagawa-Sakata (PMNS) mixing matrix 
\cite{05-Pontecorvo:1957cp,05-Maki:1962mu,05-Pontecorvo:1967fh} and $\nu_m^T \equiv (\nu_1, \nu_2, \nu_3)$
is the vector of mass eigenstates with masses $m_i$ (i = 1, 2, 3). 
We will use the standard parameterization of the PMNS matrix,  
\be
U_{PMNS} = U_{23}(\theta_{23}) I_\delta U_{13}(\theta_{13}) I_\delta^* 
U_{12} (\theta_{12}), 
\label{eq-05:upmns}
\ee 
which is the most suitable for describing usual matter effects. 
In Eq. (\ref{eq-05:upmns}) $U_{ij} (\theta_{ij})$ are the matrices of rotations in the 
$ij-$planes with angles $\theta_{ij}$ and $I_\delta \equiv \diag (1, 1, e^{\delta})$. 

In vacuum the flavor evolution of these neutrinos is described by the
the Schr\"{o}dinger-like equation 
\be
i \frac{d \nu_f}{dt} =
\frac{M M^\dagger}{2 E}\nu_f\, ,  
\label{eq-05:evolution}
\ee
where $M$ is the neutrino mass matrix in the flavor basis and $E$ is the neutrino energy. 
\eq~(\ref{eq-05:evolution}) is essentially a generalization of the equation $E \approx p + m^2/2E$ for a single ultra relativistic particle. 
According to \eq~(\ref{eq-05:evolution}), the Hamiltonian in vacuum 
can be written as 
\be
H_0 =  \frac{1}{2 E}  U_{PMNS} M^2_{diag} U_{PMNS}^\dagger,
\label{eq-05:ham3f}
\ee
where $M^2_{diag} \equiv M^{\dagger} M = \diag 
\left(m_1^2, m_2^2,  m_3^2\right)$ and we take the masses $m_i$ to be real~\footnote{The term $p I$ is omitted in (\ref{eq-05:ham3f})
since it does not produce phase difference.}.

\subsection{Refraction and matter potentials}
\label{sec-05:refraction} 

The effective potential for a neutrino in medium $V_f$  
can be computed as a forward scattering matrix element 
$V_f = \langle \Psi| H_{int}| \Psi\rangle.$
Here $\Psi$ is the wave function of the system of neutrino and medium,
and $H_{int}$ is the Hamiltonian of interactions.

At low energies,
the Hamiltonian $H_{int}$ is the effective four fermion Hamiltonian
due to exchange of the $W$ and $Z$ bosons:
\be
H_{int} = \frac{G_F}{\sqrt{2}}\bar{\nu} \gamma^{\mu}(1 - \gamma_5) \nu
\left\{\bar{e} \gamma_{\mu}(g_V + g_A \gamma_5) e + 
\bar{p} \gamma_{\mu}(g_V^p + g_A^p \gamma_5) p +
\bar{n} \gamma_{\mu}(g_V^n + g_A^n \gamma_5) n \right\},
\ee
where $g_V$ and  $g_A$ are the vector and axial
vector coupling constants. 

In the Standard Model the matrix of the potentials in the flavor basis,
is  diagonal: $V_f = \diag(V_e, V_{\mu}, V_{\tau}, 0 ...)$. 

For medium the matrix elements of vectorial components of vector current 
are proportional to velocity  of particles of medium. 
The matrix elements of the axial vector current are proportional to 
spin vector.  Therefore 
for non-relativistic and unpolarized medium (as well as for an isotropic distribution
of ultra relativistic electrons) only the 
$\gamma^0$ component of the vector current gives a non-zero result, which is proportional to the
number density of the corresponding particles. Furthermore, due to conservation of the vector current (CVC), the couplings $g_{V}^{p}$ and $g_{V}^{n}$
can be computed using the neutral current couplings of
quarks. Thus, taking into account that, in the Standard Model,
the neutral current couplings of electrons and protons are equal and of opposite sign, the NC contributions
from electrons and protons cancel in electrically
neutral medium. As a result, the potential for neutrino flavor $\nu_a$ is
\be
V_a = \sqrt{2} G_F \left( \delta_{ae}  n_e  -  \frac{1}{2} n_n \right),
\label{eq-05:VeVa}
\ee
where $n_e$ and $n_n$ are the densities of electrons and neutrons,
respectively.

Only the difference of potentials has a physical meaning. 
Contribution of the neutral current scattering to $V$  
is the same for all active neutrinos.  Since $V_a$ ($a = \mu, \tau$ or a combination  thereof) is due to the neutral current scattering, 
in a normal medium composed of protons neutrons (nuclei) and electrons,  
$V_\mu - V_\tau = 0$. 
Furthermore, the difference of the  potentials for $\nu_e$ and $\nu_a$
is due to  the charged current scattering
of  $\nu_e$ on electrons  ($\nu_e e \rightarrow \nu_e e$) \cite{05-Wolfenstein:1977ue}:
\be
V = V_e - V_a  = \sqrt{2} G_F n_e~.
\label{eq-05:Vea}
\ee

The difference of potentials leads to the appearance of an
additional phase difference in the neutrino system:
$\phi_{matter} \equiv (V_e - V_a) t \approx Vx$.
This determines the \emph{refraction length}, the distance over which an additional ``matter'' phase  equals $2\pi$,:
\be
l_0 \equiv \frac{2\pi}{V_e - V_a} = \frac{\sqrt{2}\pi} {G_F n_e}. 
\ee
Numerically, 
\be
l_0 = 1.6 \cdot 10^9~{\rm cm} \  \frac{1~{\rm g/cm^3}}{n_e m_N},
\ee
where $m_N$ is the nucleon mass.
The corresponding column density $d \equiv l_0 n_e = \sqrt{2}\pi/{G_F}$
is given by the Fermi coupling constant only. 

{}For antineutrinos the potential has an opposite sign.  
{}Being zero  in the lowest order the difference of potentials  in the $\nu_{\mu} - \nu_{\tau}$ system  appears at a level of $10^{-5} V$ 
due to the radiative corrections~\cite{05-Botella:1986wy}.
Thus in the flavor basis  in the lowest order in EW interactions 
the effect of medium on neutrinos  is described by 
$\hat{V} = diag (V_e, 0, 0)$ with $V_e$ given in \eq~(\ref{eq-05:Vea}).

The potential has been computed for neutrinos in different type of media, such as polarized or heavily degenerate electrons, in~\cite{05-Esposito:1995db,05-Nunokawa:1997dp,05-Lobanov:2001ar}.

\subsection{Evolution equation, effective Hamiltonian, and mixing in matter}
\label{sec-05:evolution} 

\subsubsection{Wolfenstein equation}

In the flavor basis, the Hamiltonian in matter can be obtained 
by adding  the interaction term to the vacuum Hamiltonian in vacuum   
\cite{05-Wolfenstein:1977ue,05-Wolfenstein:1978ui,05-Wolfenstein:1979ni,05-Mikheev:1986gs,05-Mikheev:1986wj,05-Mikheev:1986if}: 
\be
H_f =  \frac{1}{2E} U_{PMNS} M^2_{diag} U_{PMNS}^\dagger + \hat{V} . 
\label{eq-05:ham3}
\ee
In \eq~(\ref{eq-05:ham3}) we have omitted irrelevant parts of the Hamiltonian 
proportional to the unit matrix. 
The Hamiltonian for antineutrinos can be obtained by the substitution
\begin{equation}
U \rightarrow U^*, \quad V \rightarrow -V 
\,.
\label{eq-05:tr-antinu}
\end{equation}
%

There are different derivations of the neutrino evolution equation
in matter, in particular, strict derivations starting from the Dirac equation 
or derivation in the context of quantum field theory  (see~\cite{05-Akhmedov:2012mk} 
and references therein). 

Although the Hamiltonian $H_f$ describes evolution in time, with the connection 
$x=vt \approx x = ct$, \eq~(\ref{eq-05:matt2}) can be rewritten 
as $i d\nu_f /dx = (H_0 + \hat V) \nu_f$ with $V = V(x)$, so it can be used as 
an evolution equation in space. 

Due to the strong hierarchy of $\Delta m^2$ and the smallness of 1-3 mixing, 
the results can be qualitatively understood and in many cases quantitatively described by 
reducing $3\nu$-evolution to $2\nu$-evolution.  
The reason is that the third neutrino effectively decouples and its effect  
can be considered as a perturbation. Of course, there are 
genuine $3\nu-$ phenomena such as 
CP-violation, but even in this case the dynamics of evolution can be reduced 
effectively to the dynamics of evolution of $2\nu-$systems.  
The evolution equation for two flavor states,  
$\nu_f^T = (\nu_e, \nu_a)$,  in matter
is
\be
i\frac{d\nu_f}{dt} =   {\left[ \frac{\Delta m^2}{4E}
\left(\begin{array}{cc}
- \cos 2 \theta  &  \sin 2 \theta \\
\sin 2 \theta &  \cos 2 \theta
\end{array}
\right) + 
\left(\begin{array}{cc}
\frac{1}{2}V_e  &  0 \\
0      &  - \frac{1}{2}V_e
\end{array}
\right) \right]}\nu_f 
\label{eq-05:matt2}
\ee
where the Hamiltonian is written in symmetric form.

\subsection{Mixing and eigenstates in matter}
\label{sec-05:mixing} 

The mixing in matter is defined with respect to $\nu_{im}$ -  
the eigenstates of the Hamiltonian in matter $H_f$. 

As usually, the eigenstates are obtained from the equation   
\be
H_f \nu_{im} = H_{im} \nu_{im} ,
\label{eq-05:eigenst}
\ee
where $H_{im}$ are the eigenvalues of $H_f$. 
If the density, and therefore $H_f$, are constant, $\nu_{im}$
correspond to the eigenstates of propagation. 
Since $H_f \neq H_0$, the states $\nu_{im}$ differ from the mass states, 
$\nu_{i}$. For low density, $n \rightarrow 0$,
the vacuum eigenstates are recovered: $\nu_{im} \rightarrow \nu_{i}$. 
If the density, and thus $H_f$, changes during neutrino propagation, 
$\nu_{im}$ and  $H_{im}$ should be considered as the eigenstates and eigenvalues 
of the instantaneous Hamiltonian: 
$H_f  = H_f(x)$,  $\nu_{im} = \nu_{im}(x)$  
and $H_{im} = H_{im}(x)$. For $n \rightarrow 0$ we have  
$H_{im} \rightarrow m_i^2/2E$.

The mixing in matter is a generalization of the mixing
in vacuum (\ref{eq-05:pmns}). Recall that the mixing matrix in vacuum connects 
the flavor neutrinos, $\nu_f$, and the massive neutrinos, $\nu_{\rm mass}$. 
The latter  are the eigenstates 
of Hamiltonian in vacuum: $\nu_H = \nu_{\rm mass}$. 
Therefore, the mixing matrix in matter is defined as the matrix which relates  
the flavor states with the eigenstates of the Hamiltonian in matter 
$\nu_H^T = (\nu_{1m}, \nu_{2m}, \nu_{3m})$: 
\be
\nu_f = U^m \nu_H.
\label{eq-05:mixmat}
\ee  

{}From \eq~(\ref{eq-05:eigenst}) we find that 
\be 
\nu_{jm}^\dagger H_f \nu_{im}  = H_{i m} \delta_{ji}. 
\label{eq-05:diagmass} 
\ee 
Furthermore, the Hamiltonian can be represented in the flavor basis as 
\be
H_f = \sum_{\alpha \beta} H_{\alpha \beta} 
\nu_{\alpha} \nu_\beta^\dagger. 
\ee
Inserting this expression as well as 
the relation $\nu_{jm} = U_{\alpha j}^{m*} \nu_{\alpha}$, which follows from \eq~(\ref{eq-05:mixmat}),
into \eq~(\ref{eq-05:diagmass}) one obtains 
\be
\sum_{\alpha \beta} U_{\alpha j}^{m*} H_{\alpha \beta} U^m_{ \beta i}  
=  H_{i m} \delta_{ji} 
\label{eq-05:diagonal}
\ee
or in matrix form $U^{m\dagger} H_f U^m = H^{diag} = \diag (H_{1m}, H_{2m}, H_{3m})$.
Thus, the mixing matrix $U^m $ can be found diagonalizing the full Hamiltonian.  
The columns of the mixing matrix, 
$U_i \equiv  (U_{ei}^m, U_{\mu i}^m, U_{\tau i}^m)$,  
are the eigenstates of the Hamiltonian $H_f$ which correspond 
to the eigenvalues $H_{im}$. Indeed, it follows from Eq. (\ref{eq-05:diagonal}) that  
$H_f U^m = U^m H^{diag}$.

Equation (\ref{eq-05:mixmat}) can be inverted to $\nu_H = U^{m\dagger} \nu_f$,
or in components $\nu_{i m} = U_{\alpha i}^{m*} \nu_\alpha, \quad \alpha = e, \mu, \tau$. 
According to this, the  elements of mixing matrix determine 
the flavor content of the mass eigenstates so that  $|U^m_{\alpha i}|^2$ gives the 
probability to find $\nu_\alpha$ in a given eigenstate $\nu_{im}$. 
Correspondingly, the elements of the PMNS matrix 
determine the flavor composition of the mass eigenstates in vacuum. 

\subsection{Mixing in the two neutrino case}

In the $2\nu$-case, there is single mixing angle in matter $\theta_m$ 
and the relations between the eigenstates in matter 
and the flavor states reads
\be
\nu_e = \cos\theta_m \nu_{1m} + \sin \theta_m \nu_{2m}, ~~
\nu_a = \cos\theta_m \nu_{2m} - \sin \theta_m \nu_{1m}.
\label{eq-05:2num}
\ee
The angle $\theta_m$ is obtained by diagonalization
of the Hamiltonian (\ref{eq-05:matt2}) (see previous section): 
\be
\sin^2 2\theta_m =
\frac{1}{R} \sin^2 2\theta,~~~~
R \equiv {\left(\cos 2\theta - \frac{2VE}{\Delta m^2} \right)^2 
+  \sin^2 2\theta}, 
\label{eq-05:angmatt}
\ee
where $R$ is the \emph{resonance factor}. In the limit $V \rightarrow 0$,  
the factor $R \rightarrow 1$ and the vacuum mixing is recovered. 
The difference of eigenvalues $H_{im}$ equals
\be
\omega_m \equiv  H_{2m} - H_{1m} = 
\frac{\Delta m^2}{2E} \sqrt{R}
\label{eq-05:omegam}
\ee
This difference is also called the level splitting. 
or oscillation frequency, which 
determines the oscillation length: $l_m = 2\pi/\omega_m$ (see Sect.~\ref{sec-05:constant}).  

The matter potential and $\Delta m^2$ always enter the mixing angle and other 
dimensionless quantities in the combination 
\be
\frac{2EV}{\Delta m^2} = \frac{l_{\nu}}{l_0},
\label{eq-05:ratio-x}
\ee
where $l_0$ is the refraction length. 
This is the origin of the ``scaling'' behavior of various characteristics 
of the flavor conversion probabilities.  
In terms of the mixing angle in matter the Hamiltonian can be rewritten 
in the following symmetric form  
\be
H_f = \frac{\omega^m}{2}
\left( 
\begin{array}{cc}
- \cos 2 \theta_m  & \sin 2\theta_m\\
\sin 2\theta_m  & \cos 2\theta_m  
\end{array}
\right). 
\label{eq-05:matterHam}
\ee

\subsubsection{Resonance and level crossing}
\label{sec-05:resonance}

\begin{figure}[h!!]
\begin{center}
\vspace{1cm}
\includegraphics[height=5cm,angle=0]{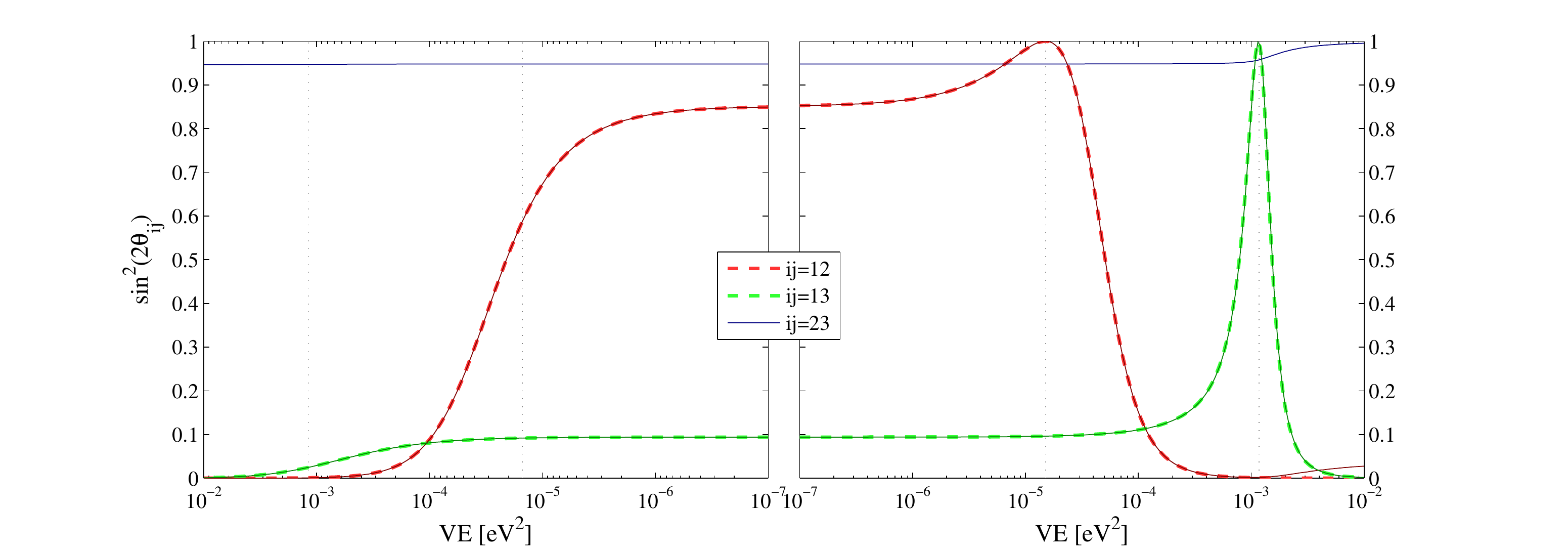} \\
\includegraphics[height=5cm,angle=0]{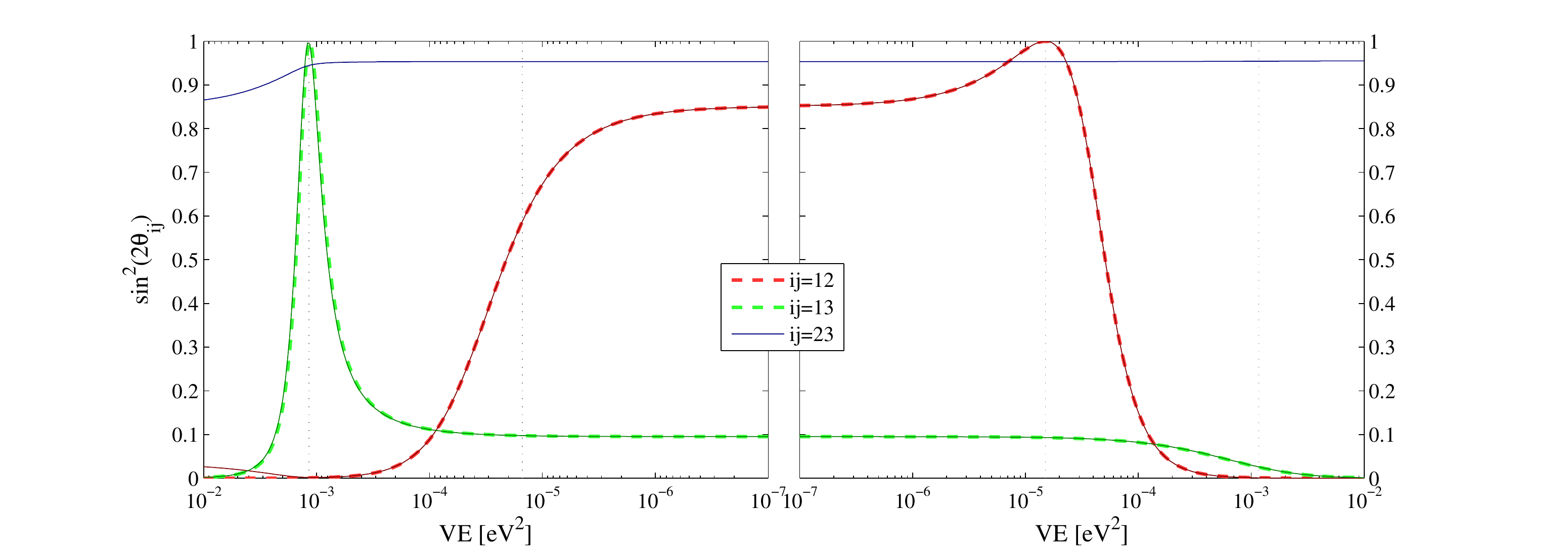}
\caption{Resonance in neutrino mixing. The dependence of 
$\sin^2 2\theta_{mij}$ on
the product $V E$ for vacuum mixing: $\sin^2 2\theta_{12} = 0.851$,  
$\Delta m^2_{21} =  7.59 \cdot 10^{-5}$~eV$^2$  
(red) and  $\sin^2 \theta_{13} = 0.0241$,  
$\Delta m^2_{31} =  2.47 \cdot 10^{-3}$~eV$^2$ (green). The left semi-plane   
corresponds to antineutrinos. The behavior of $\theta_{23}$ with vacuum value 
$\sin^2 2\theta_{23} = 0.953$ is included for completeness. 
The dashed lines are the predictions from a strict two-flavor approximation 
while the solid thin lines are the results of numerical diagonalization 
of the full three-flavor system. The upper panels show  the case of the normal 
mass hierarchy and the lower panels -- the inverted hierarchy.
\label{fig-05:angles_normal}}
\end{center}
\end{figure}

According to \eq~(\ref{eq-05:angmatt}) the effective mixing parameter in
matter,  $\sin^2 2\theta_m$, depends on the electron density 
and neutrino energy through the ratio (\ref{eq-05:ratio-x})  of
the oscillation and refraction lengths,
$x = l_\nu/l_0 \propto E V$. 
The dependence  $\sin^2 2\theta_m (VE)$ for two different values 
of the vacuum mixing angle, 
corresponding to angles from the full three flavor framework, is shown 
in \Fig~\ref{fig-05:angles_normal}.
%
The dependence of $\sin^2 2\theta_m$ on $E$ 
has a resonant character \cite{05-Mikheev:1986gs}. At
\be
l_{\nu} = l_0 \cos 2\theta~
\label{eq-05:res}
\ee
the mixing becomes maximal: $\sin^2 2\theta_m = 1$ ($R = \sin^2 2\theta$). 
The equality in (\ref{eq-05:res})  is called the \emph{resonance condition} and 
it can  be rewritten as $2EV = \Delta m^2 \cos 2\theta$.
For small vacuum mixing the condition reads:
${\rm Oscillation~~ length} \hskip 0.2cm \approx \hskip 0.2cm 
{\rm Refraction~~ length}$.  
The physical meaning of the resonance is that 
the eigenfrequency, which characterizes a system of mixed neutrinos, 
$\omega = 2\pi/l_{\nu} = \Delta m^2 /2E$,    
coincides with the eigenfrequency of the medium, $2\pi/l_0 = 1/V$.
The resonance condition (\ref{eq-05:res}) determines the resonance density
\be
n_e^{R} = \frac{\Delta m^2}{2E} \frac{\cos 2\theta}{\sqrt{2} G_F}~.
\label{eq-05:resonance}
\ee
The width of resonance on the half of height (in the density scale) is given by
$2 \Delta n_e^R = 2 n_e^R \tan 2\theta$.
Similarly, for fixed $n_e$ one can introduce the resonance energy and the width of
resonance in the energy scale. The width  can be rewritten as $\Delta n_e^R = n_0 \sin 2\theta$, where $n_0 \equiv {\Delta m^2}/{2 \sqrt{2} E G_F}$.
When the vacuum  mixing  approaches maximal value, $\theta \rightarrow \frac {\pi}{4}$
the resonance shifts to  zero density: $n_e^R \rightarrow 0$,
the width of resonance increases
converging to fixed value: $\Delta n_e^R \rightarrow   n_0$.

In a medium with varying density, the layer in which
the density changes in the interval $n_e^R \pm \Delta n_e^R$
is called the resonance layer. In this layer 
the angle  $\theta_m$ varies in the interval from $\pi/8$ to 
$3\pi/8$. 

For $V \ll V_R$, the mixing angle is close to the vacuum angle: $\theta_m \approx \theta$, 
while for $V \gg V_R$ the angle  becomes $\theta_m \approx \pi/2$  
and the mixing is strongly suppressed. 
In the resonance region, the level splitting is minimal \cite{05-cabbibo,05-Bethe:1986ej}, 
therefore the oscillation length, as the function of density, is maximal.

\subsection{Mixing of 3 neutrinos in matter}
\label{sec-05:3mixing}
 
To a large extent, knowledge of the eigenstates (mixing parameters) and 
eigenvalues of the instantaneous Hamiltonian in matter allows the determination of flavor evolution in most of the realistic situations 
(oscillations in matter of constant density, adiabatic conversion, 
strong breaking of adiabaticity). 
The exact expressions for the eigenstates and eigenvalues~\cite{05-Bueno:2000fg,05-Freund:2001pn}
are rather complicated and difficult to analyze. 
Therefore approximate expressions for the mixing angles and 
eigenvalues are usually used. They can be obtained  performing   
an approximate diagonalization of $H_f$ 
which relies on the strong hierarchy of the mass squared differences: 
\be
r_\Delta \equiv \frac{\Delta m_{21}^2}{\Delta m_{31}^2} \approx 0.03. 
\label{eq-05:r-delta}
\ee
Without changing physics, the factor  $I_{-\delta}$ 
in the mixing matrix can be eliminated by permuting it 
with $U_{12}$ and redefining the state $\nu_3$. Therefore, in what 
follows, we use  $U_{PMNS} = U_{23} I_\delta U_{13} U_{12}$. 
Here we will here describe the case of normal mass hierarchy:   
$ \Delta m_{31}^2 > 0, \Delta m_{32}^2 > 0$. Subtracting 
from the Hamiltonian the matrix 
proportional to the unit matrix $m_1^2/2E {\bf I}$, we obtain
\be
M_{diag}^2 = \Delta m_{31}^2 \diag(0, ~ r_\Delta, ~ 1 ). 
\ee

\subsubsection{Propagation basis} 

The propagation basis, $\tilde{\nu} = (\nu_e, \tilde{\nu}_{2}, \tilde{\nu}_{3})^T$,
which is most suitable for consideration of the 
neutrino oscillations in matter is defined through the relation 
\be
\nu_f = U_{23}I_{\delta} \tilde{\nu}\, . 
\label{eq-05:basisrel}
\ee 
Since the potential matrix
is invariant under 2-3 rotations the matrix of the potentials is unchanged and 
the Hamiltonian the propagation basis becomes
\be
\tilde{H} =  \frac{1}{2E} U_{13} U_{12} M^2_{diag}
U^\dagger_{12} U^\dagger_{13} ~ + ~ \hat{V} ~. 
\label{eq-05:ham-tilde}
\ee
It does not depend on the 2-3 mixing or CP-violation phase,    
and so the dynamics of the flavor evolution does not depend on 
$\delta$ and $\theta_{23}$. These parameters appear in the final amplitudes 
when projecting the flavor states onto propagation basis states 
and back (\ref{eq-05:basisrel}) at the neutrino production and detection.

Explicitly, the Hamiltonian $\tilde{H}$ can be written
\be
\tilde{H} =
\frac{\Delta m_{31}^2}{2E} \times 
\left(\begin{array}{ccc}
s_{13}^2 + s_{12}^2\, c_{13}^2\,r_\Delta + \frac{2V_e E}{\Delta m_{31}^2} &
s_{12}\,c_{12}\,c_{13}\,r_\Delta & s_{13}\,c_{13}(1 - s_{12}^2\,r_\Delta)
\\
\ldots & c_{12}^2\,r_\Delta   &   - s_{12}\,c_{12}\,s_{13}\,r_\Delta    \\
\ldots &  \ldots  &  c_{13}^2 + s_{12}^2\,s_{13}^2\,r_\Delta
\end{array}\right) .
\label{eq-05:matr1}
\ee
Here all the off-diagonal elements contain small parameters 
$r_\Delta$ and/or $s_{13}$. Notice that, for the measured oscillation 
parameters, $s_{13}^2 \sim  r_\Delta$.

\subsubsection{Mixing angles in matter}
 
The Hamiltonian in  \eq~(\ref{eq-05:matr1}) can be diagonalized 
performing several consecutive rotations which 
correspond to developing the perturbation theory in $r_\Delta $. 
After a 1-3 rotation 
\be
\tilde{\nu} = U_{13}(\theta_{13}^m) \nu'
\label{eq-05:primebasis}
\ee
over the angle $\theta_{13}^m$ determined by 
\be
\tan 2\theta_{13}^m = 
\frac{\sin 2 \theta_{13}}{\cos 2 \theta_{13} 
 - \frac{2EV^\prime}{\Delta m_{31}^2}},
\label{eq-05:13mix-matt2}
\quad {\rm where} \quad
V^{\prime} =  \frac{V}{1 - s_{12}^2 r_\Delta},
\ee
the 1-3 element of (\ref{eq-05:matr1}) vanishes.
The expression (\ref{eq-05:13mix-matt2}) differs from that 
for $2\nu$ mixing in matter by a factor  
$(1 - s_{12}^2 r_\Delta)$, which increases the potential and  
deviates from 1 by 
$$
\xi \equiv s_{12}^2 r_\Delta \approx 10^{-2}.
$$
After this rotation the Hamiltonian in the $\nu'$ basis (\ref{eq-05:primebasis}) 
becomes 
\be
H' = \frac{\Delta m_{31}^2}{2E} \times 
\left(\begin{array}{ccc}
h_{11} & s_{12}\,c_{12}\, r_\Delta \cos(\theta_{13}^m - \theta_{13}) & 0
\\
\ldots & c_{12}^2\,r_\Delta   &  s_{12}\,c_{12} r_\Delta \sin(\theta_{13}^m 
- \theta_{13}) \\
\ldots &  \ldots  &  h_{33}
\end{array}\right) ,
\label{eq-05:matr2aa}
\ee
where 
\be
h_{11,33} = \frac{1}{2} \left[(1 + \xi + x) \mp 
\sqrt{[\cos 2\theta_{13} (1 - \xi) - x]^2 +  \sin^2 2 \theta_{13} (1 - \xi)^2}
\right], 
\label{eq-05:h13int}
\ee
and $x \equiv  2EV/\Delta m_{31}^2$. 
For $\xi = 0$, these elements are reduced to the standard $2\nu$ expressions. 
In the limit  of zero density,  $x \rightarrow 0$, 
$h_{11} = \xi = s_{12}^2 r_\Delta$ and consequently  
the 11 element of the Hamiltonian equals $H_{11}' = s_{12}^2 \Delta m_{12}^2/2E$.\\  

In the lowest $r_\Delta$ approximation one can neglect the non-zero 2-3 element in \eq~(\ref{eq-05:matr2aa}). The state $\nu_3'$ then decouples 
and the problem is reduced to a two neutrino problem for $(\nu_1', \nu_2')$. 
The eigenvalue of this decoupled state equals
\be
H_{3m} \approx \frac{\Delta m_{31}^2}{2E} h_{33}, \quad h_{33} \geq 1.
\label{eq-05:h-3m}
\ee
The diagonalization of the remaining 1-2 sub-matrix is given by rotation
\be
\nu' = U_{12}(\theta_{12}^m) \nu_m, 
\ee
where $\theta_{12}^m$ is determined  by 
\be
\tan 2\theta_{12}^m = 
\frac{\sin 2 \theta_{12} r_\Delta \cos(\theta_{13}^m - \theta_{13})}
{c_{12}^2 r_\Delta - h_{11}}.
\label{eq-05:12mix-matt}
\ee
Here $h_{11}$ and $\theta_{13}^m$ are defined in \eqs~(\ref{eq-05:h13int}) and 
(\ref{eq-05:13mix-matt2}), respectively. 
The eigenvalues equal
\be 
H_{1m, 2m} = \frac{\Delta m_{31}^2}{4E} \left[c_{12}^2 r_\Delta + h_{11}
\mp  \sqrt{\left(c_{12}^2 r_\Delta - h_{11}\right)^2  + 
\sin^2 2 \theta_{12} r_\Delta^2  \cos^2(\theta_{13}^m - \theta_{13})}
\right].  
\label{eq-05:12eigenval}
\ee

According to this diagonalization procedure in the lowest order in  
$r_\Delta$ the mixing matrix in matter is given by
\be
U^{m} = 
U_{23}(\theta_{23}) I_\delta U_{13}(\theta_{13}^m) U_{12}(\theta_{12}^m), 
\ee
where mixing angles 
$\theta_{12}^m$ and $\theta_{13}^m$ are determined in \eqs~(\ref{eq-05:12mix-matt}) and (\ref{eq-05:13mix-matt2}), respectively. 
The 2-3 angle and the CP-violation phase are not modified by matter 
in this approximation.  
The eigenvalues $H_{1m}$ and $H_{2m}$ are given in \eq~(\ref{eq-05:12eigenval}) 
and $H_{3m}$ is determined by \eq~(\ref{eq-05:h-3m}). 

The 2-3 element of matrix (\ref{eq-05:matr2aa}) vanishes  after  
additional 2-3 rotation by an angle $\theta_{23}^\prime \sim r_\Delta$:  
\be
\tan 2\theta_{23}' = 
\frac{\sin 2\theta_{12}\,r_\Delta \sin(\theta_{13}^m - \theta_{13})}{
h_{33} - c_{12}^2 r_\Delta} \, ,
\label{eq-05:23rot-add}
\ee
which produces corrections of the next order in $r_\Delta$.
With an additional 2-3 rotation the mixing matrix becomes 
\be
U^{m} =
U_{23}(\theta_{23}) I_\delta U_{13}(\theta_{13}^m) U_{12}(\theta_{12}^m) 
U_{23}(\theta_{23}^\prime) \approx 
U_{23}(\theta_{23}^m) I_{\delta^m} U_{13}(\theta_{13}^m) 
U_{12}(\theta_{12}^m),
\label{eq-05:mixinmatter}
\ee
where 
\be
U_{23}(\theta_{23}^m) I_\delta^m  
= U_{23}(\theta_{23}) I_\delta  U_{23}(\bar{\theta}_{23}) 
\label{eq-05:add23matr} 
\ee
and  the last  2-3 rotation  is on the angle 
$\bar{\theta}_{23}$ determined through  
$\sin \bar{\theta}_{23} = {\sin \theta_{23}'}/{\cos \theta_{13}^m}$.
The expression on the RH of \eq~(\ref{eq-05:mixinmatter}) is obtained 
by reducing the expression on the LH side to the standard form by permuting the correction matrix $U_{23}(\theta_{23}^\prime)$. 
According to \eq~(\ref{eq-05:add23matr}), it is this matrix that leads to the modification of 2-3 mixing and CP phase in matter. 
From \eq~(\ref{eq-05:add23matr}) one finds 
$$
\sin \delta^m  \sin 2 \theta_{23}^m = 
\sin \delta \sin 2 \theta_{23},
$$
{\it i.e.}, the combination  $\sin \delta \sin 2 \theta_{23}$
is invariant under inclusion of matter effects. 
Furthermore, $\theta_{23}^m \approx \theta_{23} $ 
and $\delta^m \approx \delta$ up to corrections of the order $O(r_\Delta)$. 
The results described here allow to understand behavior of the mixing 
parameters $\sin^2 2\theta_{m ij}$ in the $EV$ region of the 1-3 resonance 
and above it (see \Fig~\ref{fig-05:angles_normal}).

\begin{figure}[h!!]
\begin{center}
\vspace{1cm}
\includegraphics[height=5.5cm,angle=0]{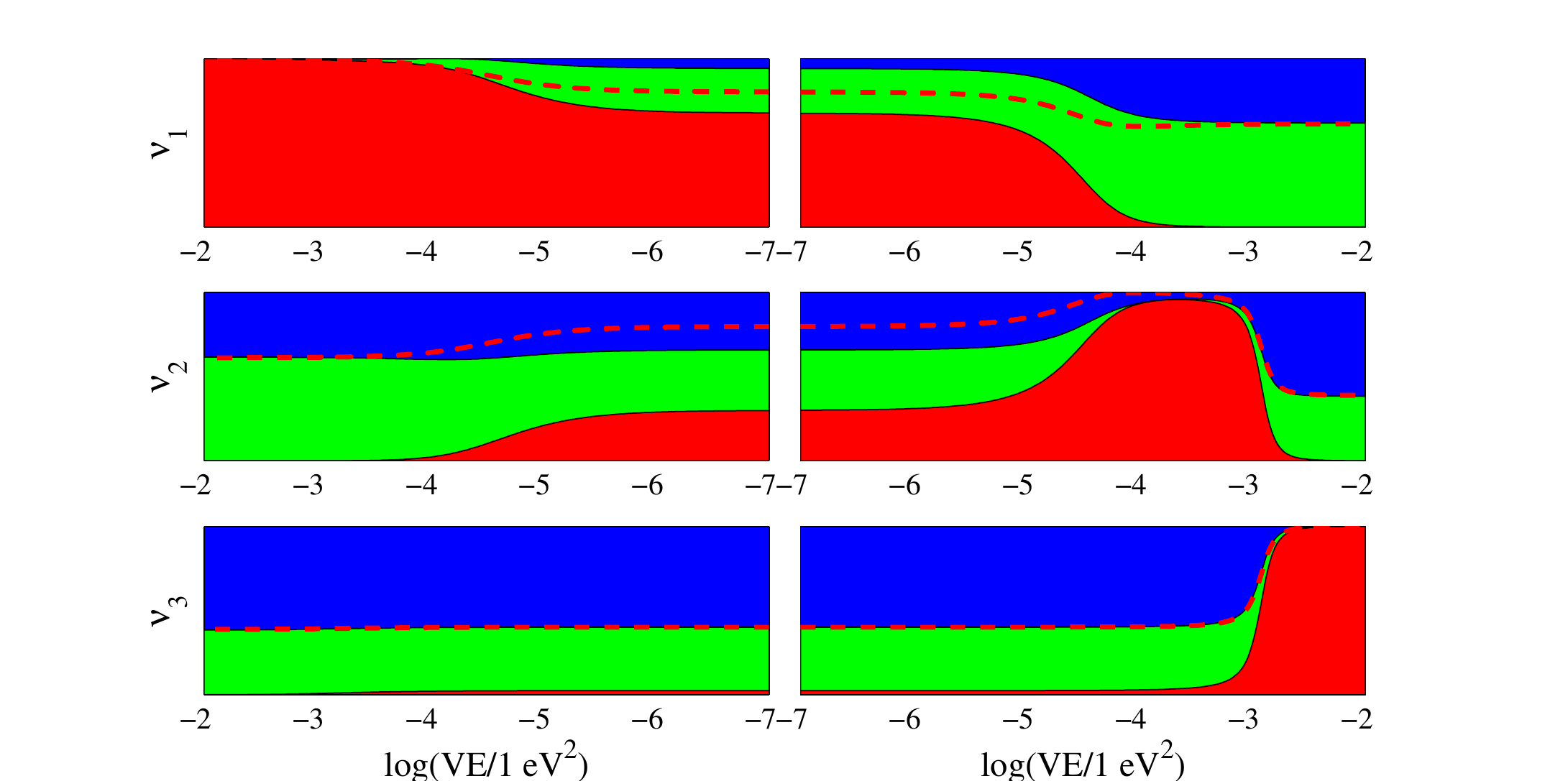} \\
\includegraphics[height=5.5cm,angle=0]{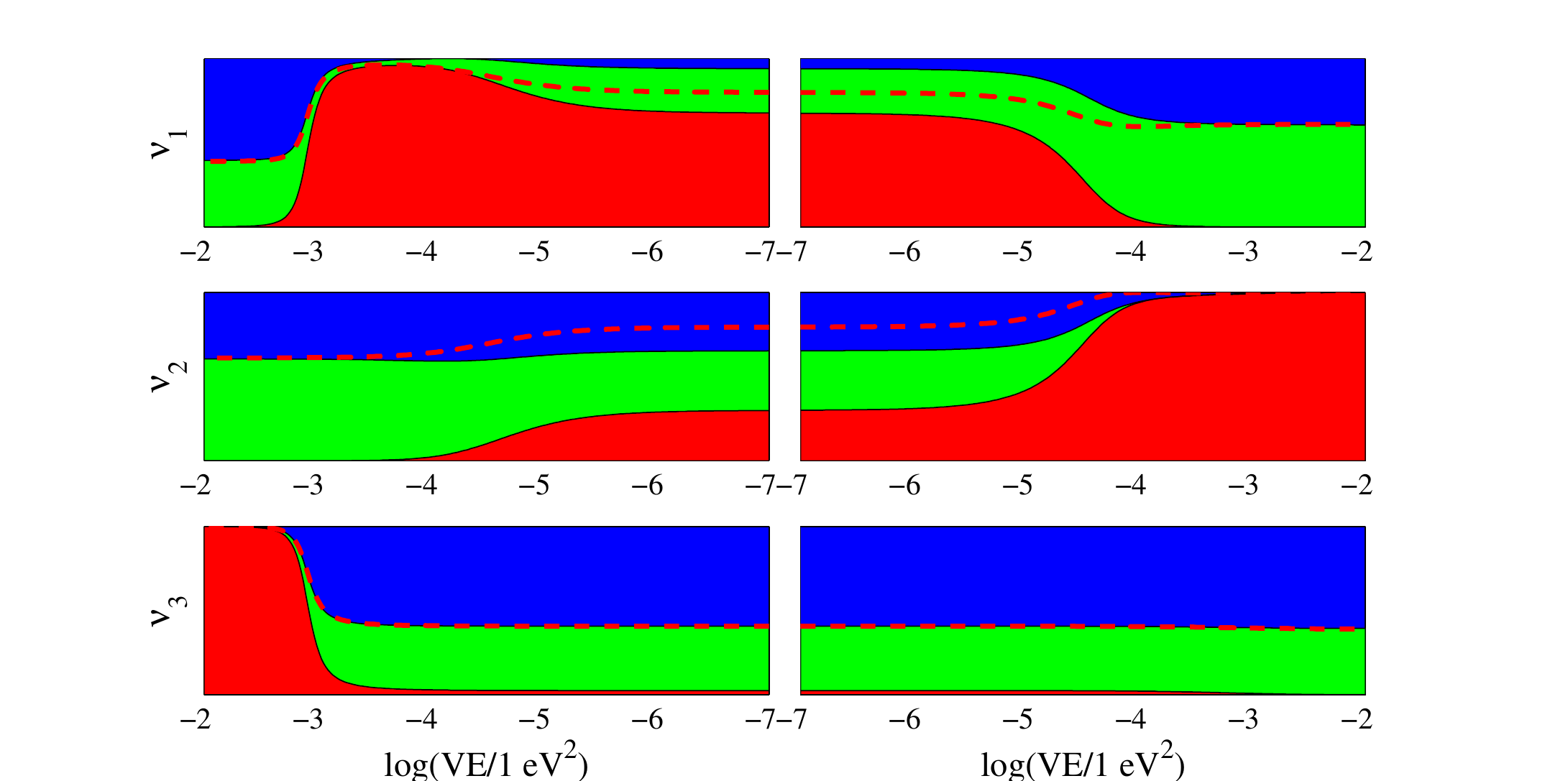}
\caption{The flavor contents of the eigenstates of the 
Hamiltonian in matter as functions of $EV$. The vertical width of the 
band is taken to be 1, then the vertical sizes of the colored parts give  
$|U_{ei}|^2$ (red), $|U_{\mu i}|^2$ (green), $|U_{\tau i}|^2$ (blue). 
The right and left panels correspond to neutrinos and anti-neutrinos, respectively. 
We take the best fit values of \cite{05-Fogli:2012ua} with $\delta = 0$. 
Variations of  $\delta$ change the relative $\nu_\mu-$ and $\nu_\tau-$ content. 
The dashed red line shows a shift of  border between $\nu_\mu-$ and $\nu_\tau-$ 
flavors for $\delta = \pi$. The upper (lower) panel corresponds to normal (inverted) mass ordering.\label{fig-05:nucontent}}
\end{center}
\end{figure}

\begin{figure}[h!!]
\begin{center}
\vspace{1cm}
\includegraphics[height=5cm,angle=0]{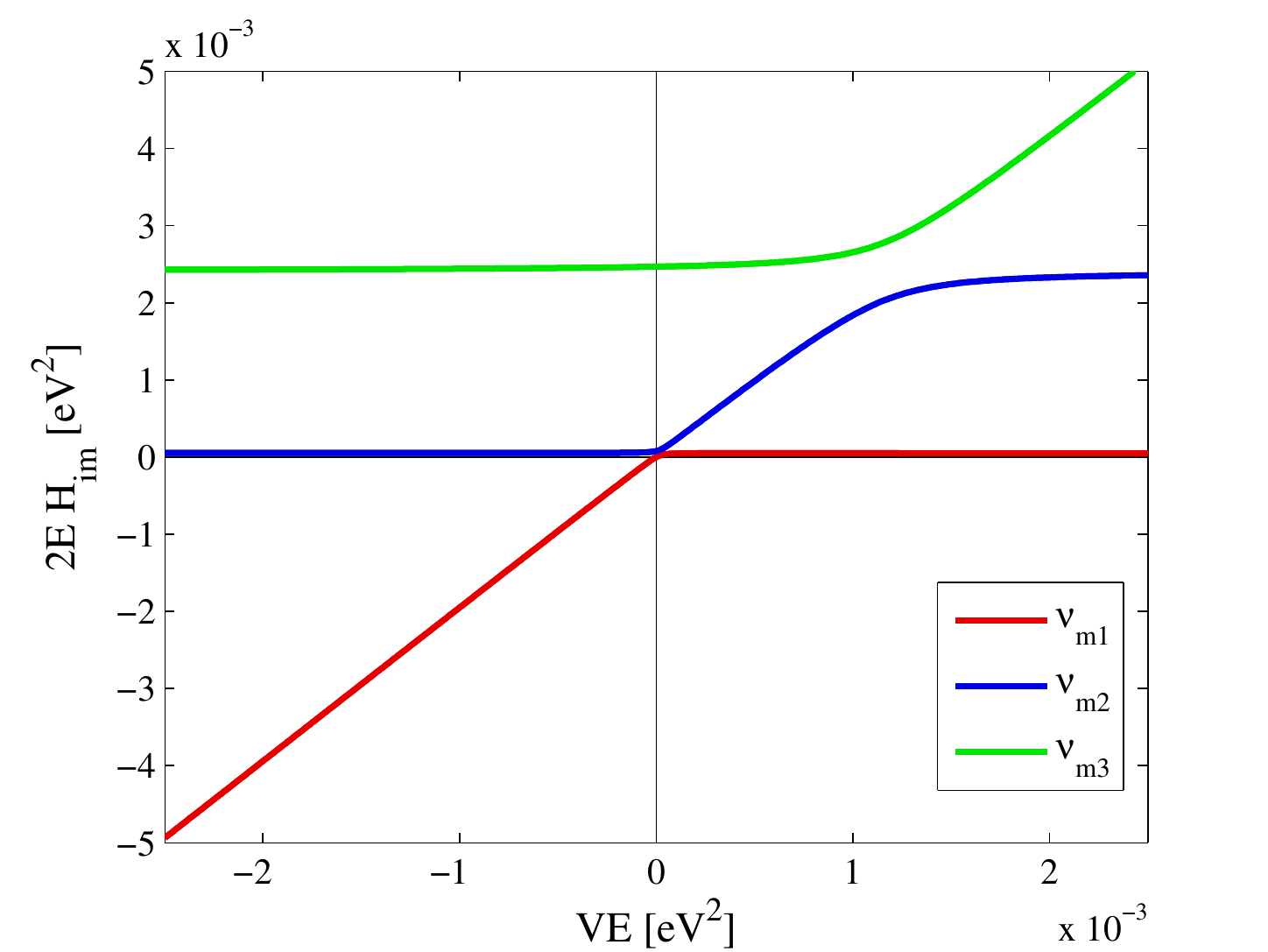}
\includegraphics[height=5cm,angle=0]{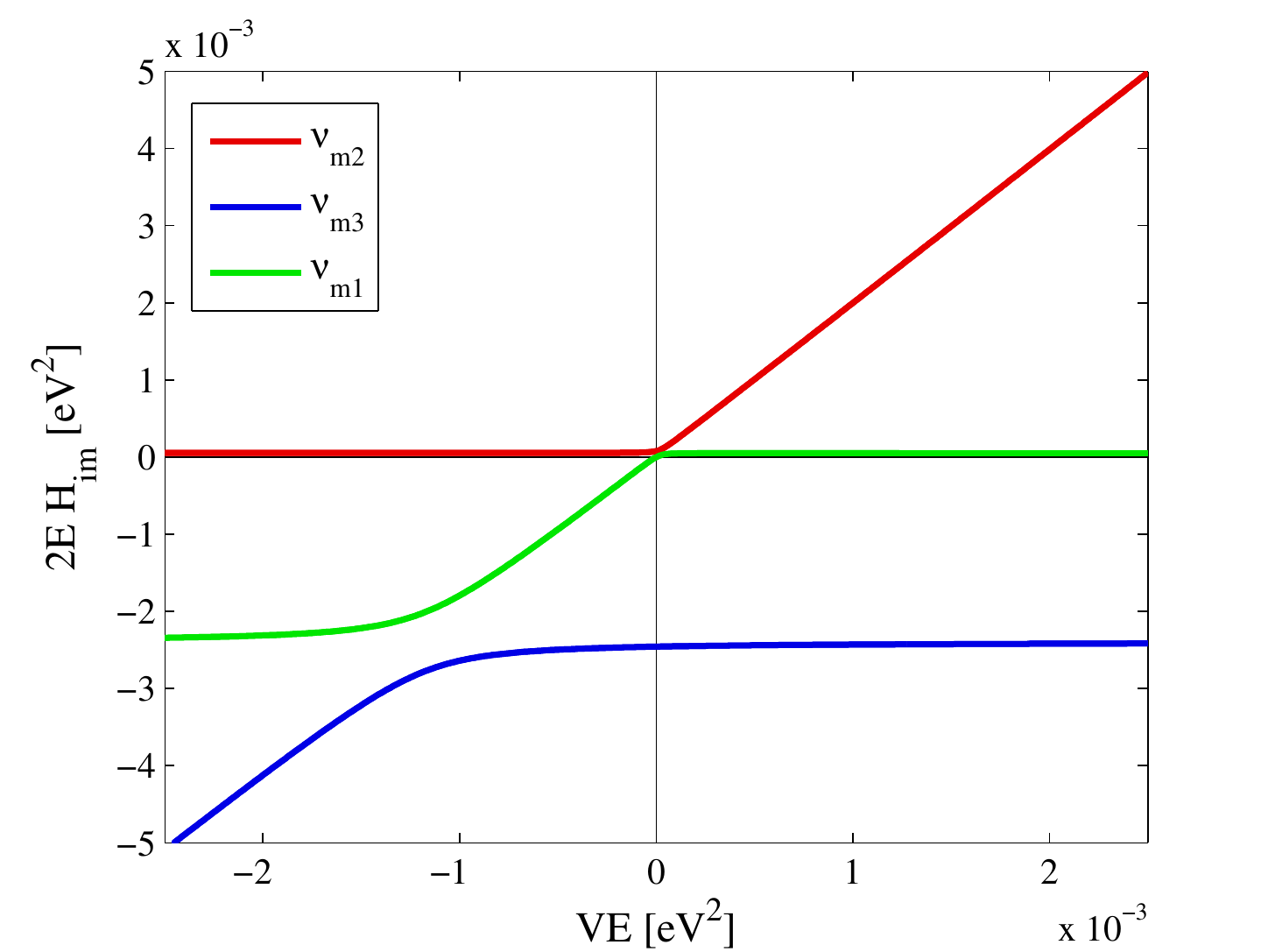}
\caption{The energy level scheme. We here show the dependence of the eigenvalues 
of the Hamiltonian in matter on $EV$. Note that we are plotting $2EH_{im}$, which goes to $\Delta m_{i1}^2$ for low $VE$.
The left (right) panel corresponds to normal (inverted) mass ordering.
\label{fig-05:levelcrossing}}
\end{center}
\end{figure}

In \Fig~\ref{fig-05:nucontent} we present dependence of the flavor content of the neutrino 
eigenstates on the potential. 
The energy level scheme, the dependence of the eigenvalues $H_{im}$ 
on matter density, is shown in \Fig~\ref{fig-05:levelcrossing}. 
The energy levels in matter do not depend on 
$\delta$ or $\theta_{23}$, but they do depend on the 1-3 and 1-2 mixing.

In the case of normal mass hierarchy, there are two resonances (level crossings).  
whose location is defined as the density (energy) at which 
the mixing in a given channel becomes maximal. 

1. The H-resonance, in the $\nu_e - \nu_\tau '$ channel, 
is associated to the 1-3 mixing and large mass splitting. According to \eq~(\ref{eq-05:13mix-matt2}) 
$\theta_{13}^m = \pi/4$ at 
\be
V_{13}^R =  \cos 2 \theta_{13}(1 - s_{12}^2 r_\Delta) 
\frac{\Delta m_{31}^2}{2E} \,.
\label{eq-05:13mix-res}
\ee 

2. The L-resonance at low densities  is associated to the 
small mass splitting and  1-2 mixing 
It appears in the $\nu_e' - \nu_\mu'$ channel, where 
$\nu_e'$ and  $\nu_e$ differ
by small (at low densities) rotation given by an angle $\sim \theta_{13}$ 
(see eq. (\ref{eq-05:13mix-matt2})). 
According to \eq~(\ref{eq-05:12mix-matt}) the position of the L-resonance, $\theta_{12}^m = \pi/4$ is given by $c_{12}^2 r_\Delta = h_{11}$,
where $h_{11}$ is defined in \eq~(\ref{eq-05:h13int}). This leads to  
\be
V^R_{12} = \cos 2\theta_{12} \frac{\Delta m^2_{21}}{2E} \frac{1}{c_{13}^2}.
\label{eq-05:12mix-res}
\ee 

For antineutrinos ($VE < 0$  in Fig. \ref{fig-05:levelcrossing}), 
the oscillation parameters in matter can be obtained from the 
neutrino parameters taking
$V \rightarrow -V$ and $\delta \rightarrow - \delta$. 
The mixing pattern and level scheme for neutrinos 
and antineutrinos are different 
both due to the possible fundamental violation of CP-invariance and 
the  sign of matter effect. Matter violates CP-invariance 
and the origin of this violation stems from the fact that 
usual matter is CP-asymmetric: in particular,  
there are electrons in the medium but no positrons.

In the case of normal mass hierarchy there is no antineutrino 
resonances (level crossings), and with the increase
of density (energy) the eigenvalues have the following asymptotic
limits:
\begin{equation}
    \label{eq-05:anlevels}
    H_{1m} \to -V \,, \qquad
    H_{2m} \to \frac{\Delta m^2_{21} c_{12}^2}{2E_\nu} \,, \qquad
    H_{3m} \to \frac{\Delta m^2_{31} c_{13}^2}{2E_\nu} \,.
\end{equation}
%

\section{Effects of neutrino propagation in different media}
\label{sec-05:neutrinopropagationindifferentmedia}

\subsection{The evolution matrix}
\label{sec-05:evolutionmatrix} 

The evolution matrix, $S(t, t_0)$, is defined as the matrix 
which gives the wave function of the neutrino  
system $\nu(t)$ at an arbitrary moment $t$ once it is known in the 
initial moment $t_0$:
\be
\nu(t) = S(t, t_0) \nu(t_0) . 
\label{eq-05:evolmat}
\ee
Inserting this expression in the evolution equation (\ref{eq-05:matt2}),
we find that $S(t, t_0)$ satisfies the same evolution equation as 
$\nu(t)$: 
\be
i\frac{d S }{dt} = HS. 
\label{eq-05:eqevmatr}
\ee
The elements $S(t, t_0)_{\alpha\beta}$ of this matrix are the amplitudes of 
$ \nu_\beta    \rightarrow \nu_\alpha$  transitions: $S(t, t_0)_{\alpha\beta} \equiv A(\nu_\beta \rightarrow \nu_\alpha)$.
The transition probability equals $P_{\alpha\beta} = |S(t, t_0)_{\alpha\beta}|^2$.
The unitarity of the evolution matrix, $S^{\dagger} S = I$,  
leads to the following relations between the amplitudes 
(matrix elements) 
\be
|S_{\alpha \alpha}|^2 + |S_{\beta \alpha}|^2 = 1, \quad 
|S_{\beta \beta}|^2 + |S_{\alpha \beta}|^2 = 1, \quad
S_{\alpha \alpha}^* S_{\alpha \beta} + 
S_{\beta \alpha}^* S_{\beta \beta} = 0, \quad
S_{\alpha \beta}^* S_{\alpha \alpha} + 
S_{\beta \beta}^* S_{\beta \alpha} = 0.
\ee
The first and the second equations express 
the fact that the total probability of transition 
of $\nu_\alpha$ to everything is one, and the same holds for 
$\nu_\beta$. The third and fourth equations are satisfied if 
\be
S_{\alpha \alpha} = S_{\beta \beta}^*, ~~~ 
S_{\beta \alpha} = - S_{\alpha \beta}^* . 
\label{eq-05:relationss}
\ee
With these relations the evolution matrix can be parametrized as   
\be
S = 
\left(
\begin{array}{cc}
\alpha  & \beta \\
- \beta^*  &   \alpha^*
\end{array}
\right), ~~~~
|\alpha|^2 + |\beta|^2 =1. 
\label{eq-05:Hvacuum2}
\ee

The Hamiltonian for a $2\nu$ system is T-symmetric  in vacuum as well as in medium 
with constant density. In medium with varying density the T-symmetry 
is realized if the potential is symmetric. 
Under T-transformations 
$
S_{\beta \alpha} \rightarrow S_{\alpha \beta}, 
$
and the diagonal elements $S_{\alpha \alpha}$ do not change. 
Therefore according to (\ref{eq-05:relationss}) the T-invariance implies that  
$
S_{\beta \alpha} = - S_{\beta \alpha}^*, 
$ 
or 
$
{\rm Re}~ S_{\beta \alpha} = 0,$  i.e., the off-diagonal elements of the $S$ matrix are pure imaginary.

\subsection{Neutrino oscillations in matter with constant density}
\label{sec-05:constant} 

In a medium with constant density and therefore constant potential the mixing is constant: $\theta_m (E, n) = {\rm const}$.  Consequently,  the flavor composition of the eigenstates do 
not change and the eigenvalues $H_{im}$ of the full Hamiltonian are
constant.
The two neutrino evolution equation in matter of constant density can be written in the matter eigenstate basis as
\be
i \frac{d\nu_m}{dx} = H^{diag} \nu_m , 
\label{eq-05:ham-eig-const}
\ee
where $H^{diag} \equiv \diag(H_{1m}, H_{2m})$. This system of equations splits and 
the integration is trivial, $\nu_{im}(t) = e^{- iH_{im} t} \nu_{im}(0)$. The corresponding $S$-matrix is diagonal:
\be 
\tilde{S}(x,0) = 
\left(
\begin{array}{cc}
e^{i \phi_m (x)}  & 0 \\
0             &  e^{-i \phi_m (x) } 
\end{array}
\right), 
\label{eq-05:s-tild-cs}
\ee
where $\phi_m \equiv \frac{1}{2}\omega^m x$
is the half-oscillation phase in matter and a matrix proportional to the 
unit matrix has been subtracted from the Hamiltonian.  

\begin{figure}[h!!]
\begin{center}
\vspace{1cm}
 \includegraphics[height=6cm,angle=0]{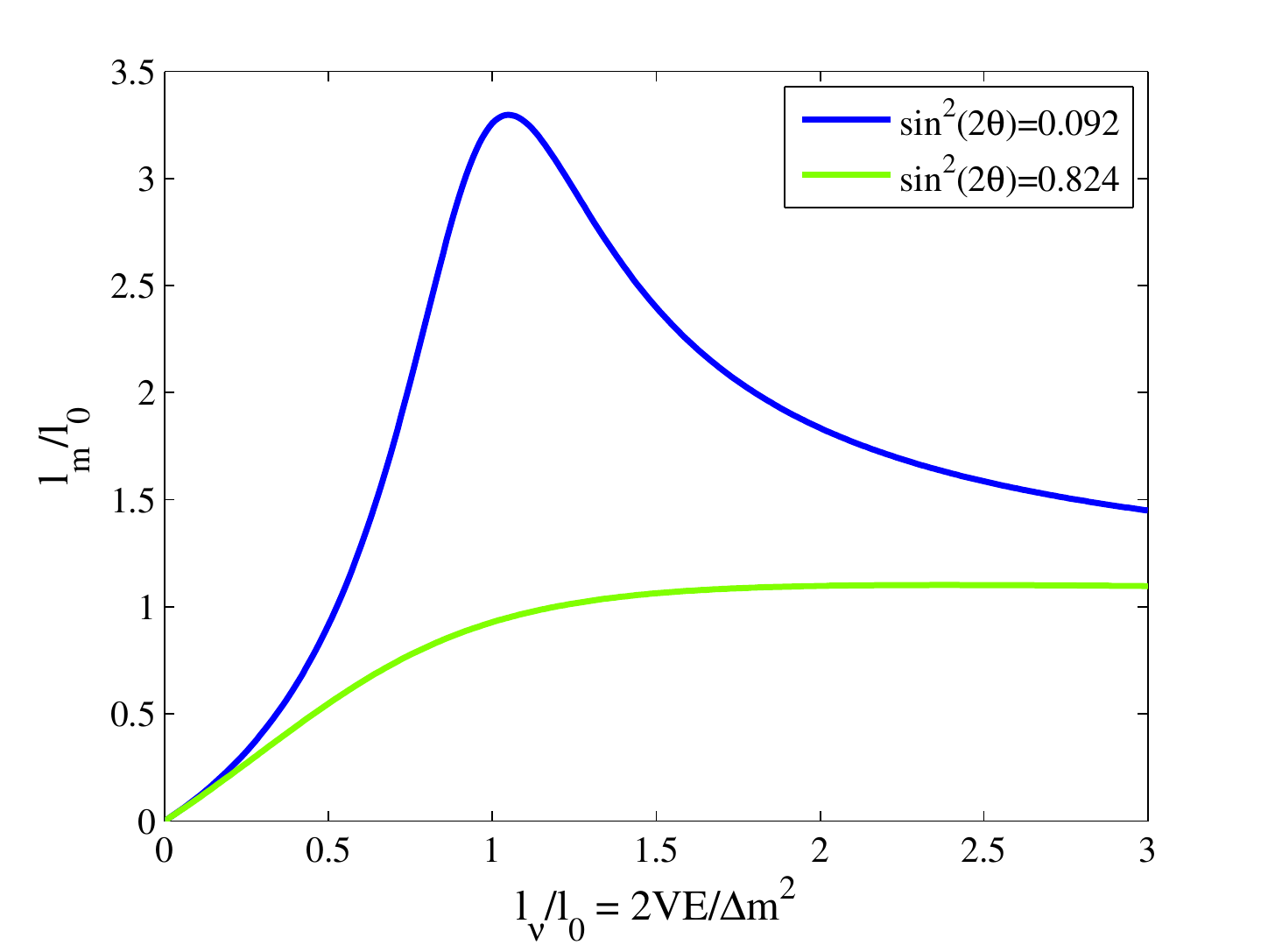}
\caption{Dependence of the oscillation length in matter in units of the refraction length  
on neutrino energy for two different mixing angles in vacuum.  
\label{fig-05:lmdependence}}
\end{center}
\end{figure}

The $S$ matrix in the flavor basis $(\nu_e, \nu_a)$ is therefore
\be 
S(x, 0) = U^m \tilde{S}(x,0)U^{m\dagger} =  
        \begin{pmatrix}
        \cos\phi_m + i \cos 2 \theta_m \, \sin \phi_m
        & -i \sin2 \theta_m \, \sin\phi_m \cr
        -i \sin2 \theta_m \, \sin \phi_m
        & \cos\phi - i \cos2 \theta_m \, \sin \phi_m
    \end{pmatrix}. 
\label{eq-05:Ubar}
\end{equation}
Then, for the transition probability, we can immediately deduce
\be
P_{ea} = |S_{ea}|^2 = \sin^2 2\theta_m \sin^2 \phi_m,  
\label{eq-05:matosc}
\ee
where $\phi_m = {\pi x}/{l_m}$
with 
\be
l_{m} = \frac{2\pi}{H_{2m} - H_{1m}} = \frac{l_\nu}{\sqrt{R}}
\ee
being the oscillation length in matter. 
The dependence of $l_m$ on the neutrino energy is shown 
in \Fig~\ref{fig-05:lmdependence}. For small energies, 
$VE \ll \Delta m^2$, the length $l_m \simeq l_\nu$. 
It then increases with energy and for small $\theta$ reaches the maximum  
$l_m^{max} = l_0/\sin 2\theta$ at $E^{max} = E_R/\cos^2 2\theta$, 
{\it i.e.}, above the resonance energy. 
For $E \rightarrow \infty$  the oscillation length 
converges to the refraction length $l_m \rightarrow l_0$.
%
%

A useful representation of the $S$ matrix for a layer with constant density   
follows from \eq~(\ref{eq-05:Ubar}):  
\be
S(x, 0) = \cos \phi_m I - i \sin  \phi_m ({\bf \sigma} \cdot {\bf n}), 
\label{eq-05:sone-lay}
\ee 
where ${\bf \sigma}$ is a vector containing the Pauli matrices 
and ${\bf n} \equiv (\sin 2\theta_{m }, 0, - \cos 2\theta_{m})$.  

The dynamics of neutrino flavor evolution in uniform matter are the same as in vacuum, 
{\it i.e.}, it has a  character of oscillations.
However, the oscillation parameters (length and depth) differ from those in vacuum.
They are now determined by the mixing 
and effective energy splitting in  matter:
$\sin^2 2\theta   \rightarrow  \sin^2 2\theta_m$, 
$l_{\nu} \rightarrow l_m $.

\subsection{Neutrino polarization vectors and graphic representation}
\label{sec-05:polarizationvectors} 

It is illuminating to  consider dynamics of transitions in different
media using graphic representation \cite{05-Smirnov:1986ij,05-Bouchez:1986kb,05-Ermilova:1986ab}.
Consider the two flavor neutrino state,  
$\psi^T = (\psi_e, \psi_a)$.  The corresponding Hamiltonian can be written as 
\be
H = ({\bf H} \cdot {\bf \sigma}),
\label{eq-05:Hvector}
\ee
where  ${\bf \sigma} = (\sigma_1, \sigma_2, \sigma_3)$,  ${\bf H}$ is the Hamiltonian vector 
${\bf H} \equiv ({2 \pi}/{l_m}) \cdot (\sin 2 \theta_m, 0, \cos 2 \theta_m)$ and 
$l_m = 2\pi/\Delta H_m$ is the oscillation length. 
The evolution equation then becomes
\be
i\dot{\psi} = ({\bf H} \cdot {\bf \sigma})~ \psi .
\label{eq-05:evolpsivec}
\ee
Let us define the polarization vector ${\bf P}$ 
\be
{\bf P} \equiv \psi^{\dagger} \frac{\bf \sigma}{2} \psi .
\label{eq-05:polvector}
\ee
In terms of the wave functions, the components of ${\bf P}$ equal
\be
(P_x, P_y, P_z) = \left({\rm Re}~ \psi_e^* \psi_a,\ {\rm Im}~  \psi_e^* \psi_a,\  
\frac{1}{2}\left(|\psi_e|^2 - |\psi_a|^2\right)\right).
\ee
The $z$-component can be rewritten as $P_z = |\psi_e|^2 - 1/2$, therefore 
$P_{e} \equiv  |\psi_e|^2  = P_z + 1/2$ and from unitarity $P_{a} \equiv  |\psi_a|^2  = 1/2 - P_z$. 
Hence, $P_z$ determines the probabilities to find the neutrino of 
in a given flavor state. The  flavor evolution of the neutrino state corresponds to  a motion
of the polarization vector in the flavor space.
The evolution equation for ${\bf P}$ can be obtained by
differentiating \eq~(\ref{eq-05:polvector}) with respect to time and inserting $\dot{\psi}$
and $\dot{\psi}^{\dagger}$ from evolution equation (\ref{eq-05:evolpsivec}).
As a result, one finds that
\be
\frac{d}{dt}{\bf P} = {\bf H} \times {\bf P}. 
\label{eq-05:evoleqpol}
\ee
If ${\bf H}$ is identified with the strength of a magnetic field, 
the equation of motion (\ref{eq-05:evoleqpol}) coincides with the equation 
of motion for the spin of electron in the magnetic field. 
According to this equation ${\bf P}$ precesses around ${\bf H}$.  

\begin{figure}[h!!]
\begin{center}
\vspace{1cm}
\includegraphics[height=4.5cm,angle=0]{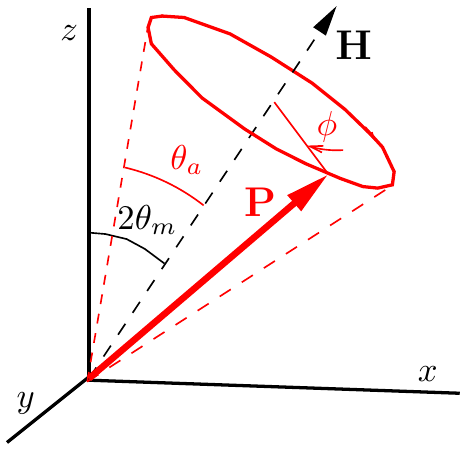}
\caption{Graphic representation of neutrino oscillations. 
Neutrino polarization vector ${\bf P}$ precesses around the Hamiltonian vector 
${\bf H}$ (or the vector of eigenstates of the Hamiltonian). 
The angle between ${\bf P}$ and ${\bf H}$  is given by the 
cone angle $\theta_a$, and the direction of 
axis of the cone is determined by the mixing angle 
in matter $2\theta_m$. 
\label{fig-05:polarization}}
\end{center}
\end{figure}

With an increase of the oscillation phase $\phi$ (see \Fig~\ref{fig-05:polarization})  
the vector ${\bf P}$  moves on the surface of the cone having  axis ${\bf H}$.
The cone angle $\theta_a$, the angle between ${\bf P}$ and
${\bf H}$ depends both on the mixing angle and  on the initial state, and in general, changes
in process of evolution, e.g., if the neutrino evolves through several layers of different density. 
If the  initial state is $\nu_{e}$, 
the angle equals  $\theta_{a} =  2\theta_m$ in the initial moment.

The components of the polarization vector ${\bf P}$
are nothing but the elements of the density matrix
$\rho =  {\bf \sigma} \cdot {\bf P}$. The evolution equation for $\rho$ can be obtained
from (\ref{eq-05:evoleqpol})
\be
i\frac{d\rho}{dt} = [H, \rho].
\ee
The diagonal elements of the density matrix give the probabilities to 
find the neutrino in the corresponding flavor state.

\subsection{Resonance enhancement of oscillations}

\begin{figure}[h!!]
\begin{center}
\vspace{1cm}
\includegraphics[height=6cm,angle=0]{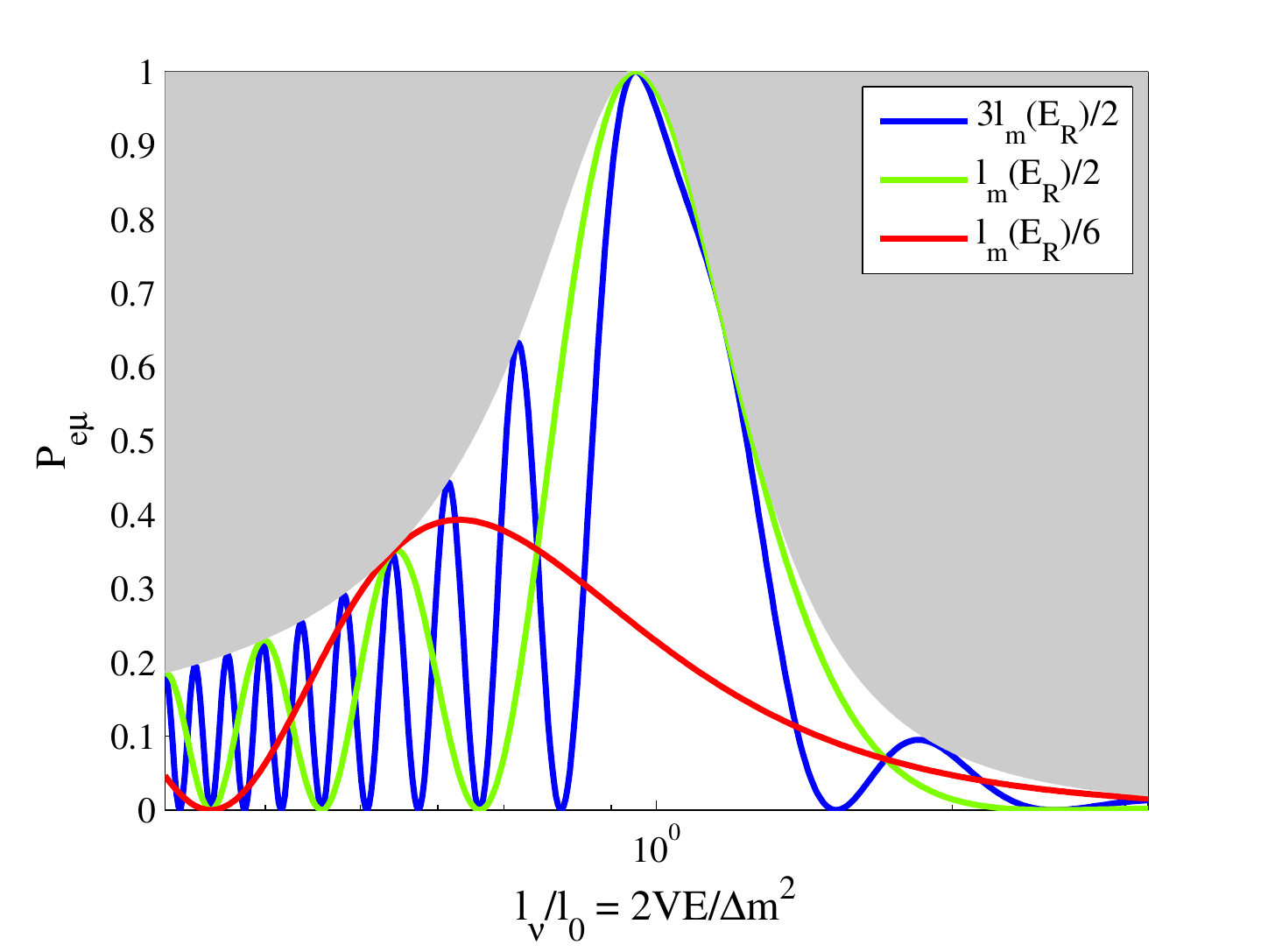}
\caption{Resonance enhancement of neutrino 
oscillations in matter with constant density. Shown is the dependence 
of the transition probability $\nu_e \rightarrow \nu_\mu$  on 
energy for $\sin^2 \theta_{13} = 0.0241$ 
for three different sizes of  layers: $L = 3 l_m (E_R)/2$, $l_m(E_R)/2$ and 
$l_m (E_R)/6$. The shaded area shows the resonance envelope: $\sin^2 2\theta_m (E)$. 
\label{fig-05:resonance}}
\end{center}
\end{figure}

Suppose a source produces flux of neutrinos in the flavor state $\nu_\mu$ 
with continuous energy spectrum.  This flux  
then traverse a layer of length $L$ with constant density $n_e$. 
At the end of this layer a detector measures the $\nu_e$ component 
of the flux, so that  oscillation effect is given by 
the transition probability $P_{\mu e}$. In \Fig~\ref{fig-05:resonance} we show  dependence of this probability on energy for thin and thick layers.
The oscillatory curves are inscribed in to the resonance envelope $\sin^2 2\theta_m$.
The period of the oscillatory curve  
decreases with the length $L$. At the resonance energy,
\be
E_R = \frac{\Delta m^2 \cos 2\theta }{2V} = 
\frac{\Delta m^2 \cos 2\theta }{2\sqrt{2} G_F n_e}, 
\ee
oscillations proceed with maximal depths.
Oscillations are enhanced up to $P > 1/2$ in the resonance range 
$(E_R \pm \Delta E_R)$ where  $\Delta E_R = \tan 2 \theta E_R$ 
(see \Sec~\ref{sec-05:resonance}). This effect was called the resonance enhancement of oscillations. 

\subsection{Three neutrino oscillations in matter with constant density}

The oscillation probabilities in matter with constant density 
have the same form as oscillation probabilities in vacuum and
the generalization of \eq~(\ref{eq-05:s-tild-cs}) is straightforward. In the basis of the eigenstates of  the Hamiltonian the evolution matrix equals 
\be
\tilde{S}(x,0) =
\left(
\begin{array}{ccc}
e^{- 2i\phi_{1m} (x)}  & 0             &  0 \\
0             &  e^{-2 i \phi_{2m} (x)} &  0 \\
0     &  0    &  e^{-2 i \phi_{3m} (x)}
\end{array}
\right),
\label{eq-05:s-tild3-cs}
\ee
and for the elements of the $S$ matrix in the flavor basis we obtain 
$S_{\alpha \beta} = \sum_i U_{\alpha i}^{m*} U_{\beta i}^{m} e^{-2 i \phi_i^m (x)}$.
Removing $e^{-2 i \phi_{2m}}$ and using the unitarity of the mixing matrix 
in matter we have 
\be
S_{\alpha \beta} = 
\delta_{\alpha \beta} +  
2 i e^{\phi^{m}_{21}(x)} U_{\alpha 2}^{m*} U_{\beta 2}^{m} \sin \phi^{m}_{21}(x)  
- 2 i e^{- i \phi^{m}_{32}(x)} U_{\alpha 3}^{m*} U_{\beta 3}^{m} 
\sin \phi^{m}_{32}(x). 
\label{eq-05:amplinmatt}
\ee
In particular, for the amplitudes in matter involving only $\nu_e$ and $\nu_\mu$,
we obtain 
\begin{eqnarray}
    \label{eq-05:me-ampl}
    S_{e \mu}^{\rm cst} &=& 2 i \, e^{i \phi_{21}^m}
    \left[U_{e1}^{m} U_{\mu 1}^{m *} \sin\phi_{21}^m 
    - e^{-i \phi_{31}^m} U_{e3}^{m} U_{\mu 3}^{m *} \sin\phi_{32}^m
    \right] \, ,\\
    \label{eq-05:mm-ampl}
    S_{\mu\mu}^{\rm cst} &=& 1
    + 2i\, e^{i \phi_{21}^m} |U_{\mu 1}^m|^2 \sin\phi_{21}^m
    - 2i\, e^{-i \phi_{32}^m} |U_{\mu 3}^m|^2 \sin\phi_{32}^m \,.\\
    \label{eq-05:ampleee}
    S_{ee}^{\rm cst} &=& 1 + 2i\, e^{i \phi_{21}^m} \cos^2 \theta_{13}^m
    \cos^2 \theta_{12}^m \sin \phi_{21}^m
    - 2i\, e^{-i \phi_{32}^m} \sin^2 \theta_{13}^m \sin\phi_{32}^m.
\end{eqnarray}

[[do we use this? add more?]]

\subsection{Propagation in a medium with varying density and the MSW effect}

\subsubsection{Equation for the instantaneous eigenvalues and the adiabaticity condition}
In non-uniform media, the density changes along neutrino trajectory:
$n_e = n_e(t)$. Correspondingly, the Hamiltonian of  system depends on time,
$H = H(t)$, and  therefore 
the mixing angle changes during neutrino propagation:
$\theta_m = \theta_m (n_e(t))$. 
Furthermore, the eigenstates of the instantaneous Hamiltonian,
$\nu_{1m}$ and  $\nu_{2m}$, are no longer the
``eigenstates'' of propagation. Indeed, 
inserting  $\nu_f = U (\theta_m) \nu_m$ in the  equation for the
flavor states [c.f., \eq~(\ref{eq-05:evolution})] we obtain the evolution 
equation for eigenstates $\nu_{im}$ 
\be
i\frac{d\nu_m}{dt} =
\left(\begin{array}{cc}
H_{1m}  &  - i \dot{\theta}_m \\
i \dot{\theta}_m  &  H_{2m}
\end{array}
\right)\nu_m , 
\label{eq-05:eqmatt2a}
\ee 
where $\dot{\theta}_m \equiv d\theta_m/dt$. 
The Hamiltonian for 
$\nu_{im}$  (\ref{eq-05:eqmatt2a}) is non-diagonal, and consequently, the transitions 
$\nu_{1m} \leftrightarrow \nu_{2m}$ occur. 
The rate of these transitions 
is given by the speed with which the mixing angle changes with time.  
According to \eq~(\ref{eq-05:eqmatt2a})  \cite{05-Mikheev:1986gs,05-Messiah:1986fc},
$|\dot{\theta}_m|$ determines the energy of transition
$\nu_{1m} \leftrightarrow \nu_{2m}$ and  $|H_{2m}  - H_{1m}|$
gives the energy gap between the levels.

The off-diagonal elements of the evolution equation \eq~(\ref{eq-05:eqmatt2a}) can be neglected 
if  $\dot{\theta}_m$  is much smaller than other energy scales in the system. 
The difference of the diagonal elements of the Hamiltonian is, in fact, the only other energy 
quantity and therefore the criterion for smallness of $\dot{\theta}_m$ is 
\be
\dot{\theta}_m  \ll H_{2m}  - H_{1m}.
\label{eq-05:adiab0}
\ee
This inequality implies a slow enough change of density and is called the \emph{adiabaticity condition}. 
%
Defining  the adiabaticity parameter as \cite{05-Messiah:1986fc,05-Smirnov:1986ij} as 
\be
\gamma \equiv \left| \frac{\dot{\theta}_m}{H_{2m}  - H_{1m}} \right|.
\label{eq-05:adiab}
\ee
the adiabaticity condition can be written as  $\gamma \ll 1$.  

For small mixing angle, the adiabaticity condition is most crucial in the resonance layer
where the level splitting is small and the mixing angle changes rapidly.
In the resonance point, it takes the physically transparent form \cite{05-Mikheev:1986gs}: 
$\Delta r_R >  l_m^R$, where 
$l_m^R \equiv {l_{\nu}}/{\sin 2\theta}$ is the oscillation length in resonance,
and $\Delta r_R \equiv \left(\frac{n_e}{dn_e/dr}\right)_R \tan 2\theta$
is the spatial width of
the resonance layer. According to this condition at least one
oscillation length should be obtained within the resonance layer.

In the case of large vacuum mixing, the point of maximal adiabaticity
violation~\cite{05-Lisi:2000su,05-Friedland:2000rn} is shifted
to density, $n_e(av)$, larger than the resonance density:
$n_e(av) \rightarrow n_B > n_R$. Here
$n_B = \Delta m^2 /2\sqrt{2} G_F E$ is the density at  the border of 
resonance layer for maximal mixing. 
Outside the resonance and in the non-resonant channel, the adiabaticity condition has
been considered in~\cite{05-Smirnov:1993ku,05-Minakata:2000rx}.

\subsection{Adiabatic conversion and the MSW effect}

If the adiabaticity condition if fulfilled and $\dot{\theta}_m$ can be neglected, 
the Hamiltonian for the eigenstates becomes diagonal. Consequently, the equations for the 
instantaneous eigenstates $\nu_{im}$ split  
as in the constant density case. The instantaneous eigenvalues evolve independently, 
but the flavor content of the eigenstates changes according to the change of mixing in matter. 
This is the essence of the adiabatic approximation: We neglect $\dot{\theta}_m$ in evolution equation but do not neglect the dependence of $\theta_m$ on density.
The solution can be obtained immediately as 
\be
\tilde{S}(x, 0) = 
\left(
\begin{array}{cc}
e^{i \phi_{m}}  & 0 \\
0             &  e^{-i \phi_m} 
\end{array}
\right), \qquad \phi_m = \frac{1}{2} \int_0^x (H_{2m} - H_{1m}) d x'. 
\label{eq-05:s-tilde1symm}
\ee
in symmetric form.
The only difference from the constant density case is that 
the eigenvalues now depend on time and therefore integration 
appears in the phase factors.

The evolution matrix in the flavor basis can be obtained by 
projecting back from the eigenstate basis to the flavor basis with the 
mixing matrices corresponding to initial and final densities: 
\be
S_f(x, 0) = U^m(t) \tilde{S}(x,0)U^{m\dagger}(0)
=
\left(
\begin{array}{cc}
c_m c_m^0 e^{ i \phi_m} + s_m s_m^0 e^{ - i\phi_m}  &  
-  c_m s_m^0 e^{i \phi_m} + s_m c_m^0 e^{- i\phi_m}   \\
- s_m c_m^0 e^{i \phi_m} +  c_m s_m^0 e^{- i\phi_m}  & 
s_m s_m^0 e^{i \phi_m} + c_m c_m^0 e^{- i\phi_m}
\end{array}
\right). 
\label{eq-05:s-ad}
\ee
{ }From this procedure we find, e.g., the probability of 
$\nu_e - \nu_e$ transition  
\be
P_{ee}  =  |S_f (x, 0){}_{ee}|^2 = \frac{1}{2} 
\left[ 1 + \cos 2\theta_m(x) \cos 2\theta_m(0)\right] 
 + \frac{1}{2} \sin 2\theta_m(x) \sin 2\theta_m(0) \cos 2\phi_m(x). 
\label{eq-05:ee-prob-ad}
\ee
If the initial and final densities coincide, 
as in the case of neutrinos crossing the Earth, we obtain the same formulas 
as in constant density case: 
\be
P_{\alpha \beta} = 
|\sum_i U_{\alpha i}^m (0) U_{\beta i}^{m*}(0) e^{-i \phi_{im}(t, 0)}|  
\ee
with the mixing angle taken at the borders (initial or final state). 
In particular, the survival probability equals 
$P_{\alpha \alpha} = 1 - \sin^2 2\theta_m(0) \sin^2 \phi_m(x)$.

Averaging over the phase, which means that the contributions from
$\nu_1$ and $\nu_2$ add incoherently, gives
\be
P = (\cos \theta_m \cos \theta_m^0)^2 + (\sin \theta_m \sin \theta_m^0)^2
= \sin^2 \theta_m  +  \cos 2 \theta_m \cos^2 \theta_m^0.
\label{eq-05:adiabformav}
\ee
The mixing in the  neutrino production point $\theta_m^0$
is determined by  density in this point, 
$n_e^0$, and  the resonance density.
Consequently, the picture of the conversion depends on
how far from the resonance layer (in the density scale) a neutrino is produced.
Strong transitions occur if the initial and final mixings differ substantially, 
which is realized when initial density is much above the resonance density and 
final one is below the resonance density and therefore neutrinos cross
the resonance layer. 

According to \eq~(\ref{eq-05:ee-prob-ad}) the oscillation depth equals
$ D = |\sin 2\theta_m \sin 2\theta_m^0|$.   
Both the averaged probability (\ref{eq-05:adiabformav}) 
 and the depth (\ref{eq-05:ee-prob-ad}) 
are determined by the  initial and final densities and do not
depend on the density distribution along the neutrino trajectory. 
Essentially they are determined by the ratios $y \equiv n/n_R$ in the 
initial and final moments. This is a manifestation of the universality of the 
adiabatic approximation result.   

In contrast, the phase do depend on the density distribution and the
period of oscillations (the latter is given is by the oscillation length in matter). 
So, it is the phase that encodes an information about the density distribution.  

The probability depends on $t$ via the phase $\phi_m(t)$ and also via the mixing
angle $\theta_m(t)$. Two degrees of freedom are operative and  $P$
dependence on time is an interplay of two effects: oscillations, associated to
the phase $\phi_m(t)$, and the adiabatic conversion related to change of $\theta_m$.
Depending on initial condition $n_e^0$, the relative importance of the two effects is 
different. If neutrinos are produced far 
above the resonance,  $n_e^0 \gg n_e^R$, the initial mixing 
is strongly suppressed, $\theta^0_m \approx \pi /2$. 
Consequently,  the neutrino state, e.g. $\nu_e$,  consists mainly of one
eigenstate, $\nu_{2m}$, and furthermore, one flavor $\nu_e$, dominates in $\nu_{2m}$.   
Since the admixture of the second eigenstate is very small, oscillations
(interference effects) are strongly suppressed.
Thus, here the non-oscillatory flavor transition occurs
when the flavor of whole state (which nearly coincides with
$\nu_{2m}$) follows the density change.
At zero density $\nu_{2m} = \nu_{2}$,  and therefore
the probability to find the electron neutrino (survival probability) equals~\cite{05-Mikheev:1986gs} 
\be
P = |\langle \nu_e | \nu(t) \rangle|^2 \approx |\langle \nu_e |\nu_{2m}(t) \rangle|^2
= |\langle \nu_e |\nu_{2} \rangle|^2 \approx \sin^2 \theta.
\label{eq-05:surv1}
\ee
The final probability, $P = \sin^2 \theta$, is the feature
of the non-oscillatory transition (as pure adiabatic conversion).
Deviation from this value indicates the presence of oscillations, 
see \eq~(\ref{eq-05:ee-prob-ad}). 

If neutrinos are produced not too far from 
resonance, e.g. at $n_e^0 > n_e^R$,
the initial mixing is not suppressed.  Although $\nu_{2m}$ is the main component 
of the neutrino state, the second eigenstate, $\nu_{1m}$,
has appreciable  admixture;
the flavor mixing  in the neutrino eigenstates is significant, 
and the  interference effect is not suppressed.
Here we deal with the interplay of the adiabatic conversion and
oscillations.  

Production in the resonance is a special case: If $\theta_m^0 = 45^{\circ}$, the averaged 
probability equals $\bar{P} = 1/2$ independently of the final mixing. 
This feature is important for determining the oscillation parameters. 
Strong transitions ($P > 1/2$) occur when neutrinos cross resonance layer.
These features are realized for solar neutrinos when propagating from their production region inside the Sun to the surface of the Sun. 
The adiabatic propagation occurs also in a single layer of the Earth (e.g. in the mantle). 


\subsection{Adiabaticity violation}

For most of applications the adiabaticity is either 
well satisfied (neutrinos in the Sun or supernovae), or maximally broken due to sharp 
(instantaneous) density change   
(neutrinos in the Earth, neutrinos crossing the shock wave fronts in supernova). 
In the former case the evolution is described  
by the adiabatic formulas.  
In the latter case description is also simple, one just needs to match the flavor 
conditions at the borders between layers: 
find the flavor state before the density jump and then use it as an initial state 
for the evolution after the jump. 
The intermediate case of the adiabaticity breaking can be realized for neutrinos in the mantle of the Earth, for high energy neutrinos propagating in the Sun (neutrinos from annihilation of 
hypothetical WIMPs) or for sterile neutrinos with very small mixing.

If the density changes rapidly, $\dot{\theta}_m$ is not negligible in  
(\ref{eq-05:eqmatt2a}) and the adiabaticity condition (\ref{eq-05:adiab}) is not satisfied. 
The transitions $\nu_{1m} \leftrightarrow \nu_{2m}$ become noticeable
and therefore the admixtures of the eigenstates in a given propagating state
change. The $S$ matrix in the flavor basis is given by 
$$
S_f(x, 0) = U^m(t)\tilde{S}(x,0)U^{m\dagger}(0) = 
 U^m(t) \left(
\begin{array}{cc}
S_{11}  & - S_{21}^* \\
S_{21}  &  S_{11}^* 
\end{array}
\right) U^{m\dagger}(0),  
$$
where $\tilde S$ is the evolution matrix in the basis of instantaneous eigenstates.
Then the  $\nu_e - \nu_e$ transition probability
$P_{ee} \equiv |{S_f(x, 0)}_{ee}|^2$  equals
\be
P_{ee}  = \frac{1}{2} 
\left[1 + \cos 2\theta_m(t) \cos 2\theta_m(0)\right] 
- P_{21}\cos 2\theta_m(t) \cos 2\theta_m(0) + P_{int},
\label{eq-05:ee-prob-adv}
\ee
where $P_{21} \equiv |S_{21}|^2$ is the probability of $\nu_{2m} \rightarrow \nu_{1m}$ transitions
and $P_{int}$ is an interference term 
\be
P_{int} = \frac{1}{4}\sin 2\theta_m(t) \sin 2\theta_m(0)  
\left[S_{11}^2 + S_{11}^{*2} + S_{21}^2 + S_{21}^{*2} \right] +  \frac{1}{2}\sin[2\theta_m(0) - 2\theta_m(x)] \left[S_{11} S_{21}^* + S_{11}^*S_{21} \right] 
\ee
which depends on the  oscillation phase.
The averaged probability ($P_{int} = 0$) equals \cite{05-Parke:1986jy} 
\be
P_{ee} = \frac{1}{2} + \left(\frac{1}{2} - P_{21} \right) 
\cos 2\theta_m(t) \cos 2\theta_m(0). 
\label{eq-05:madform}
\ee
If the initial density is much larger than the resonance density, 
then $\theta_m(0) \approx \pi/2$ and $\cos 2\theta_m(0) = - 1$. 
In this case the averaged probability can be rewritten as 
\be
P_{ee} = \sin^2 \theta_m(t)+ P_{21} \cos 2\theta_m(t). 
\ee
Violation of adiabaticity weakens transitions if $\cos 2\theta_m(t) > 0$, 
thus leading to an increase of the survival probability.
In the adiabatic case $S_{11} = e^{i\phi_m}$, $S_{21} = 0$, and therefore  
$S_{11}^2 + S_{11}^{*2} = 2\cos 2\phi_m(x)$,  so that  \eq~(\ref{eq-05:ee-prob-adv}) 
is reduced to (\ref{eq-05:ee-prob-ad}).   

In the graphic  representation (\Fig~\ref{fig-05:polarization}),
the neutrino vector moves on the surface of the cone (phase change) and 
the axis of the cone rotates according to the density change. 
The cone angle $\theta_a$ changes  as a result of violation of the adiabaticity).

There are different approaches to compute the flop probability $P_{21}$.
In the adiabatic regime the probability 
of transition between the eigenstates is exponentially 
suppressed  $P_{12} \sim \exp{(- \pi/2\gamma)}$
with $\gamma$ given in Eq.~(\ref{eq-05:adiab}) \cite{05-Haxton:1986bc,05-Parke:1986jy}.
One can consider such a transition as penetration through a barrier of height
$H_{2m} - H_{1m}$ by a system with the kinetic energy $d\theta_m /dt$.
This leads to the Landau-Zener probability: 
\be
P_{LZ} =  
\exp(-\pi^2 \kappa_R) = \exp\left(-\frac{\pi h \Delta m^2}{4E} 
\frac{\sin^2 2\theta}{\cos 2\theta}\right), 
\ee 
where $h \equiv n (dn/dr)^{-1}$~\cite{05-Petcov:1987xd}. 
In the case of weak adiabaticity violation,
one can develop an adiabatic perturbation theory which gives the results as
a series expansion in the adiabaticity parameter~\cite{05-deHolanda:2004fd}.

\subsection{Theory of small matter effects}

\subsubsection{Minimal width condition}
If the vacuum mixing  angle is small, there exists a lower limit on  
amount of matter needed to induce significant flavor change due to matter effect. 
The amount of matter is characterized by the 
column density of electrons along the neutrino trajectory: 
\begin{equation}
 d = \int_0^L n_e(x) dx. 
\end{equation}
We can define $d_{1/2}$ as the column density for which the 
oscillation transition probability surpasses $1/2$ for the 
first time in the course of propagation. Then 
it is possible to show that~\cite{05-Lunardini:2000swa}
\begin{equation}
\label{eq-05:minres}
d_{1/2} \geq   d_{\rm min}= \frac{\pi}{2\sqrt 2 G_F \tan 2\theta} 
\end{equation}
for all density profiles. Furthermore, the minimum,  $d_{\rm min}$, is realized  
for oscillations in a medium of constant  density equal to the 
resonance density. The relation (\ref{eq-05:minres}) is known as the minimal width condition.  
This condition originates from an interplay between matter effects 
and vacuum mixing: The acquired matter phase, $\sqrt 2 G_F d$, must be large. 
At the same time, since matter effects by themselves are flavor conserving,  
also vacuum mixing is required in order to induce flavor conversion. The smaller vacuum 
mixing, the large width is required.

\subsubsection{Vacuum mimicking}

Vacuum mimicking \cite{05-Minakata:2000ee}, 
which states that regardless of  the matter density, the initial flavor evolution of 
neutrino state is similar to that of vacuum oscillations. Consequently for 
small baselines, $L$, it is not possible to see matter effect and any such effect appears  
in higher order of $L$. Indeed,  consider the evolution matrix   
\begin{equation}
 S = \mathcal{T}\left[
  \exp\left( - i \int_0^L H(x) dx \right)
 \right],
\end{equation}
where $\mathcal T$ denotes time ordering  of the exponential. 
For small values of  $L$, it can be expanded as
\begin{equation}
 S = 1 - i \int_0^L H(x) dx + \mathcal O(L^2).
\end{equation}
If initial neutrino state  has definite flavor, 
the amplitude of flavor transition  is given by the off-diagonal 
element of $H(x)$  which does not depend on matter potential.  
The matter contribution to $H(x)$ is diagonal. Therefore  the 
flavor transitions  depends on the matter 
density only at higher order in $L$.  This result holds true as long as $L \ll l_m$ or 
when the phase of oscillation is small~\cite{05-Akhmedov:2000cs}.

This can be seen explicitly in the case of medium with constant density 
where expanding the oscillatory factor for small oscillation phase we have
the transition probability
\be
P = \sin^2 2\theta_m \sin^2 \phi^m = \frac{1}{R} \sin^2 2\theta 
\sin^2 \phi \sqrt{R} \approx \phi^2 \sin^2 2\theta.
\ee

Note that vacuum mimicking only occurs if the initial neutrino state is a flavor eigenstate~\cite{05-Akhmedov:2000cs}. 
If the initial neutrino is in a flavor-mixed  state, e.g. in a mass eigenstate, then 
matter will affect this state already at lowest order in $L$. 
This situation is realized in several settings involving astrophysical neutrinos propagating through the Earth, e.g., solar and supernova neutrinos, where the neutrinos arrive at the Earth as mass eigenstates.
The mimicking is not valid if there are non-standard flavor changing interactions, so that 
matter effect appears in the off-diagonal elements of the Hamiltonian.

\subsubsection{Effects of small layers of  matter}

If the minimal width condition is not satisfied,  that is  
$d = n x \ll G_F^{-1}$, the matter effect on result of evolution is small.
This inequality can be written as $V x \ll 1$ which means that 
the oscillation phase is small.  
In this case the matter effect can be considered as small  perturbation 
of the vacuum oscillation result even if  the MSW resonance 
condition is satisfied.

The reasons for the smallness of the matter effect  
are different depending  on the energy interval. 
Consider a layer of constant density with the length $x$. 
There are three possibilities

(i) $E \ll E_R$, ($E_R$ is the resonance density) -   
nearly vacuum oscillations in low density medium take place. 
Matter effect gives small corrections to  the oscillation 
depth and length which are characterized by 
$\frac{2V E}{\Delta m^2}  = \frac{Vx}{2\pi} \ll 1$ ,
here  $x \sim l_\nu$. 

(ii) $E \sim E_R$ - modification of oscillation parameters is strong, however 
$l_\nu^R \sim l_\nu /\sin2\theta \sim  2\pi/(V \sin 2\theta)$. 
Consequently, 
$x/l_\nu^R = xV \sin 2\theta/2\pi \ll 1$. Oscillations are undeveloped 
due to smallness of phase. 

(iii) $E \gg E_R$ - matter suppresses oscillation depth by a factor 
$E_R/E \ll 1$. Since the oscillation length equals $l_m \approx 2\pi /V $,  
one obtains $x/ l_m = xV/2\pi \ll 1$. Hence in this case the distance 
is very small and oscillation effect in the layer has double suppression.

\subsection{Propagation in multilayer medium}
\label{sec-05:multi} 

\subsubsection{Parametric effects in the neutrino oscillations}



The strong transitions discussed in the previous sections require the existence of
large effective mixing, either in the entire medium (constant density)
or at least in a layer (adiabatic conversion).
There is a way to get strong transition without large
vacuum or matter mixings. This can be realized with periodically or 
quasi periodically changing density 
\cite{05-Ermilova:1986ab,05-Akhmedov:1988kd} when the conditions of  
parametric resonance are satisfied. Although the flavor conversion in a layer which 
corresponds to one period is small, strong transitions can build up over several
periods. 
For large mixing even a small number of periods is enough to obtain 
strong flavor transitions.  

The usual condition of parametric resonance is that
the period of density change $T_n$ is an integer times 
the effective oscillation length $l_m$ \cite{05-Krastev:1989ix}: 
\begin{equation}
\int_{l_T} \frac{dx}{l_m} = k\ ,\quad  (k = 1, 2, 3,...) ,  
\end{equation} 
or  $l_T/\bar{l}_m = k$.  
Such an enhancement has been considered first for 
modulation of the profile by sine function~\cite{05-Ermilova:1986ph}. 
This may have some applications for intense neutrino 
fluxes when neutrino-neutrino interactions become important. 

The solvable case, which has simple physical interpretation,  
is provided by the \emph{castle wall} profile,  
for which the period $l_T$ is divided into two parts $l_1$ and $l_2$ ($l_1 + l_2 =  l_T$)
with densities $n_1$ and $n_2$, respectively  ($n_1 \neq n_2$ and, in general, $l_1 \neq l_2$).
Thus, the medium consists of alternating layers with two different 
densities~
\cite{05-Akhmedov:1988kd,05-Liu:1997yb,05-Liu:1998nb,05-Petcov:1998su,05-Akhmedov:1998ui,05-Chizhov:1998ug,05-Chizhov:1999az,05-Chizhov:1999he}.

For the ``castle wall'' profile, the simplest realization of the 
parametric resonance condition is
reduced to  equality of the oscillation phases acquired by
neutrinos over the two parts of the periods~\cite{05-Liu:1998nb}:
\begin{equation}
\Phi_1 = \Phi_2 = \pi~.
\label{eq-05:pi}
\end{equation}

The enhancement of transition depends on the number of periods and
on the amplitude of perturbation, which determines
the swing angle (the difference of the mixing angles in the two layers, 
$\Delta \theta  \equiv  2\theta_{1m} - 2\theta_{2m}$). For small $\Delta \theta$ 
a large transition probability can be achieved after many periods.
For large ``swing'' angle, even a small number of periods is
sufficient.

\subsubsection{Parametric enhancement, general consideration. }

In general the condition (\ref{eq-05:pi}) is not necessary for the enhancement 
or even for  maximal enhancement. 
First, consider the  oscillation effect over one period.  The corresponding 
evolution matrix is given by the product 
\be
S_T = S_2 S_1,   
\label{eq-05:sone-per}
\ee  
where $S_k$ (k = 1,2) is the evolution in layer $k$ given by 
\eq~(\ref{eq-05:sone-lay}). For brevity we will write it as  
$
S_k = c_k I - i s_k ({\bf \sigma} \cdot {\bf n}_k), \quad k = 1,2,
$  
where $c_k \equiv \cos \phi_k$,  $s_k \equiv \sin \phi_k$
and $\phi_k$ is the half-phase acquired in layer $k$: 
\be
\phi_k = \frac{1}{2} \Delta H_k l_k = \frac{\Delta m^2}{4E}
R(V_k)^{1/2} l_k, \quad
{\rm and}
\quad
{\bf n}_k \equiv (\sin 2\theta_{m k}, 0, - \cos 2\theta_{m k}).
\ee 
Here $\theta_{m k}$ is the mixing angle in layer $k$.  

Insertion of $S_k$ from (\ref{eq-05:sone-lay}) into (\ref{eq-05:sone-per})
gives~\cite{05-Akhmedov:1988kd}
\be
S_T =  Y {\bf I} - i  ({\bf \sigma} \cdot {\bf X}),  
\label{eq-05:sone-per2}
\ee  
where 
$$
Y  \equiv  c_1 c_2 - s_1 s_2 ({\bf n}_1 \cdot {\bf n}_2), \quad
{\bf X}  =  s_1 c_2 {\bf n}_1 + s_2 c_1 {\bf n}_2 - s_1 s_2 [{\bf n}_1 \times {\bf n}_2]. 
$$
Explicitly: $({\bf n}_1 \cdot {\bf n}_2) = \cos(2\theta_{m1} - 2\theta_{m2})$
and $[{\bf n}_1 \times {\bf n}_2] =  \sin(2\theta_{m1} - 2\theta_{m2}) {\bf e}_y$. 
Using unitarity of $S_T$, which gives $X^2 + Y^2 = 1$,  
one can parametrize $X$ and $Y$ 
with a new phase $\Phi$ as $Y \equiv \cos \Phi$ and $X \equiv \sin \Phi$.
Then the  evolution matrix $S_T$ can be written in the form
$S_T =  \cos \Phi - i  \sin \Phi ({\bf \sigma} \cdot {\bf \hat{X}}) = 
e^{- i ({\bf \sigma} \cdot {\bf \hat{X}})\Phi}$,  
where ${\bf \hat{X}} \equiv {\bf X}/X$. 
Consequently, the evolution matrix after $n$ periods equals  
\be
S^n  =  (S_T)^n = e^{- i ({\bf \sigma} \cdot {\bf \hat{X}}) n \Phi} 
   =  \cos n \Phi - i  ({\bf \sigma} \cdot {\bf \hat{X}})\sin n \Phi ,
\label{eq-05:stot-n}
\ee 
It is simply accounted for by an increase of the phase: $\Phi \rightarrow n \Phi$. 
This is the consequence of the fact that the evolution matrices over all 
periods are equal and therefore commute. 
If the evolution ends at some instant $t$ which does not coincide 
with the end of a full period, 
{\it i.e.}, $t = nT + t'$, then $S (t) = S(t') S_n$.  

The transition probability computed with \eq~(\ref{eq-05:stot-n}) is
\be
P^n_{e\mu} = |S^n_{e\mu}|^2 = \frac{X_1^2 + X_2^2}{X^2} \sin^2 n \Phi. 
\label{eq-05:prob-nper}
\ee
It has the form of the usual oscillation probability with 
phase $n \Phi$ and depth 
$(X_1^2 + X_2^2)/X^2$. 
The oscillations described by \eq~(\ref{eq-05:prob-nper}) are 
called the parametric oscillations. Under condition 
\be
- X_3 = s_1 c_2 \cos 2\theta_{m1}  + s_2 c_1 \cos 2\theta_{m2} = 0,  
\label{eq-05:parcond}
\ee
which is called the parametric resonance condition, 
the depth of oscillations (\ref{eq-05:prob-nper}) becomes 1
and the transition probability is maximal when $n \Phi = \pi/2 + \pi k$, where $k$ is an integer.  There are different realizations of the condition (\ref{eq-05:parcond}) 
which imply certain  correlations among the mixing angles and phases. 
The simplest one, $c_1 = c_2 = 0$, coincides with \eq~(\ref{eq-05:pi}).

\subsubsection{Parametric enhancement in three layers}

For small number of layers an enhancement of flavor transition 
can occur due to certain relations between the phases and mixing angles 
in different layers. This in turn impose certain conditions 
on the parameters of the layers: their densities 
and widths.  The conditions are the similar to 
the parametric resonance condition and  
this enhancement is called the \emph{parametric enhancement} of flavor transitions. 
These conditions can be satisfied for certain energies and baselines 
for neutrinos propagating in the Earth. 

Consider conditions for maximal enhancement of oscillations 
for different number of layers. 
It is possible to show \cite{05-Akhmedov:2006hb} that they are generalizations 
of the conditions in one layer which require that (i) the depths of oscillations is 1
we call it the {\it amplitude condition} and the oscillation phase  is 
$\phi = \pi/2 + \pi k$ - the {\it phase condition}. 

Consider first the case of one layer 
with (in general) varying density (it can correspond to the mantle crossing 
trajectories in the Earth). 
The resonance condition for constant density case, $\cos 2 \theta_m = 0$, can
be written according to \eqs~(\ref{eq-05:matterHam}) and~(\ref{eq-05:Hvacuum2}) as
$\alpha = \alpha^*$, i.e., $S_{11}^{(1)} = S_{22}^{(1)}$,
or equivalently, ${\rm Im} S_{11}^{(1)} = 0$,
where the superscript indicates the number of layers. This
generalization goes beyond the original MSW-resonance
condition (even for constant density). 
The phase condition  can be 
rewritten in terms of the elements of the evolution matrix, 
[c.f., \eq~(\ref{eq-05:Ubar})] as ${\rm Re}~\alpha \equiv {\rm Re}~ S_{11}^{(1)} = 0$.
The absolute maximum of the transition probability occurs when
these conditions are satisfied simultaneously, {\it i.e.}, when $S_{11}^{(1)} = 0$. 

The parametric resonance condition~(\ref{eq-05:parcond}) can be
generalized to the case of non-constant densities in the layers 
although the generalization is not unique.
Indeed, according to \eq~(\ref{eq-05:sone-per2}) the condition $X_3 = 0$ can be
written in terms of the elements of the evolution matrix for the two layers as
the equality of the diagonal elements $S_{11}^{(2)} = S_{22}^{(2)}$.
Let us find the conditions for extrema
for density profiles consisting of two layers. We have
$S^{(2)} =  S_2 \, S_1$, 
where $S_{11}^{(2)} = \alpha_2 \, \alpha_1 - \beta_2 \, \beta_1^* \,,
    \
    S_{12}^{(2)} = \alpha_2 \, \beta_1 + \beta_2 \, \alpha_1^* \,$,
and $\alpha_i$, $\beta_i$ for each layer have been defined in
\eq~(\ref{eq-05:Hvacuum2}). The sum of the two complex numbers in the
transition amplitude $S_{12}^{(2)}$ has the
largest possible result if they have the same phase:
$\arg(\alpha_2 \beta_1) = \arg(\beta_2 \alpha_1^*)$,
which can also be rewritten as
\begin{equation} \label{eq-05:collin2}
    \arg(\alpha_1 \alpha_2 \beta_1) = \arg(\beta_2) \,.
\end{equation}
This condition is called the \emph{collinearity condition}~\cite{05-Akhmedov:2006hb}. It is
an extremum condition for the two-layer transition probability
under the constraint of fixed transition probabilities in the
individual layers. In other words, if the absolute values $|\beta_i|$
of the transition amplitudes are fixed while their arguments are
allowed to vary, then the transition probability reaches an extremum when
these arguments satisfy \eq~(\ref{eq-05:collin2}). 

The conditions for maximal transition probability 
for three layers can be found in the following 
way.  The 1-2 elements of the evolution matrix $S^{(3)}$ equals 
\be
    \label{eq-05:def-b}
    S_{12}^{(3)} = \alpha_3\, S_{12}^{(2)} + \beta_3\, S_{11}^{(2)*}
    =  \alpha_3\, \alpha_2\, \beta_1 + \alpha_3\, \beta_2\, \alpha_1^*
    + \beta_3\, \alpha_2^*\, \alpha_1^*
    - \beta_3\, \beta_2^*\, \beta_1 \,.
\ee
In the case of neutrino oscillations in the Earth, the third layer is
just the second mantle layer, and its density profile is the reverse
of that of the first layer. The evolution matrix for the third layer
is therefore the transpose of that for the first one~\cite{05-Akhmedov:2001kd},
{\it i.e.},  $\alpha_3=\alpha_1$, $\beta_3=-\beta_1^*$, and the expression for
$S_{12}^{(3)}$ can be written as
\begin{equation} \label{eq-05:res2}
    S_{12}^{(3)} = \alpha_1 \alpha_2 \beta_1
    - \alpha_1^* \alpha_2^* \beta_1^*
    + |\alpha_1|^2 \beta_2 + |\beta_1|^2 \beta_2^* \,.
\end{equation}
Note that $\beta_2$ is pure imaginary because the core density profile
is symmetric. Therefore the amplitude $S_{12}^{(3)}$ in
\eq~(\ref{eq-05:res2}) is also pure imaginary, as it must be because the
overall density profile of the Earth is symmetric as well. 
If the collinearity condition for two
layers~(\ref{eq-05:collin2}) is satisfied, then not only the full
amplitude $S_{12}^{(3)}$, but also \emph{each} of the four terms on
the right hand side of \eq~(\ref{eq-05:res2}) is pure imaginary. 
If the collinearity condition is satisfied for
two layers, then it is automatically satisfied for three layers. 
This is a consequence of the facts
that the density profile of the third layer is the reverse of that of
the first layer and that the second layer has a symmetric profile.
The conditions described here allow to reproduce very precisely 
all main structures of the oscillograms of the Earth 
(see sect. \ref{sec-05:propthroughearth}).

\subsection{Oscillations of high energy neutrinos}

At high energies or in high density medium when  
$V > \Delta m^2/2E$, we can use $\Delta/V \equiv \Delta m^2/4EV$ as a small parameter  
and develop a perturbation theory using its smallness. 
However,
in most situations of interest, the neutrino path length in matter $L$ is so large that
$\Delta \cdot L \aprge  1$. Therefore the vacuum part of the Hamiltonian cannot be considered
as a small perturbation in itself and the effect of $\Delta$ on the
neutrino energy level splitting
should be taken into account. For this reason we split the Hamiltonian as
$H = \tilde{H}_0 + H_I$ with
\begin{equation}
\label{eq-05:Hsum}
\tilde{H}_0 = \frac{\omega^m}{2}
\left( \! \! \begin{array}{cc}
~1 & ~~0 ~\\
~0 & -1
\end{array}  \! \! \right), ~~~
H_I = \sin 2\theta \, \Delta \, \left(  \! \! \begin{array}{cc}
- \epsilon & ~1 \\
~1 & ~\epsilon
\end{array}  \! \! \right)\, ,
\end{equation}
where $\omega^m$ is the oscillation frequency~(\ref{eq-05:omegam})
and $\epsilon \equiv ({2\Delta \cos 2\theta- V + \omega^m })/
{2\Delta \sin 2\theta  } \approx \frac{\Delta }{V} \sin 2\theta \ll 1$. 
The ratio of the second and the first terms in the Hamiltonian (\ref{eq-05:Hsum})
is given by the mixing angle in matter $\theta_m$:
$2\Delta \sin 2\theta/\omega^m = \sin 2\theta_{m}$.
Therefore for $\sin 2\theta_{m} \ll 1$ the term
$H_{I}$  can be considered as a perturbation.
Furthermore, $\epsilon \sim \sin 2\theta_m$, so
 the diagonal terms in $H_I$ can be neglected in the lowest approximation.

The solution for $S$ matrix can be found in the form $S=S_0 \cdot S_I$,
where $S_0$ is the solution of the evolution equation with $H$ replaced by
$H_0$ [see \eq~(\ref{eq-05:s-tilde1symm})].
The
matrix $S_{I}$ then satisfies the equation
\be
\label{eq-05:eqSdelta}
i \frac{d S_{I}}{dx} = S_0^{-1} H_{I} S_0 \, S_{I} = \tilde{H}_I
S_{I}\,,
\ee
where $\tilde{H}_I \equiv S_0^{-1} H_{I} S_0$ is the perturbation Hamiltonian in the ``interaction'' representation. 
\eq~(\ref{eq-05:eqSdelta}) can be solved by
iterations: $S_I = I + S_{I}^{(1)} + ...$,  which leads to the
standard perturbation series for the $S$ matrix.  
For neutrino propagation
between  $x = 0$ and $x= L$ we have, to the lowest non-trivial order,
\be
\label{eq-05:A}
S(L) =  S_0(L)
\left[I  - i \Delta \sin 2\theta  \int_0^L dx
\left(\begin{array}{cc}
  0 &  e^{i 2 \phi(x)}~\\
e^{-i 2 \phi (x)}  &   0~
\end{array}
\right)
\right].
\ee
The  $\nu_{e} \leftrightarrow \nu_a$ transition probability 
$P_2 = [S(L)]_{a e}$ is given by 
\be
\label{eq-05:prob}
P_2 = \Delta^2 \sin^2 2\theta \left| \int_0^L dx  ~e^{-i 2 \phi(x)}
\right|^2.
\ee

For density profiles that are symmetric with respect to the
center of the neutrino trajectory, $V(x) = V(L - x)$, \eq~(\ref{eq-05:prob}) gives
\be
\label{eq-05:probsym}
P_2 = 4\left(\frac{\Delta m^2 }{4E}\right)^2
\sin^2 2\theta \left[ \int_0^{L/2}dz \cos 2\phi(z)\right]^2,
\ee
where $z = x - L/2$ is the distance
from the midpoint of the trajectory and $\phi(z)$ is the phase acquired
between this midpoint and the point $z$.
The transition probability $P_2$ decreases with the increase of neutrino energy essentially
as $E^{-2}$. The accuracy of \eq~(\ref{eq-05:prob}) also improves with energy as
$E^{-2}$.  

Inside the Earth, the accuracy of the analytic formula is extremely good already
for $E\aprge 8$~GeV.
When neutrinos do not cross the Earth's core ($\cos\Theta>
-0.837$), and so experience a slowly changing potential $V(x)$, the accuracy of
the approximation (\ref{eq-05:prob}) is very good even in the MSW resonance region
$E\sim $(5 -- 8)~GeV.

The above formalism applies 
in the low energy case as well, with only minor modifications: the sign of
$H_0$ in \eq~(\ref{eq-05:Hsum}) has to be flipped, and correspondingly one has to
replace $\omega^m\to -\omega^m$ in the definition of $\epsilon$.
The expressions for the transition  probability
in \eqs~(\ref{eq-05:prob}) and (\ref{eq-05:probsym}) remain unchanged.

\subsection{Effects of small density perturbations}
\label{sec-05:smalldensityperturbations}

Let us consider perturbation around smooth profile for which  
exact solution is known.  The simplest possibility 
that has implications for the Earth matter profile is the constant 
density with additional perturbation: $V(x) = \bar{V} + \Delta V(x)$.
Correspondingly, the Hamiltonian of the system can be written as the
sum of two terms:
\begin{equation}
    H(x) = \bar{H} + \Delta H(x)\,,
    \label{eq-05:smalldensityH}
\end{equation}
where
\begin{equation}
    \bar{H} \equiv \bar\omega
    \begin{pmatrix}
        -\cos 2\bar\theta & \sin 2\bar\theta \cr
        \hphantom{-}\sin 2\bar\theta & \cos 2\bar\theta
    \end{pmatrix},
    \qquad
    \Delta H \equiv \frac{\Delta V(x)}{2}
    \begin{pmatrix}
        1 & \hphantom{-}0 \cr
        0 & -1
    \end{pmatrix}\,.
\end{equation}
Here, $\bar\theta = \theta_m(\bar V)$ is the mixing angle in matter and
$\bar\omega = \omega^m(\bar V)$ is half of the energy splitting
(half-frequency) in matter, both with the average potential $\bar V$.
We will denote by $\bar{S}(x)$ the evolution
matrix of the system for the constant density case $H(x) = \bar{H}$.
The expression for $\bar{S}(x)$ is given in \eq~(\ref{eq-05:Ubar}) 
with  $\theta_m = \bar \theta$ and  $\phi_m(x) = \phi (x) \equiv \bar\omega \, x $, 
$\bar\omega = \omega^m (\bar{V})$.

The solution of the evolution
equation with Hamiltonian (\ref{eq-05:smalldensityH})~\cite{05-Akhmedov:2006hb} is of the form
\begin{equation} 
\label{eq-05:exp2}
    S(x) = \bar{S}(x) + \Delta S(x), \qquad
    \Delta S(x) = -i \, \bar{S}(x) \, K_1(x)\,,
\end{equation}
where $K_1(x)$ satisfies $|K_1(x)_{ab}| \ll 1$. Inserting
\eq~(\ref{eq-05:exp2}) into the evolution equation,  
one finds the following
equation for $K_1(x)$ to the first order in $\Delta H(x)$ and
$K_1(x)$:
\begin{eqnarray} 
\label{eq-05:diff}
    \frac{dK_1(x)}{dx} =
    \bar{S}^\dagger \, \Delta H(x) \, \bar{S} =
    \frac{\Delta V}{2} \left[
    -\cos2\bar\theta
    \begin{pmatrix}
        -\cos2\bar\theta & \sin2\bar\theta \cr 
        \hphantom{-}\sin2\bar\theta & \cos2\bar\theta
        \end{pmatrix}
    + \sin2\bar\theta \cos2\phi ~G(\bar\theta) + \sin2\bar\theta 
\sin2\phi~ \sigma_2 \right] \,.
\end{eqnarray}
where $G(\bar\theta) \equiv 
\cos2 \bar\theta ~\sigma_1 + \sin2\bar\theta ~\sigma_3$.
The first term in \eq~(\ref{eq-05:diff}) does not 
contribute to $S \equiv S(L)$ since $\langle
\Delta V \rangle \equiv \int \Delta V(x) dx = 0$, and
\eq~(\ref{eq-05:diff}) can be immediately integrated:
\begin{eqnarray}
    K_1(L) = \frac{1}{2} \sin2\bar\theta \left[
G(\bar\theta) \int_0^L \Delta V(x) \, \cos2\phi(x) \, dx
+ \sigma_2 \int_0^L \Delta V(x) \, \sin2\phi(x) \, dx
    \right] \,.
    \label{eq-05:K1}
\end{eqnarray}
Introducing  the distance from the midpoint 
of the neutrino trajectory $z \equiv x - L/2$, one obtains  
from \eq~(\ref{eq-05:K1})
\begin{equation}
    \label{eq-05:DeltaU}
    \Delta S \equiv \Delta S(L) = -i \, \sin2\bar\theta
    \left[ G(\bar\theta)
    \Delta I + \sigma_2 \Delta J \right] \,,
\end{equation}
where
$\Delta I \equiv \frac{1}{2}
    \int_{-L/2}^{L/2} \Delta V(z) \, \cos(2\bar\omega z) \, dz \,,
    \
    \Delta J \equiv \frac{1}{2}
    \int_{-L/2}^{L/2} \Delta V(z) \, \sin(2\bar\omega z) \, dz \,$.
In these integrals, $\Delta V(z) \equiv \Delta V(x(z))$ and $x(z) = z
- L/2$. The integral $\Delta J$ vanishes if the perturbation $\Delta
V(z)$ is symmetric with respect to the midpoint of the trajectory.
Analogously, $\Delta I$ vanishes if $\Delta V(z)$ is antisymmetric.
The expression for $S$ defined in \eq~(\ref{eq-05:exp2}) is equivalent
to \eqs~(13--16) obtained in Ref.~\cite{05-Lisi:1997yc} in the context of
solar neutrino oscillations.

For practical purposes it is useful to
have an expression for $S$ which is \emph{exactly} unitary regardless
of the size of the perturbation.  
For this we rewrite \eq~(\ref{eq-05:DeltaU}) as follows:
\begin{equation}
    \Delta S = \varepsilon \, S' \,,
    \qquad
    S' = -i
    \left[G(\bar{\theta})
    \cos\xi + \sigma_2 \sin\xi \right] \,,
\end{equation}
where $\sin \xi = \frac{\Delta J}{\sqrt{(\Delta J)^2+(\Delta I)^2}}$
and $\epsilon = \sin2 \bar\theta \cdot {\sqrt{(\Delta J)^2 +(\Delta I)^2}}$.
Thus, $S = \bar S + \varepsilon \, S'$ and we  
replace it by 
\begin{equation} \label{eq-05:Uimproved}
S = \cos\varepsilon \, \bar S + \sin\varepsilon \, S' \,.
\end{equation}
Here both $S'$ and $\bar S$ are unitary matrices, and  due to
their specific form the combination on the right-hand-side of
\eq~(\ref{eq-05:Uimproved}) is exactly unitary.

For a symmetric density profile 
with respect to the midpoint of the trajectory, the term $\Delta J$ is
absent. From \eqs~(\ref{eq-05:Ubar}), (\ref{eq-05:DeltaU})
and~(\ref{eq-05:Uimproved}) we immediately get the transition probability
\begin{equation} \label{eq-05:probab}
    P = \left[ \cos\varepsilon \, \sin2\bar\theta \, \sin\phi
    + \sin\varepsilon \, \cos2\bar\theta \right]^2
    \approx \sin^22\bar\theta \, \left[ \sin\phi + \Delta I \,
    \cos2\bar\theta \right]^2,
\end{equation}
where $\varepsilon \equiv \sin2\bar\theta \, \Delta I$ and $\phi
\equiv \phi(L) = \bar\omega L$. Here the first term in the square
brackets describes oscillations in constant density matter with
average potential $\bar{V}_1$.

\subsection{Oscillation probabilities and their properties} 
\label{sec-05:oscprobproperties}

It is convenient to consider the neutrino flavor evolution in the 
propagation basis  
$\tilde{\nu} = (\nu_e, \tilde{\nu}_{2}, \tilde{\nu}_{3})^T$,
defined in \eq~(\ref{eq-05:basisrel}). 
In this basis propagation is not affected by the 2-3 mixing and CP-violation.  
The dependence on these parameters appears when
one projects the initial flavor state on the propagation basis and the
final state back onto the original flavor basis.
The propagation basis states are related to the mass states as 
\be 
\tilde{\nu} = U_{13} \, I_{-\delta} \, U_{12} \, \nu . 
\ee 
Since the transformations which connect  $\tilde{\nu}$ and $\nu_f$, do not depend on 
matter potential and therefore distance, the states $\tilde{\nu}$ 
satisfy the the evolution equation 
$i \frac{d \tilde{\nu}}{dt} =  \tilde{H}  \tilde{\nu}$, with the Hamiltonian $\tilde{H}$ defined in \eq~(\ref{eq-05:ham-tilde}).

\subsubsection{S-matrix and oscillation amplitudes}
\begin{figure}[h!!]
\begin{center}
\vspace{1cm}
\includegraphics[height=4.5cm,angle=0]{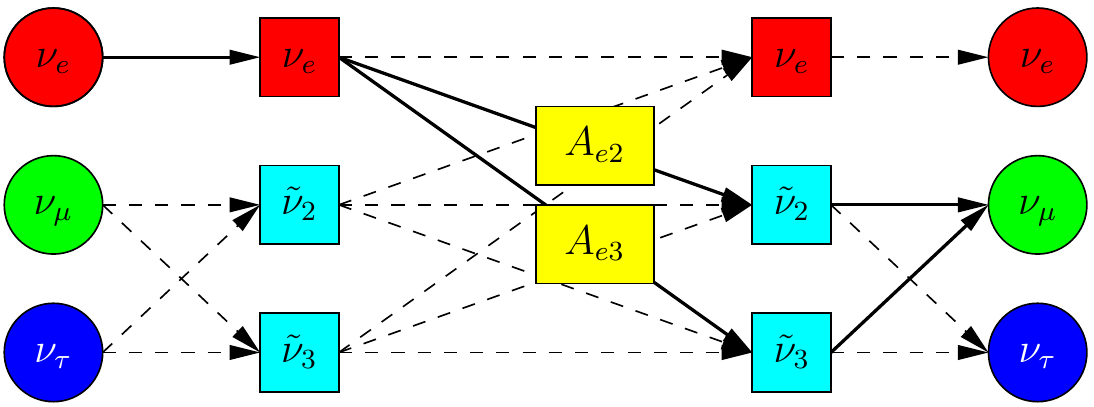}
\caption{Scheme of transitions between the flavor states. 
Evolution is considered in the propagation basis $\tilde{\nu}$. 
The lines which connect the flavor states and the propagation 
basis states  indicated projection of one basis onto another. 
The lines connecting the states  of propagation basis 
$\tilde{\nu}$  show transitions between them. 
\label{fig-05:schematic}}
\end{center}
\end{figure}

A number of properties of the oscillation probabilities can be obtained 
from general consideration of matrix of the oscillation amplitudes. 
We introduce the evolution matrix (the matrix of 
amplitudes) in the propagation basis as
\be
    \tilde{S} =
\left(\begin{array}{ccc}
	A_{ee} & A_{e\tilde{2}} & A_{e\tilde{3}} \\
	A_{\tilde{2}e} & A_{\tilde{2}\tilde{2}} & A_{\tilde{2}\tilde{3}} \\
	A_{\tilde{3}e} & A_{\tilde{3}\tilde{2}} & A_{\tilde{3}\tilde{3}}
\end{array}\right).
\label{eq-05:matr2}
\ee
Then according to \eq~(\ref{eq-05:basisrel}) the  $S$ matrix in the flavor basis
equals
\be
S = \tilde{U} \tilde{S} \tilde{U}^{\dagger}, ~~~\tilde{U} \equiv U_{23}I_{\delta}.
\ee
In this part, we use the notation $A_{ij}$ for the amplitudes in the propagation basis and 
$S_{ij}$ for the amplitudes in the flavor basis.
In terms of the propagation-basis amplitudes (\ref{eq-05:matr2}) the $S$ matrix in the flavor basis 
can be written as 
\be
  S =
\left(\begin{array}{ccc}
	A_{ee}
	& c_{23} A_{e\tilde{2}} + s_{23} e^{-i\delta} A_{e\tilde{3}}
	& - s_{23} A_{e\tilde{2}} + c_{23}e^{-i\delta} A_{e\tilde{3}}
	\\
	c_{23} A_{\tilde{2}e} + s_{23} e^{i\delta} A_{\tilde{3}e}
	& c_{23}^2 A_{\tilde{2}\tilde{2}} + s_{23}^2 A_{\tilde{3}\tilde{3}} + K_{\mu\mu}
	& - s_{23} c_{23} (A_{\tilde{2}\tilde{2}} - A_{\tilde{3}\tilde{3}}) + K_{\mu\tau}
	\\
	- s_{23} A_{\tilde{2}e} + c_{23}e^{i\delta} A_{\tilde{3}e}
	& - s_{23}c_{23} (A_{\tilde{2}\tilde{2}} - A_{\tilde{3}\tilde{3}}) + K_{\tau\mu}
	& s_{23}^2 A_{\tilde{2}\tilde{2}} + c_{23}^2 A_{\tilde{3}\tilde{3}} + K_{\tau\tau}
\end{array}\right),
\label{eq-05:matr2s}
\ee
where
\bea
    K_{\mu\mu}
    & \equiv &s_{23} c_{23} (e^{-i\delta}
    A_{\tilde{2}\tilde{3}} + e^{i\delta} A_{\tilde{3}\tilde{2}}) \, ~~~~
    K_{\mu\tau} \equiv  c_{23}^2 e^{-i\delta} A_{\tilde{2}\tilde{3}}
    - s_{23}^2 e^{i\delta} A_{\tilde{3}\tilde{2}} 
    \nonumber \\ 
K_{\tau\mu} & = & K_{\mu\tau}(\delta \to -\delta,\,
    \tilde{2} \leftrightarrow \tilde{3}) \, ~~~~
    K_{\tau\tau} =  -K_{\mu\mu} \,.
\eea
The scheme of transitions is shown in \Fig~\ref{fig-05:schematic}.
There is certain hierarchy of the amplitudes which can be obtained 
immediately from the form of the Hamiltonian in the propagation basis (\ref{eq-05:matr1}): 
\begin{equation}
    A_{e\tilde{3}}, A_{\tilde{3}e} \sim s_{13}, \qquad
    A_{e\tilde{2}}, A_{\tilde{2}e} \sim r_\Delta \sim s_{13}^2, \qquad
    A_{\tilde{3}\tilde{2}}, A_{\tilde{2}\tilde{3}} \sim s_{13} r_\Delta \sim s_{13}^3,
\end{equation}
{\it i.e.}, $A_{\tilde{2}\tilde{3}}$ and $A_{\tilde{3}\tilde{2}}$ are the
smallest amplitudes. 
%
%
In the propagation basis there is no fundamental CP- or T- violation. 
Therefore for a symmetric density profile with respect to the middle point of trajectory
(as in the case of the Earth) the neutrino evolution is T-invariant which yields  
\be
    \label{eq-05:sym}
    A_{\tilde{2}e} = A_{e\tilde{2}} \,, \qquad
    A_{\tilde{3}e} = A_{e\tilde{3}} \,, \qquad
    A_{\tilde{3}\tilde{2}} = A_{\tilde{2}\tilde{3}} \,.
\ee 
Consequently, for $K_{\alpha \beta}$ we obtain
\be
\label{eq-05:K}
    K_{\mu\tau}
    = A_{\tilde{2}\tilde{3}} (\cos 2\theta_{23} \cos\delta - i \sin\delta), \quad
    K_{\tau\mu} = K_{\mu\tau} (\delta \rightarrow - \delta), \quad
    K_{\mu\mu} = -K_{\tau\tau} = A_{\tilde{2}\tilde{3}}
    \sin 2\theta_{23} \cos\delta \,.
\ee
These terms proportional to small
amplitudes $A_{\tilde{2}\tilde{3}}$ and
$A_{\tilde{3}\tilde{2}}$ are of the order $O(s_{13}^2)$.


For a symmetric density profile, from \eqs~(\ref{eq-05:matr2s}), (\ref{eq-05:sym})
and~(\ref{eq-05:K}) one finds for the probabilities 
$P_{\alpha\beta} \equiv |S_{\beta\alpha}|^2$: 
\bea
\label{eq-05:Pee}
    P_{ee} &= & |A_{ee}|^2= 1- |A_{e\tilde{2}}|^2-|A_{e\tilde{3}}|^2 \\
\label{eq-05:Pmue}
    P_{\mu e} &= & c_{23}^2 |A_{e\tilde{2}}|^2
    + s_{23}^2 |A_{e\tilde{3}}|^2 + 2\, s_{23}\, c_{23}\,
    {\rm Re}( e^{-i \delta} A_{e\tilde{2}}^* A_{e\tilde{3}} ) \,,
    \\
    \label{eq-05:Ptaue}
    P_{\tau e} &= & s_{23}^2 |A_{e\tilde{2}}|^2
    + c_{23}^2 |A_{e\tilde{3}}|^2 - 2\, s_{23}\, c_{23}\,
    {\rm Re}(e^{-i\delta} A_{e\tilde{2}}^* A_{e\tilde{3}}) \,,
    \\
    \label{eq-05:Pmumu}
    P_{\mu\mu} &= &|c_{23}^2 A_{\tilde{2}\tilde{2}}
    + s_{23}^2 A_{\tilde{3}\tilde{3}}
    + 2\, s_{23}\, c_{23}\, \cos\delta A_{\tilde{2}\tilde{3}}|^2 \,,
    \\
    \label{eq-05:Pmutau}
    P_{\mu\tau} &=&
    |s_{23}\, c_{23} (A_{\tilde{3}\tilde{3}} - A_{\tilde{2}\tilde{2}})
    + (\cos 2\theta_{23}\, \cos\delta + i \sin\delta)
    A_{\tilde{2}\tilde{3}}|^2 \,.
\eea
For antineutrinos the amplitudes can be obtained  
from the results presented above substituting
\begin{equation}
    \label{eq-05:pranti}
    \delta \to -\delta, \quad A_{ij} \to \bar{A}_{ij}, 
    ~~{\rm where}~~
    \bar{A}_{ij} \equiv A_{ij}(V \to -V).
\end{equation}
Notice that the amplitudes of transitions (\ref{eq-05:Pmue}) and
(\ref{eq-05:Ptaue}), that involve $\nu_e$, are given by linear combinations
of two propagation-basis amplitudes. The other flavor amplitudes depend on
three propagation-basis amplitudes.

\subsubsection{Factorization approximation and amplitudes for constant density}  
As follows immediately from the form of
the Hamiltonian $\tilde{H}$ in \eq~(\ref{eq-05:matr1}), in the limits
$\Delta m^2_{21} \to 0$ or/and $s_{12} \to 0$ the state $\tilde{\nu}_2$
decouples from the rest of the system, and consequently, the amplitude
$A_{e\tilde{2}}$ vanishes. In this limit, $A_{e\tilde{3}}$ (as well as
$A_{\tilde{3}\tilde{3}}$ and $S_{ee}$) is reduced to a $2\nu$
amplitude which depends on the parameters $\Delta m^2_{31}$ and
$\theta_{13}$: $A_A (\Delta m^2_{31}, \theta_{13}) \equiv
    A_{e\tilde{3}} (\Delta m^2_{21} = 0)$.
The corresponding probability equals $P_A \equiv |A_A|^2$.

In the limit $s_{13}\to 0$ the state $\tilde{\nu}_3$ decouples while the
amplitude $A_{e\tilde{3}}$ vanishes and the amplitude
$A_{e\tilde{2}}$ reduces to a $2\nu$ amplitude depending on the
parameters of the 1-2 sector, $\Delta m^2_{21}$ and $\theta_{12}$. Denoting
this amplitude by $A_S$ we have $ A_S (\Delta m^2_{21}, \theta_{12}) \equiv A_{e\tilde{2}}(\theta_{13} = 0)$.
We will use the notation $P_S \equiv |A_S|^2$.

This consideration implies that to the leading non-trivial order in
the small parameters $s_{13}$ and $r_\Delta$ the amplitudes
$A_{e\tilde{2}}$ and $A_{\tilde{2}e}$ are reduced to two neutrino probabilities and
depend only on the ``solar'' parameters, whereas the amplitudes
$A_{e\tilde{3}}$ and $A_{\tilde{3}e}$ -- 
only on the ``atmospheric'' parameters:
\be
    \label{eq-05:factor}
A_{e\tilde{2}}  \simeq  A_{\tilde{2}e} \simeq  A_S(\Delta m^2_{21}, \theta_{12}) \,,
\quad
A_{e\tilde{3}}  \simeq  A_{\tilde{3}e}
     \simeq A_A(\Delta m^2_{31}, \theta_{13}). 
\ee
The approximate equalities in 
\eq~(\ref{eq-05:factor}) are called the factorization approximation.

Due to the level crossing phenomenon the factorization approximation
(\ref{eq-05:factor}) is not valid in the energy range  
of the 1-3 resonance where the 1-3 mixing in matter is enhanced. 
In the case of a matter with an arbitrary density profile, one
can show, using simple power counting arguments, that the corrections
to the factorization approximation for the amplitude $A_{e\tilde{2}}$
are of order $s_{13}^2$, whereas the corrections to the
``atmospheric'' amplitude $A_{e\tilde{3}}$ are of order
$r_\Delta$~\cite{05-Takamura:2005df}, in agreement with our consideration
for constant density. The amplitude $A_{e\tilde{3}}$ does not in general  
have a 2-flavor form, once the corrections to the
factorization approximation are taken into account.\\

Using the expressions for $U_{e i}^{m}$ and $U_{\mu i}^m$ in terms of
the mixing angles in the standard parametrization, we can rewrite
\eq~(\ref{eq-05:me-ampl}) as
\begin{equation}
    S_{e \mu}^{\rm cst} = \cos\theta_{23}^m A_{e\tilde{2}}^{\rm cst}
    + \sin\theta_{23}^m e^{- i\delta^m} A_{e\tilde{3}}^{\rm cst} \,,
\end{equation}
where
\be
    \label{eq-05:ample2}
    A_{e\tilde{2}}^{\rm cst}
    \equiv  -i\, e^{i \phi_{21}^m} \cos\theta_{13}^m \,
    \sin 2\theta_{12}^m \, \sin\phi_{21}^m \,, \quad
    A_{e\tilde{3}}^{\rm cst}
    \equiv  -i \, e^{i \phi_{21}^m} \sin 2\theta_{13}^m
    \left[\sin \phi_{32}^m \, e^{-i\phi_{31}^m}
    + \cos^2 \theta_{12}^m \sin\phi_{21}^m \right] \,.
\ee
Here $\phi_{31}^m = \phi_{32}^m + \phi_{21}^m$. Since to a good
approximation $\theta_{23}^m \approx \theta_{23}$ and $\delta^m \approx \delta$  
(see \Sec~\ref{sec-05:3mixing}) \cite{05-Toshev:1991ku,05-Freund:2001pn}, the amplitudes
$A_{e\tilde{2}}^{\rm cst}$ and $A_{e\tilde{3}}^{\rm cst}$ can be identified with
$A_{e\tilde{2}}$ and $A_{e\tilde{3}}$ in \eq~(\ref{eq-05:Pmue}) and
(\ref{eq-05:Ptaue}).

In terms of mixing angles, $U_{\mu 1}^m = -s_{12}^m c_{23}^m  - c_{12}^m s_{13}^m s_{23}^m e^{i\delta^m}$,
$U_{\mu 3}^m = c_{13}^m s_{23}^m$,
the amplitude $S_{\mu\mu}^{\rm cst}$ can be rewritten as
\begin{equation}
    \label{eq-05:mm-ampl-a}
    S_{\mu\mu}^{\rm cst} = \cos^2 \theta_{23}^m A_{\tilde{2}\tilde{2}}^{\rm cst} 
    + \sin^2 \theta_{23}^m A_{\tilde{3}\tilde{3}}^{\rm cst}
    + \sin 2\theta_{23}^m \cos\delta^m A_{\tilde{2}\tilde{3}}^{\rm cst} \,,
\end{equation}
where
\bea
\label{eq-05:ample22}
    A_{\tilde{2}\tilde{2}}^{\rm cst}
    &\equiv & 1 + 2i\, e^{i \phi_{21}^m} \sin^2 \theta_{12}^m \sin\phi_{21}^m \,,
    \\
    \label{eq-05:ample33}
    A_{\tilde{3}\tilde{3}}^{\rm cst}
    &\equiv & 1 - 2i\, e^{-i \phi_{32}^m} \cos^2 \theta_{13}^m \sin\phi_{32}^m
    + 2i\, e^{i \phi_{21}^m} \sin^2\theta_{13}^m
    \cos^2\theta_{12}^m \sin\phi_{21}^m \,,
    \\
    \label{eq-05:ample23}
    A_{\tilde{2}\tilde{3}}^{\rm cst}
    &\equiv & i\, e^{i \phi_{21}^m} \sin\theta_{13}^m
    \sin 2\theta_{12}^m \sin\phi_{21}^m \,.
\eea
Notice that $A_{\tilde{2}\tilde{2}}^{\rm cst}$ has exactly the form of the
corresponding $2\nu$ amplitude driven by the solar parameters. The
amplitude $A_{\tilde{3}\tilde{3}}^{\rm cst}$ also coincides to a very good
approximation with the corresponding $2\nu$ amplitude driven by the
atmospheric parameters. In the approximation $\theta_{23}^m \approx
\theta_{23}$ and $\delta^m \approx \delta$ the amplitudes
(\ref{eq-05:ample22}), (\ref{eq-05:ample33}) and (\ref{eq-05:ample23}) can be
identified with the corresponding amplitudes in the propagation basis.

\subsubsection{Properties of the flavor oscillation probabilities}

1). $\nu_e - \nu_e$ channel.  The total probability of the $\nu_e$ disappearance equals 
\be
1 - P_{ee} = P_{e \mu} + P_{e \tau} =  P_{e\tilde{2}} + P_{e\tilde{3}}. 
\label{eq-05:3-tran}
\ee
The probability $P_{ee}$ does not depend on the
CP-violating phase  and the 2-3 mixing in the standard parametrization. 
The interference of the solar and atmospheric modes in $P_{ee}$ 
originates mainly from $P_{e\tilde{3}} \equiv |A_{e\tilde{3}}|^2$. 
The survival probability then equals
$P_{ee} = 1 - P_{e \mu} - P_{e \tau} = 1 - P_A - P_S$.
At high energies, where the effects of the 1-2 mixing 
and mass splitting in $P$ are suppressed, the probability is
$P_{ee} \approx 1 - P_{e \tau} \approx 1 - P_A$.  \\

2).  $\nu_e - \nu_\mu$ and $\nu_e - \nu_\tau$ channels. 
The transition probability $P_{\mu e} \equiv P(\nu_\mu \to \nu_e)$
(see (\ref{eq-05:Pmue})) can be rewritten as
\begin{equation}
    \label{eq-05:Pmue1}
    P_{\mu e} = c_{23}^2 |A_{e\tilde{2}}|^2
    + s_{23}^2 |A_{e\tilde{3}}|^2
    + \sin 2\theta_{23} |A_{e\tilde{2}}^* A_{e\tilde{3}}|
    \cos(\phi - \delta) \,,
\end{equation}
where $\phi \equiv \arg(A_{e\tilde{2}}^* A_{e\tilde{3}})$.
Unlike $1-P_{ee}$, this probability contains the interference term
between $A_{e\tilde{2}}$ and $A_{e\tilde{3}}$ which depends on the
CP-violation phase. 
 
Since the amplitude $A_{e\tilde{2}}$ is suppressed at high energies
due to the smallness of the 1-2 mixing in matter, in the lowest
approximation we have
\begin{equation}
    P_{\mu e} \approx \sin^2 \theta_{23} |A_{e\tilde{3}}|^2
    \approx \sin^2 \theta_{23} |A_A|^2.
\end{equation}
The maximal value of the probability equals $P_{\mu e} \simeq s_{23}^2$.

According to \eqs~(\ref{eq-05:Pmue}) and (\ref{eq-05:Ptaue}) the oscillation
probabilities $P_{\tau e}$ and $P_{e\tau}$ can be obtained from the
corresponding probabilities $P_{\mu e}$ and $P_{e\mu}$ through the
substitution $s_{23} \to c_{23}$, $c_{23} \to
-s_{23}$~\cite{05-Akhmedov:2004ny}. The interference term has the
opposite signs for channels including $\nu_\tau$ as compared with
those with $\nu_{\mu}$, which can be obtained from the unitarity
condition $P_{ee} + P_{\mu e} + P_{\tau e} = 1$ and the fact that
$P_{ee}$ does not depend on $\delta$.\\

3). The $\nu_\mu$ survival probability, $P_{\mu\mu}$, for symmetric
density profiles, \eq~(\ref{eq-05:Pmumu}), can be rewritten as
\be
\label{eq-05:pmumutot}
    P_{\mu\mu}  = 
    |c_{23}^2 A_{\tilde{2}\tilde{2}} + s_{23}^2 A_{\tilde{3}\tilde{3}}|^2
   +  2\sin 2\theta_{23} \cos\delta\, {\rm Re} \left[
    A_{\tilde{2}\tilde{3}}^* (c_{23}^2 A_{\tilde{2}\tilde{2}}
    + s_{23}^2 A_{\tilde{3}\tilde{3}}) \right]
  +  \sin^2 2\theta_{23} \cos^2\delta |A_{\tilde{2}\tilde{3}}|^2 \,.
\ee
Since
$A_{\tilde{2}\tilde{3}} = \mathcal{O}(r_\Delta s_{13})$ is a small
quantity, to a good approximation one can neglect the term $\sim
\cos^2\delta$ in \eq~(\ref{eq-05:Pmumu}), which is proportional to
$|A_{\tilde{2}\tilde{3}}|^2$.

For high energies in the limit $\Delta m^2_{21} \to 0$ we have $A_{\tilde{2}\tilde{2}} = 1$,
$A_{\tilde{2}\tilde{3}} = 0$. Then, 
parameterizing the 33-amplitude as $A_{\tilde{3}\tilde{3}} = \sqrt{1 - P_A} e^{-i\phi_{\tilde{3}\tilde{3}}^m}$ we obtain from (\ref{eq-05:pmumutot})
\begin{equation}
    \label{eq-05:mm-lim}
    P_{\mu\mu} (\Delta m^2_{21} = 0)
    = 1 - \sin^2 2\theta_{23} \sin^2 \phi_x
    - s_{23}^4 P_A -  0.5 \sin^2 2\theta_{23}
    \cos 2\phi_X (1 - \sqrt{1 - P_A}) \, , 
\end{equation}
where  $\phi_X = 0.5 arg[A_{\tilde{3}\tilde{3}}^* A_{\tilde{2}\tilde{2}}] =
\phi_{\tilde{2}\tilde{2}}^m - \phi_{\tilde{3}\tilde{3}}^m$. 
The probability can be rewritten as 
\begin{equation}
    \label{eq-05:mm-lim1}
    P_{\mu\mu} 
    = 1 -  0.5 \sin^2 2\theta_{23} 
    - s_{23}^4 P_A + 0.5 \sin^2 2\theta_{23} (\sqrt{1 - P_A}) \cos 2\phi_X\, .  
\end{equation}

4). $\nu_\mu - \nu_\tau$ channel. 
For symmetric matter density profiles the probability 
of $\nu_\mu \to \nu_\tau$ oscillations is given in \eq~(\ref{eq-05:Pmutau}). It can be
rewritten as
\bea
\label{eq-05:mmtt}
    P_{\mu\tau} & = & \frac{1}{4}\sin^2 2\theta_{23} |A_{\tilde{2}\tilde{2}}
    - A_{\tilde{3}\tilde{3}}|^2
    + \sin 2\theta_{23} \cos 2\theta_{23} 
    \cos\delta\, {\rm Re} \left[ (A_{\tilde{3}\tilde{3}}^* - A_{\tilde{2}\tilde{2}}^*)
    A_{\tilde{2}\tilde{3}} \right]
\nonumber    \\
    & - &\sin 2\theta_{23} \sin\delta \,
    {\rm Im}  \left[ A_{e\tilde{2}}^* A_{e\tilde{3}} \right]
    + (1 - \sin^2 2\theta_{23} \cos^2 \delta) |A_{\tilde{2}\tilde{3}}|^2 \,.
\eea
The amplitude depends on $\delta$ through the
terms proportional to $\cos\delta$ and $\sin\delta$, and therefore
$P_{\mu\tau}$ contains both CP- and T-even and odd terms.  
One can show that the $\delta$-dependent interference terms, which are
proportional to $\sin\delta$ and $\cos\delta$, satisfy the
relation
$P_{\mu\tau}^{\delta} = - P_{\mu e}^{\delta} - P_{\mu\mu}^{\delta}$. 
In the limit $\Delta m^2_{21} \to 0$ we obtain
\begin{equation}
    \label{eq-05:mt-lim1}
    P_{\mu\tau} 
    = 0.5 \sin^2 2\theta_{23} 
    - s_{23}^2 c_{23}^2 P_A -  0.5 \sin^2 2\theta_{23} (\sqrt{1 - P_A}) \cos 2\phi_X\, . 
\end{equation}

\section{Matter effects and determination of neutrino mass hierarchy}
\label{sec-05:masshierarchy}

\subsection{Propagation of neutrinos through the Earth}
\label{sec-05:propthroughearth}

Flavor neutrino evolution in the Earth is essentially oscillations 
in a multi-layer medium with slowly changing density in the individual layers 
and sharp density change on the borders of layers. 
For energies $E > 0.1$~GeV, possible short-scale
inhomogeneities of the matter  distribution 
can be neglected  and the density profile experienced by neutrinos 
is symmetric with respect to the midpoint of the trajectory:  
\begin{equation}
    V(x) = V(L - x).
\end{equation}
Here $L = 2 R_\oplus |\cos\theta_z|$ 
is the length of the trajectory inside the Earth,  
$R_\oplus = 6371$~km is the Earth radius  
and  $\theta_z$ is the zenith angle related to the nadir angle as 
$\Theta_\nu = \pi - \theta_z$. 
For $0\le \Theta_\nu \le 33.1^\circ$
neutrinos cross both the mantle and the core of the Earth, whereas for
larger values of the nadir angle they only cross the Earth's mantle.
The column density of the Earth along the diameter equals $d_{Earth} = 
\int n(x) dx$, which is bigger than  the minimal width;  
the size of the Earth is comparable with the neutrino refraction length. 

For the 1-2 channel, the adiabaticity 
is well satisfied  for all energies. We can therefore use the adiabatic approximation.  
The results of the evolution are determined by the mixing at the surface of the Earth 
and by the adiabatic phase. 
In the 1-3 channel the adiabaticity is  broken at the resonance. Thus, the constant 
density approximation with the average density works well in this regime. 
For energies below the resonance the matter effect becomes small and 
the constant density approximation and the adiabatic approximation 
give very similar results. 

For the core crossing trajectories, the profile consists 
of three layers in the first approximation: (i) mantle (with increasing density); 
(ii) core (with a symmetric profile) 
and (iii) second mantle layer (with decreasing density). 
This second mantle layer is T-inverted with respect to the first. 
In this approximation the profile can be considered as three layers of constant 
effective densities. As such, it looks like a part 
(1.5 period) of the castle wall profile. 
Consequently,  the parametric enhancement of oscillations,  
and in particular,  the parametric resonance  can be realized.

\subsubsection{Neutrino oscillograms of the Earth}
\label{sec-05:oscillograms}

\begin{figure}[h!!]
\begin{center}
\vspace{1cm}
\includegraphics[height=6cm,angle=0]{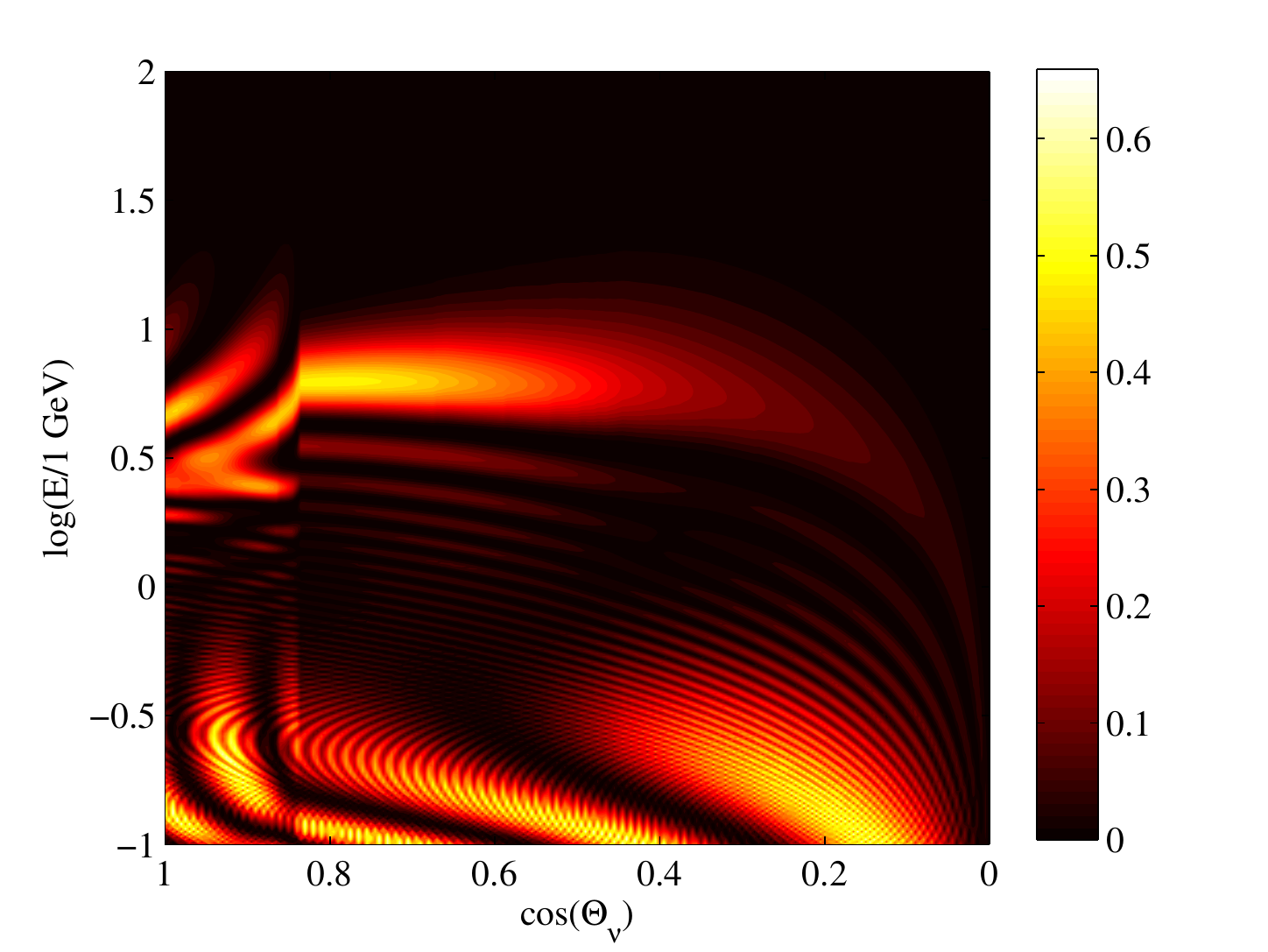}
\includegraphics[height=6cm,angle=0]{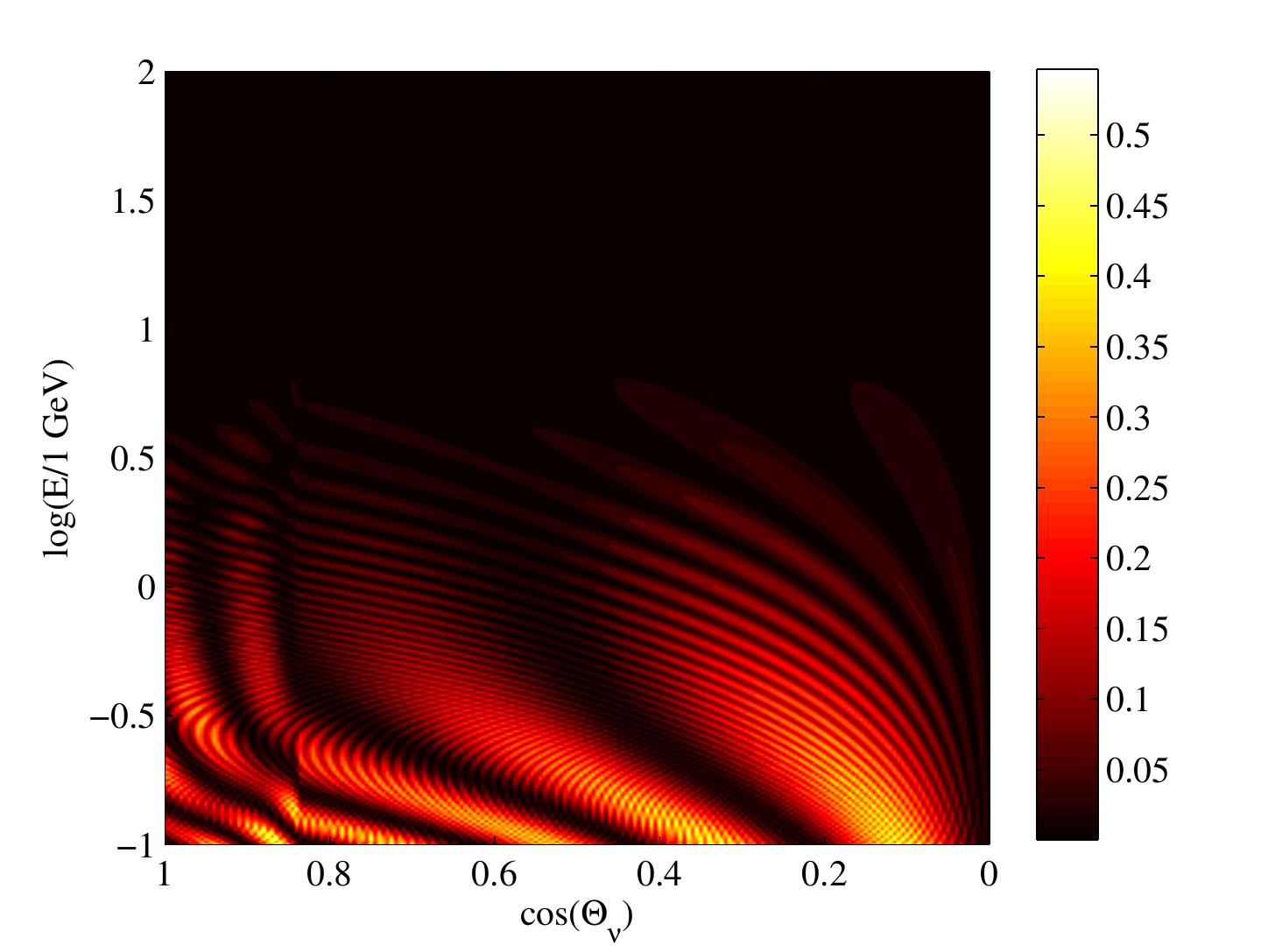}\\
\includegraphics[height=6cm,angle=0]{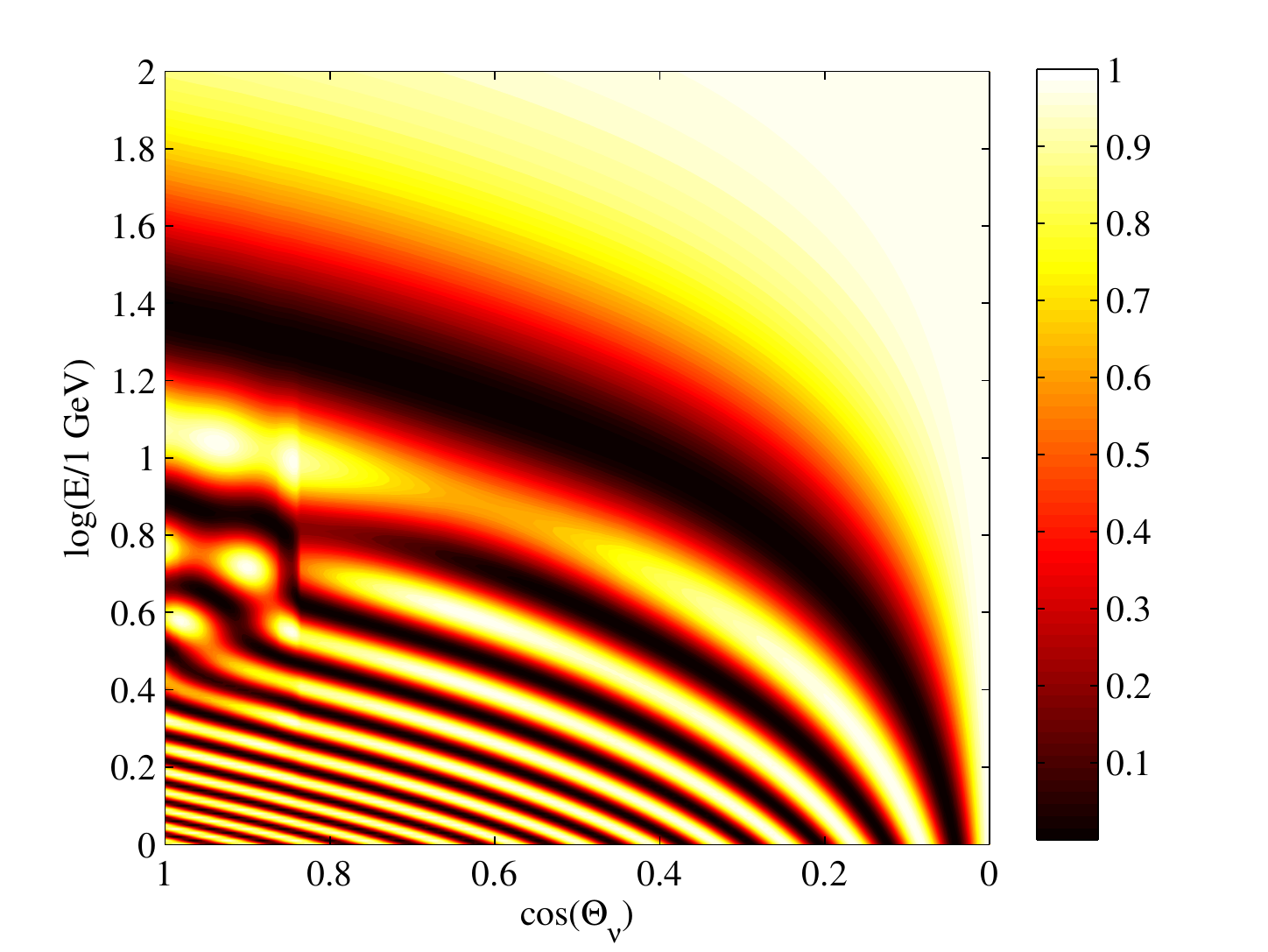}
\includegraphics[height=6cm,angle=0]{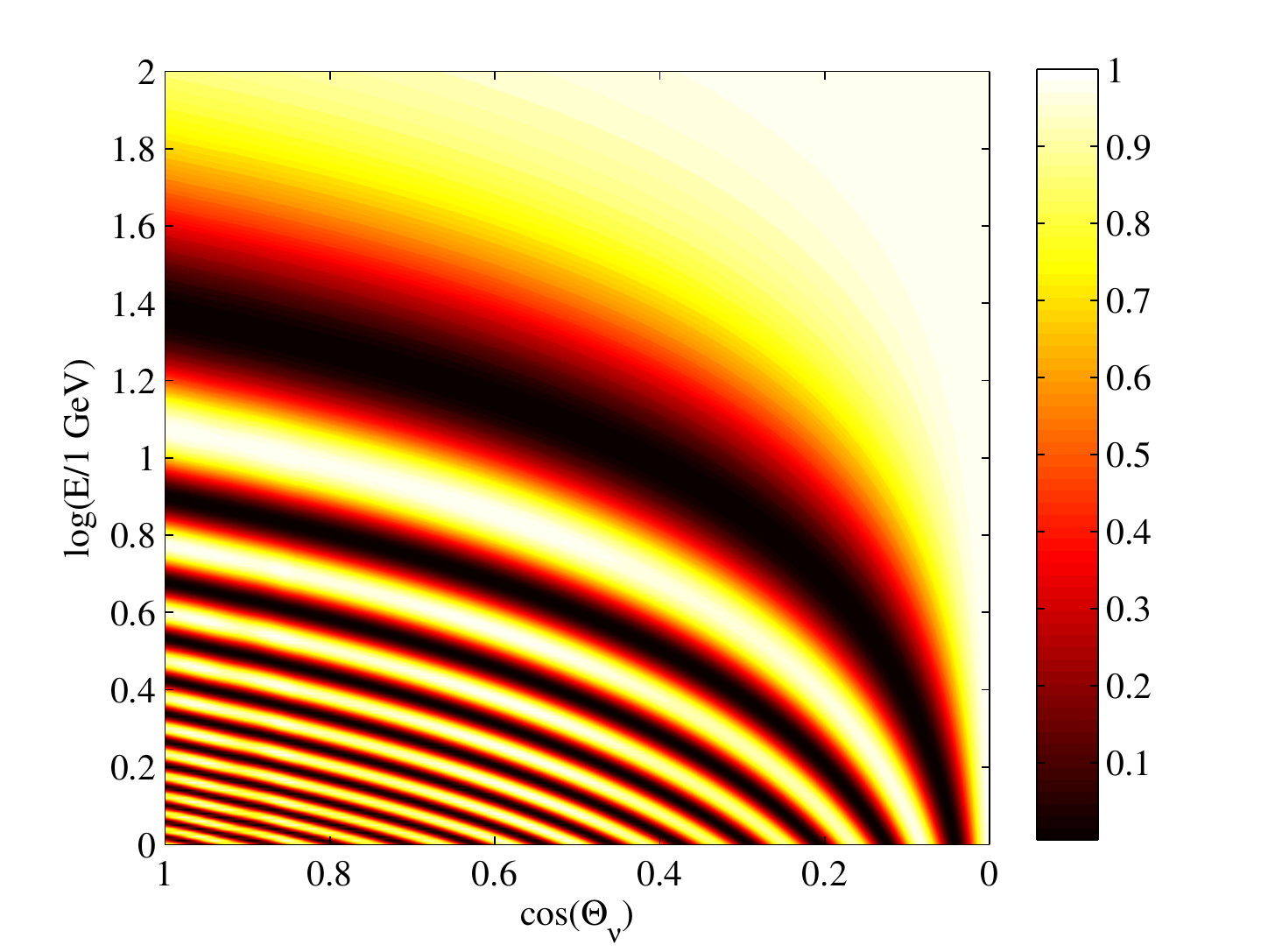}
\caption{Neutrino oscillograms of the Earth. Shown are the lines 
of equal flavor conversion  probability in the $E_\nu - \cos \Theta_\nu$ 
plane.  Upper panel: $\nu_e \rightarrow \nu_\mu$  (left) and 
$\bar{\nu}_e \rightarrow \bar{\nu}_\mu$ (right); 
bottom panel: $\nu_\mu \rightarrow \nu_\mu$  (left) and 
$\bar{\nu}_\mu \rightarrow \bar{\nu}_\mu$ (right). 
Normal hierarchy is assumed. 
\label{fig-05:oscillograms}}
\end{center}
\end{figure}

A comprehensive description of effects of neutrino passage through the Earth 
can be obtained in terms of neutrino oscillograms. 
The oscillograms are defined as lines of equal probabilities 
(or certain combinations of probabilities)  
in the $E_\nu - \cos \theta_z$ plane. 
In \Fig~\ref{fig-05:oscillograms},
we show the oscillograms for the oscillation probabilities $P_{e\mu}$ and $P_{\mu\mu}$, as well as the corresponding probabilities for antineutrinos~
\cite{05-Lipari:unp,05-Chizhov:1998ug,05-Ohlsson:1999um,05-Jacobson:2003wc,05-Kajita:2004ga,05-Akhmedov:2006hb}.

The structure of the oscillograms is well defined and unique, and reflects 
the structure of the Earth as well as the properties of the neutrinos themselves. 
In a sense, the oscillograms are the neutrino images of the Earth.
In contrast to usual light, there are several different images in different 
flavors as well as in neutrinos and antineutrinos.

\begin{figure}[h!!]
\begin{center}
\vspace{1cm}
\includegraphics[height=4.5cm,angle=0]{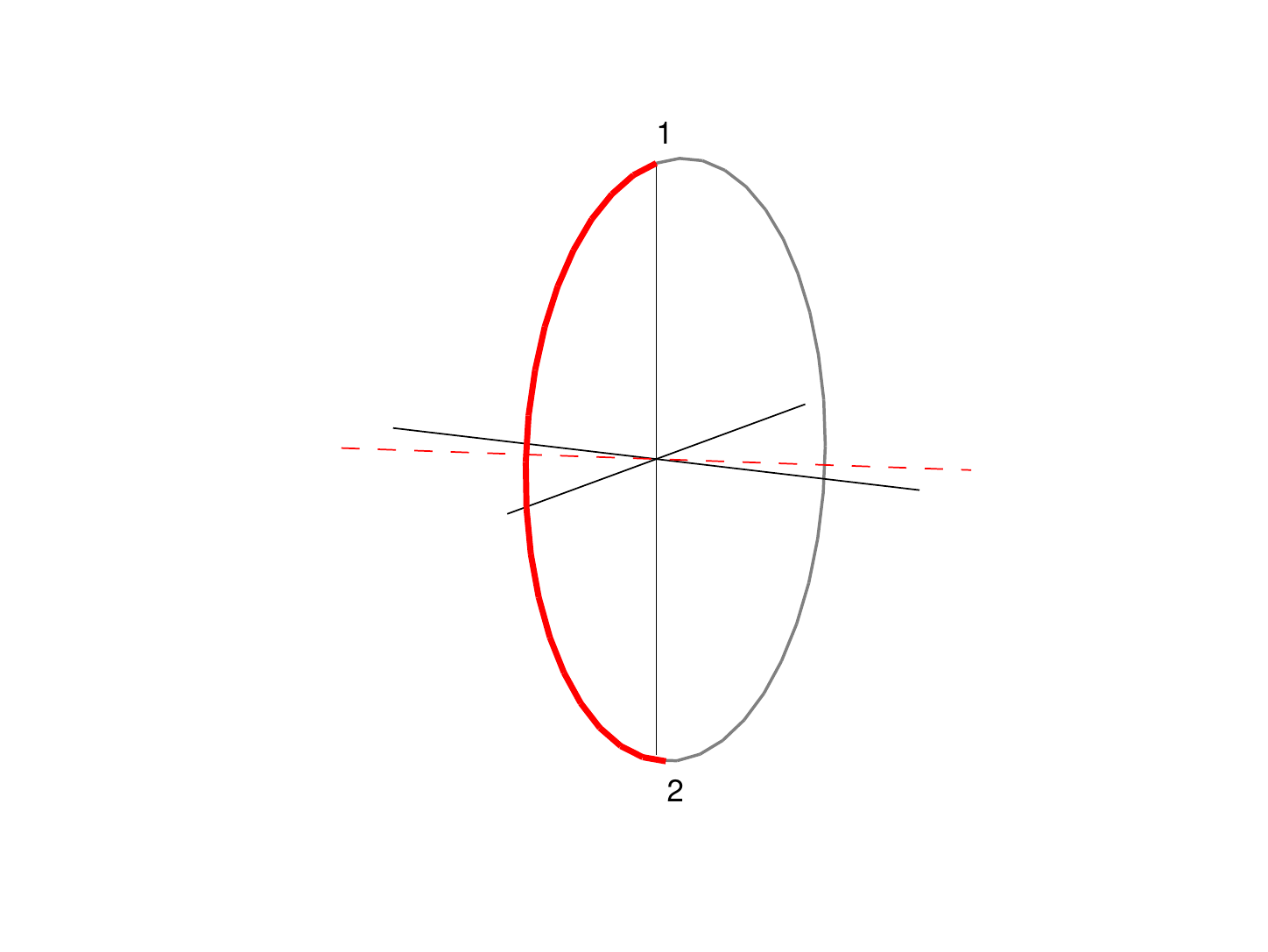} \hspace{-1.5cm}
\includegraphics[height=4.5cm,angle=0]{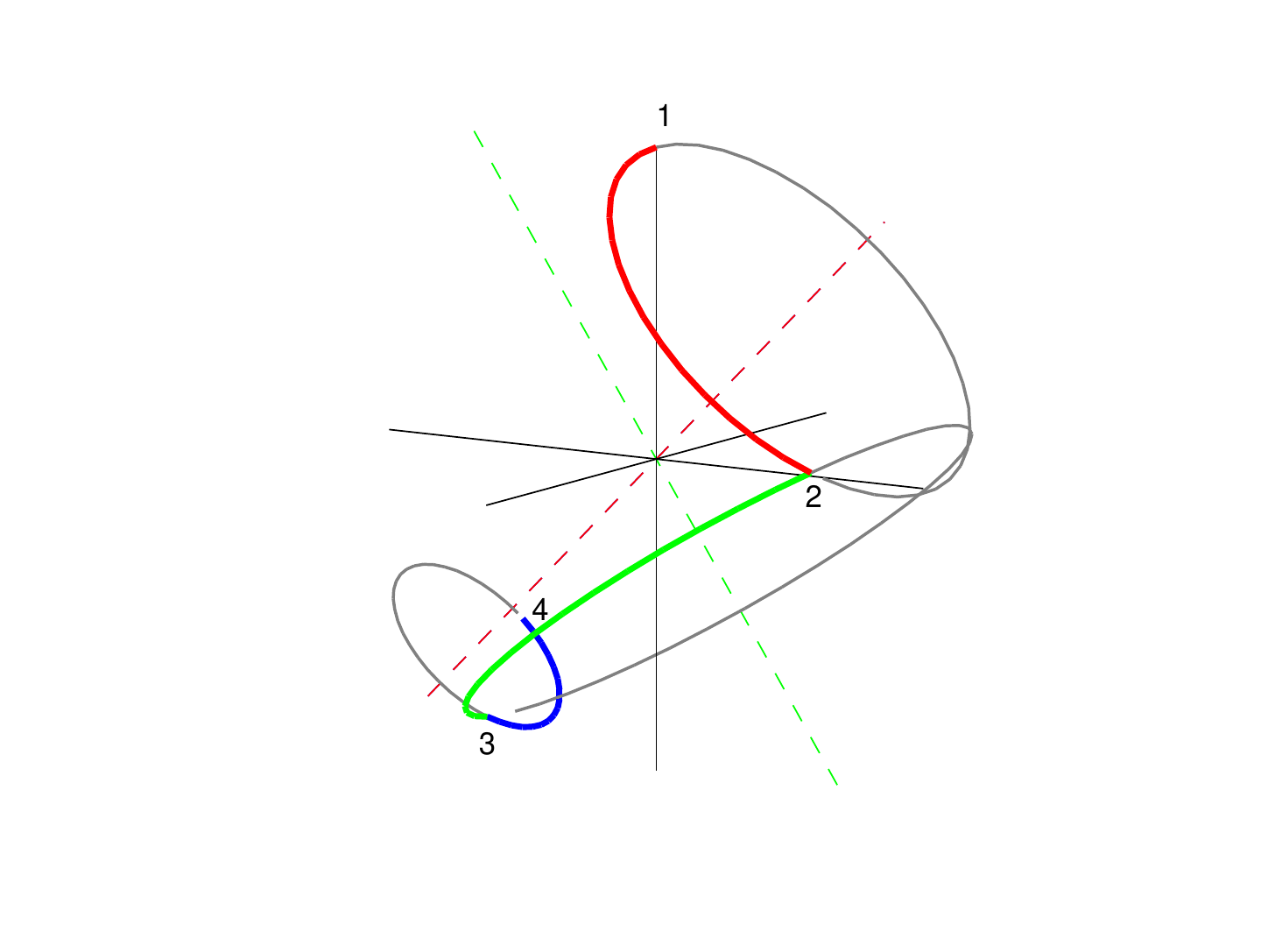} \hspace{-1.5cm}
\includegraphics[height=4.5cm,angle=0]{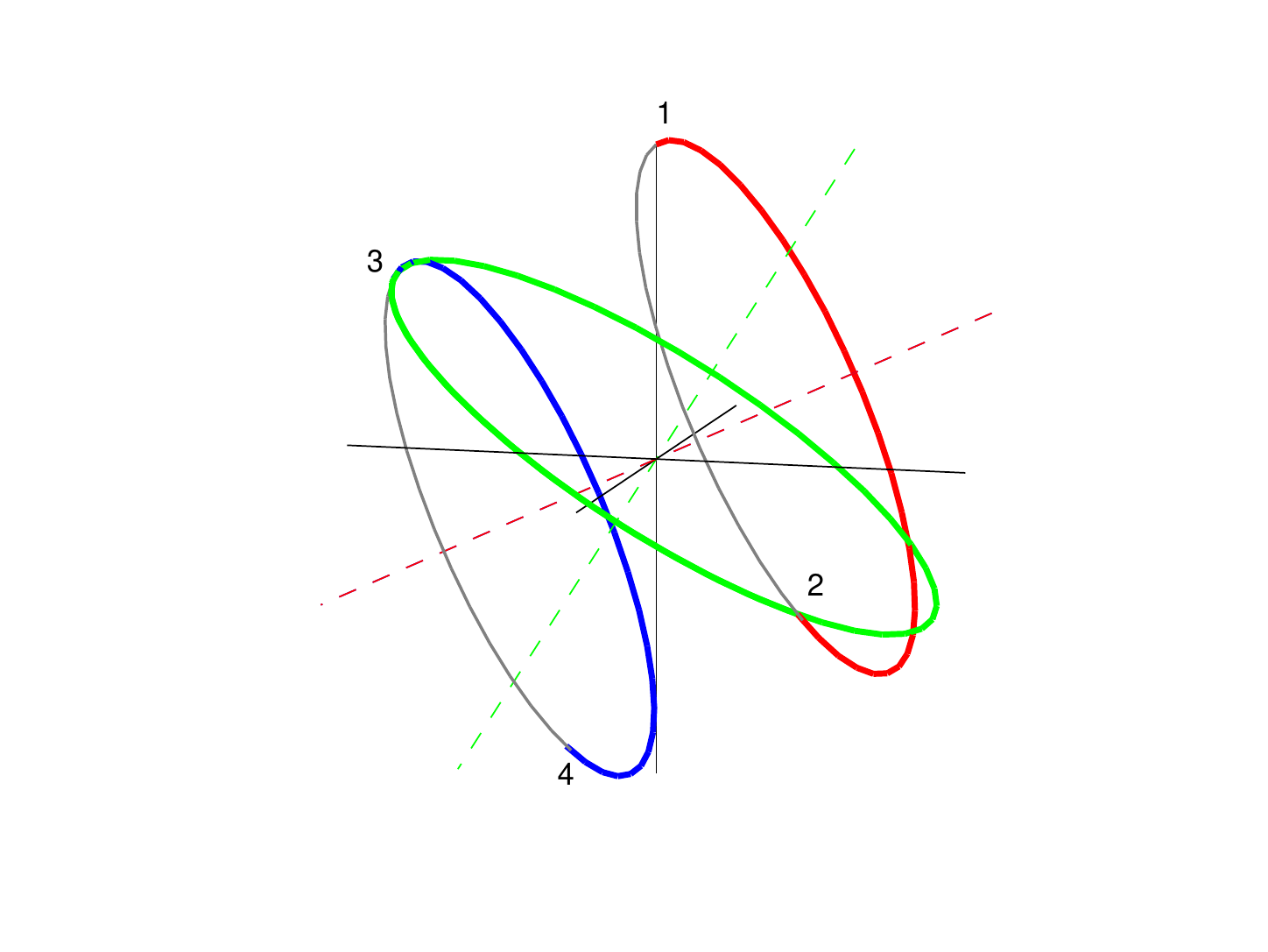}
\caption{Graphic representation of transition in different points of 
oscillogram: peak due to MSW resonance in the mantle (left),  
peak due to parametric enhancement of transition driven by 1-3 mixing (middle),
peak due to parametric enhancement of transition driven by 1-2 mixing (right).
In the left panel, neutrinos traverse only the mantle layer. In the right panel, neutrinos traverse the mantle (red), the core (green), and again the mantle (blue). The dashed lines correspond to the Hamiltonian vector ${\bf H}$ for the mantle (red) and core (green), respectively.
\label{fig-05:oscillogramGraphics}}
\end{center}
\end{figure}

The positions of all main structures of the oscillograms   
are determined by different realizations of
the amplitude condition and the phase condition.  
These are generalizations of the condition for maximal flavor 
transitions in the case of vacuum oscillations or oscillation in uniform matter.  
Recall that, in the latter case,   
$P = 1$ requires   
\begin{itemize}
\item $\sin^2 2\theta_m = 1$; the amplitude condition,
which is nothing but the MSW resonance condition, and 
\item $\phi = \pi/2 + \pi k$; the phase condition. 
\end{itemize}
At $E > 1$ GeV  the main structures of oscillograms 
are generated by the 1-3 mixing. 
They include:
\begin{enumerate}
\item  The MSW resonance pattern (resonance enhancement of the
 oscillations) for trajectories crossing only the mantle, with the main
peak at $E_\nu \sim (5 - 7)$ GeV.  The position of the maximum is
given by the MSW resonance condition:
\begin{equation} \label{eq-05:Eres}
    E_\nu = E_R (\Theta_\nu) =
    \frac{\Delta m ^2_{31} \cos 2 \theta_{13}}
    {2\bar{V}_1(\Theta_\nu)} \,,
\end{equation}
where $\bar{V}_1(\Theta_\nu)$ is the average value of the potential
along the trajectory characterized by $\Theta_\nu$.
The phase condition becomes $2\phi(E_\nu, \Theta_\nu) =
    2\omega(\bar{V}, E_\nu) L(\Theta_\nu) = \pi$
and the intersection of the resonance and the phase condition lines
gives the absolute maximum of $P_A$. Combining these conditions gives 
the coordinates of the peak: 
$\cos\Theta_\nu = 0.77$ and $E_R = 6$~GeV.


\item Three parametric ridges
in the domain of  core-crossing trajectories 
$|\cos \theta_z| > 0.87$
and $E_\nu > 3$~GeV. The parametric ridges differ by the oscillation phase
acquired in the core, $\phi_2$:

\begin{itemize}
\item[-] \emph{Ridge A} lies between the core resonance (at $\Theta_\nu \sim
0^\circ$) and the mantle resonance regions, $E_\nu \approx  3 - 6$~GeV. 
The phase in the core is $\phi_2 \aprle \pi$. 
This ridge merges with the MSW resonance peak in the
mantle.

\item[-] \emph{Ridge B} is situated at $E_\nu \geq 5$ GeV. For
the lowest energies in the ridge and $\Theta_\nu \sim 0$, the
half-phase in the core equals $\phi_2 \sim (1.2 - 1.3) \pi$. 

\item[-] \emph{Ridge C} is located at $E_\nu > 11$~GeV in the
matter dominated region, where the mixing, and consequently, oscillation
depth are suppressed. 
\end{itemize}

\item  The MSW resonance peak in core located at $E_\nu
\sim 2.5 - 2.8$~GeV.  

\item The regular oscillatory pattern at low energies
with ``valleys'' of zero
    probability and ridges in the mantle domain and a more complicated
    pattern with local maxima and saddle points in the core domain.
 
\end{enumerate}
In \Fig~\ref{fig-05:oscillogramGraphics}, we show graphic representations of oscillations which correspond to 
salient features of the oscillograms.

For energies $E_\nu < 1$~GeV the main structures are induced by the 1-2 mixing 
with small  
corrections due to 1-3 vacuum oscillations. 
Neglecting effect of $\theta_{13}$
we have $1 - P_{ee} = |A_{e\tilde{2}}|^2 \equiv P_S$.
The probabilities of the modes including $\nu_e$ are expressed in terms of a unique probability $P_S$. 

The 1-2 pattern differs from the pattern
for the 1-3 mixing due to the large value of the 1-2 mixing. 
  The oscillation length at the resonance is smaller than
    that for the 1-3 mixing,  
$l_m^R = l_\nu/\sin 2\theta_{12} \sim l_\nu$.
The resonance energy is shifted to smaller values both 
    due to $\Delta m^2_{21} \ll \Delta m^2_{31}$ and  because of the factor
    $\cos 2\theta_{12} \approx 0.4$:
$E_{12}^R = \frac{\Delta m^2_{21}}{2 \bar{V}} \cos 2\theta_{12}$.
    Here $\bar{V}$ is the average value of the potential.
The adiabaticity is better satisfied than for the 1-3
    mixing case and therefore the oscillation probability in the
    mantle is determined by the potential near the surface of the
    Earth $\bar{V}$ averaged over a distance of the order of the first
    oscillation length.
The oscillation length in matter $l_m$ monotonically increases with energy, approaching 
    the refraction length $l_0 \equiv
    2\pi/V$~(c.f., \Fig~\ref{fig-05:lmdependence}).
The jump of the mixing angle at the mantle-core boundary is
    small. Thus, the sudden distortion of the oscillation patterns
    at $\Theta_\nu = 33^\circ$ is not as significant as it is for the
    1-3 mixing, in particular below the 1-2 resonance energy.
These features allow to understand the structure of the
oscillograms.  In the mantle domain ($\Theta_\nu > 33^\circ$) the
oscillation pattern for neutrinos is determined by the resonance
enhancement of oscillations. There are three MSW resonance peaks above
$0.1$ GeV, which differ from each other by value of the total
oscillation phase. The outer peak ($\Theta_\nu \approx 82^\circ$)
corresponds to $\phi \approx \pi/2$, the middle ($\Theta_\nu =
60^\circ$) to $\phi \approx 3\pi/2$, and the inner ($\Theta_\nu
\approx 40^\circ$) to $\phi = 5\pi/2$. Recall that such a large phase
can be acquired due to the smaller resonance oscillation length in comparison to that of the 1-3 mixing case,
for which only one peak with $\phi = \pi/2$ can be realized. 
The resonance energy is given by \eq~(\ref{eq-05:12mix-res}), and for the surface
potential we find
\begin{equation}
    \label{eq-05:eressurf}
    E_{12}^R \approx 0.12~{\rm GeV} \,.
\end{equation}
The ratio of the 1-2 and 1-3 resonance energies equals ${E_{12}^R}/{E_{13}^R}  \approx \frac{1}{50}$.
The estimate
(\ref{eq-05:eressurf}) is valid for the two outer peaks. For the peak at
$\Theta_\nu = 40^\circ$, $\bar V$ is larger and, accordingly, the
resonance energy is slightly smaller.
The width of the 1-2 resonance is large and therefore the regions of
sizable oscillation probability are more extended in the $E_\nu$
direction as compared to the oscillations governed by the 1-3 mixing
and splitting.

The resonance energy in the core is $E_{12}^R \approx 0.04$ GeV.
Therefore for $E_\nu > (0.10-0.15)$ GeV the 1-2 mixing in the core is
substantially suppressed by matter. The peak at $E_\nu \simeq
0.2$ GeV and $\Theta_\nu \simeq 25^\circ$ is due to the parametric
enhancement of the oscillations. 
It corresponds to the realization of
the parametric resonance condition when the oscillation half-phases
equal approximately $\phi_{\rm mantle} \approx \pi/2$ and
$\phi_{\rm core} \approx 3\pi/2$ (note that the total phase is $\approx
5\pi/2$ and this parametric ridge is attached to the $5\pi/2$ MSW
peak in the mantle domain).

\subsubsection{ Oscillograms for the inverted mass hierarchy}  

The main change compared to the normal hierarchy is that 
the 1-3 resonance structure now appears in the antineutrino channel. 
The level crossing scheme is modified in comparison to NH. In the
neutrino channel there is only the 1-2 resonance. 

In the approximation of $\Delta m^2_{21} = 0$, the neutrino oscillograms for
the inverted hierarchy coincide with the antineutrino oscillograms for
the normal hierarchy and vice-versa, provided that $\Delta m^2_{31}$ is
taken to be the same in both cases~\cite{05-Akhmedov:2012ah}: $P_A^{IH} = \bar{P}_A^{NH},\ \phi_X^{IH} = - \bar{\phi}_X^{NH}$,
Therefore $P_{\alpha \beta}^{IH} = \bar{P}_{\alpha \beta}^{NH},\ 
\bar{P}_{\alpha \beta}^{IH} = P_{\alpha \beta}^{NH}$.
The inclusion of the 1-2 mixing
and mass splitting breaks this symmetry.

\subsection{CP-violation effects}
\label{sec-05:cpviol}

\subsubsection{Interference and CP-violation}

The survival probability $P_{ee}$ does not depend on the CP-violating
phase $\delta$ neither for oscillations in vacuum nor in
matter~\cite{05-Kuo:1987km,05-Minakata:1999ze}. This is the consequence of
the facts that $\delta$ can be removed by transforming to the
propagation basis and that $\nu_e$ is not affected
by this transformation. For oscillations in vacuum, or in
matter with symmetric density profiles, the other two survival
probabilities, $P_{\mu\mu}$  and $P_{\tau\tau}$, are T-even quantities dependent on 
$\delta$ only through terms proportional to $\cos\delta$ and $\cos
2\delta$~\cite{05-Yokomakura:2002av}. In
contrast to this, for oscillations in a matter with non symmetric density profiles, these probabilities also acquire terms proportional to
$\sin\delta$ and $\sin 2\delta$.

Introducing the phase $\phi \equiv \arg(A_{e\tilde{2}}^* A_{e\tilde{3}})$
and omitting small terms proportional to $|A_{\tilde{2}\tilde{3}}|^2 = \mathcal O(s_{13}^6)$  
we obtain 
\bea
    \label{eq-05:Pme-2}
    P_{\mu e}^\delta &= &\hphantom{-} \sin 2\theta_{23}
    \cos(\phi - \delta) \, |A_{e\tilde{2}} A_{e\tilde{3}}| \,,
    \\
    P_{\mu\mu}^\delta &= &-\sin 2\theta_{23} \cos\delta \cos\phi \,
    |A_{e\tilde{2}} A_{e\tilde{3}}| - D_{23} \,,
    \\
    P_{\mu\tau}^\delta & = & -\sin 2\theta_{23} \sin\delta \sin\phi \,
    |A_{e\tilde{2}} A_{e\tilde{3}}| + D_{23} \,,
\eea
where $D_{23} \equiv \frac{1}{2} \sin 4\theta_{23} \cos\delta~
    {\rm Re}\left[ A_{\tilde{2}\tilde{3}}^* (A_{\tilde{3}\tilde{3}}
    - A_{\tilde{2}\tilde{2}}) \right]$
is proportional to the small  deviation of the 2-3 mixing from maximal
one. Notice that $D_{23}$ enters  $P_{\mu\mu}^\delta$ and
$P_{\mu\tau}^\delta$ with opposite signs while $P_{\mu e}^\delta$ does
not depend on $D_{23}$ at all. $D_{23}$ is CP-even. 
The sum of these interference terms is zero. 

For the other channels,
$P_{\alpha \beta}^\delta = P_{\beta \alpha}^{-\delta}$.
For antineutrinos, according to (\ref{eq-05:pranti}), the probabilities
have the same form as the corresponding probabilities derived above
with a changed sign of $\delta$ and the amplitudes computed with the
opposite sign of the potential. 
Thus, the $\delta$ dependent parts in all the channels are expressed in
terms of two combinations of the propagation basis amplitudes,
$|A_{e\tilde{2}} A_{e\tilde{3}}|$ and $D_{23}$.

\subsubsection{Magic lines and CP-domains}

To better assess the effect of  $\delta$,
one can consider  the difference of the oscillation probabilities for two different
values of the CP-phase $\Delta P_{\alpha\beta}^{\rm CP}(\delta)
    \equiv P_{\alpha\beta}(\delta) - P_{\alpha\beta}(\delta_0) $.
In practice, this quantifies how well the phase $\delta$ fits with some assumed true value $\delta_0$.  
The structure of the oscillograms for 
$\Delta P_{\alpha\beta}^{\rm CP}(\delta)$ can be 
understood in terms of the
grids of magic lines and interference phase lines along which 
$\Delta P_{\alpha\beta}^{\rm CP}(\delta) \approx 0$. 

For the $\nu_\mu \to \nu_e$ oscillation probability, the equality
\begin{equation}
    \label{eq-05:diffpro}
    \Delta P_{\mu e}^{\rm CP}(\delta)
    \equiv P_{\mu e}(\delta) - P_{\mu e}(\delta_0)
    = P_{\mu e}^\delta(\delta) - P_{\mu e}^\delta(\delta_0)
\end{equation}
is exact and the condition $\Delta P_{\mu e}^{\rm CP} = 0$ is equivalent
to
\begin{equation}
    \label{eq-05:fact}
    |A_{e\tilde{2}} A_{e\tilde{3}}| \cos(\phi - \delta)
    = |A_{e\tilde{2}} A_{e\tilde{3}}| \cos(\phi - \delta_0) \,.
\end{equation}
This equality is satisfied if at least one of the following three
conditions is fulfilled
\be
\label{eq-05:abc}
     A_{e\tilde{2}}(E_\nu, \Theta_\nu)  = 0, \quad
      A_{e\tilde{3}}(E_\nu, \Theta_\nu) = 0, \quad
    \phi(E_\nu, \Theta_\nu) - \delta_0 =
    - \left[ \phi(E_\nu, \Theta_\nu) - \delta \right] + 2\pi l .
\ee
The last condition implies
\begin{equation}
    \label{eq-05:C}
    \phi(E_\nu, \Theta_\nu) = (\delta + \delta_0) / 2 + \pi l \,.
\end{equation}
Under the conditions (\ref{eq-05:abc}), the equality (\ref{eq-05:fact}) is 
satisfied
identically for all values of $\delta$. In these cases the transition
probability does not depend on the CP-phase.
Since the amplitudes $A_{e\tilde{2}}$ and  $A_{e\tilde{3}}$ 
are complex quantities, these conditions 
can be satisfied in isolated points of the ($\Theta_\nu, E_\nu$) plane only.
In contrast to this, in the factorization approximation 
$A_{e\tilde{2}}
= A_S$ and $A_{e\tilde{3}} = A_A$ both the conditions 
are fulfilled along certain lines in the oscillograms. 
This occurs because the amplitudes $A_S$
and $A_{A}$ take a 2-flavor form. In the basis of neutrino states where the 
corresponding $2 \times 2$ Hamiltonians are traceless, both $A_A$ and $A_S$ are pure
imaginary because of the symmetry of the Earth's density
profile~\cite{05-Akhmedov:2001kd}. 

Let us consider the equalities $A_{S} = 0$ and $A_{A} = 0$ 
using the constant density approximation:
%
\begin{figure}
 \begin{center}
  \includegraphics[trim=0mm 100.5mm 0mm 0mm, clip,width=120mm]{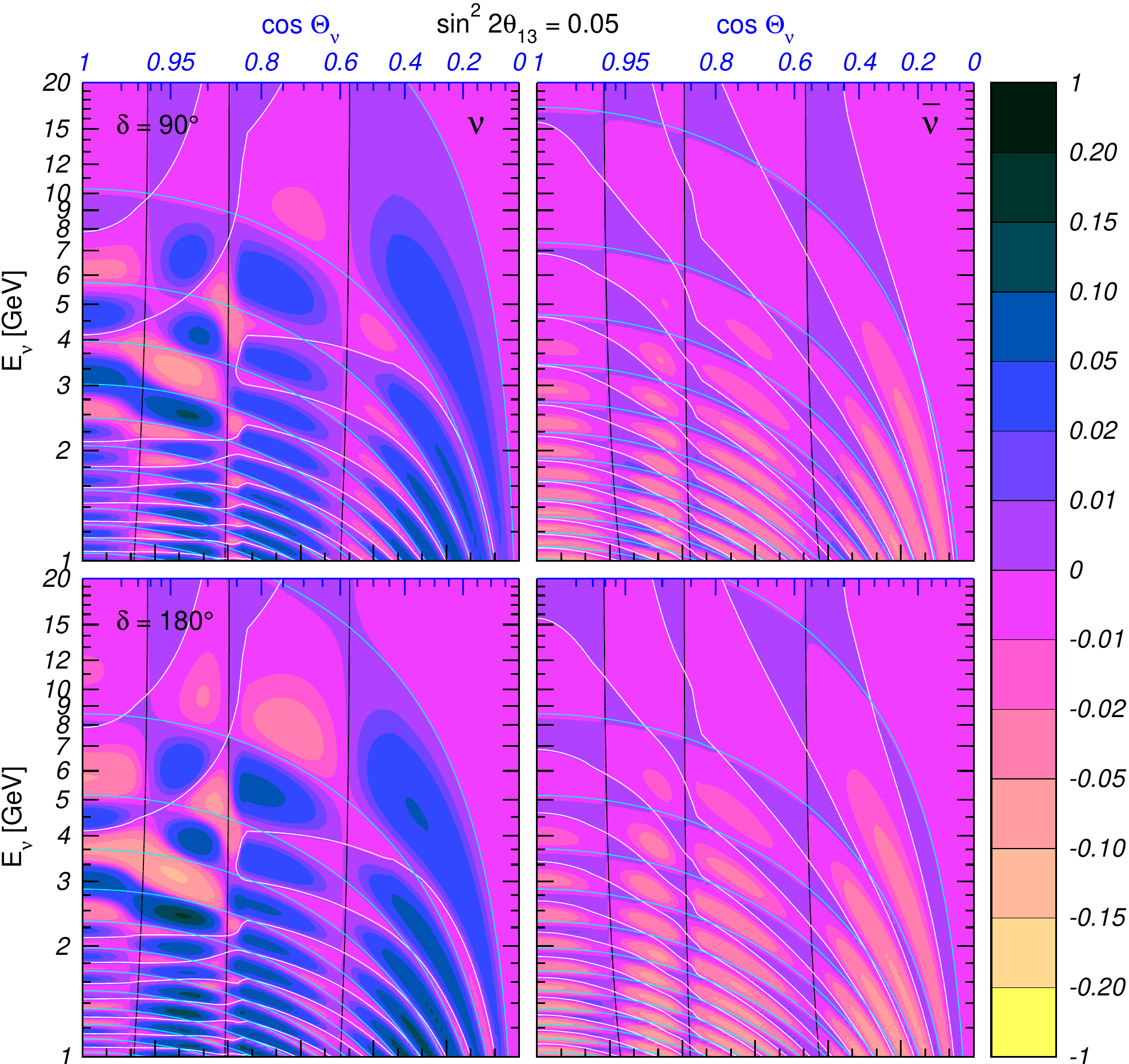}
   \label{fig-05:dcp-me}
   \caption{
    Oscillograms for the difference of probabilities $\Delta P_{\mu
    e}^{\rm CP}(\delta) = P_{\mu e}(\delta) - P_{\mu e}(\delta_0)$
    with $\delta_0 = 0^\circ$. Shown are the solar (black),
    atmospheric (white) and interference phase condition (cyan)
    curves. From~\cite{05-Akhmedov:2008qt}.}
 \end{center}
\end{figure}
%

1.  The condition $A_{S}(E_\nu, \Theta_\nu) = 0$ 
is satisfied when $\sin \phi_{S} (E_\nu, \Theta_\nu) = 0$,
which leads to 
\begin{equation}
    \label{eq-05:cond1}
    L(\Theta_\nu) \approx
    \frac{2\pi  n}{\omega^m_{21}},  \qquad n = 1,\, 2,\, \ldots
\end{equation}
At energies $E_\nu\aprge 0.5$ GeV  which are much higher than the 1-2 mixing MSW
resonance in the mantle and in the core of the Earth
one has  $\omega^m_{21} \approx V$ and the condition (\ref{eq-05:cond1}) 
becomes
\begin{equation}
    \label{eq-05:cond1a}
    L(\Theta_\nu)  \simeq \frac{2\pi n}{V} \,.
\end{equation}
This expression is energy independent and determines the baselines for
which the ``solar'' contribution to the probability vanishes~\cite{05-Smirnov:2006sm}.  In the
plane $(\Theta_\nu, E_\nu)$ it represents nearly vertical lines
at fixed $\Theta_\nu$.
There are three solar magic lines which correspond to $n = 1$ (in the
mantle domain) $\Theta_\nu \approx 54^\circ$ and $n = 2, 3$ (in the core domain)~\cite{05-Smirnov:2006sm} $\Theta_\nu \approx 30^\circ$ and $12^\circ$. 
The existence of a baseline ($L\approx 7600$ km) for which the probability of
$\nu_e\leftrightarrow \nu_\mu$ oscillations in the Earth is
approximately independent of the ``solar'' parameters ($\Delta m^2_{21}$,
$\theta_{12}$) and of the CP-phase $\delta$ was first pointed out
in~\cite{05-Barger:2001yr} and later discussed in e.g.,~\cite{05-Huber:2002uy,05-Huber:2003ak,05-Gandhi:2004bj,05-Blondel:2006su,05-Huber:2006wb,05-Smirnov:2006sm,05-Agarwalla:2007ai}. This baseline was dubbed ``magic''
in~\cite{05-Huber:2002uy}.\\

2.  The atmospheric magic lines are determined by the condition $A_A
(E_\nu, \Theta_\nu) = 0$~\cite{05-Smirnov:2006sm}. 
Along these lines, the
``atmospheric'' contribution to the amplitudes of $\nu_\mu
\leftrightarrow \nu_e$ and $\nu_\tau \leftrightarrow \nu_e$
transitions vanishes  and 
the probabilities of oscillations involving $\nu_e$ or $\bar{\nu}_e$ 
do not depend on CP-phase.
In the constant density approximation, the condition $A_A = 0$ is satisfied when
$\sin \phi_A = 0$ ($\phi_A = \pi k$, $k = 1, 2, \dots$) or explicitly
\begin{equation}
    \label{eq-05:cond2}
    L(\Theta_\nu) \approx \frac{2 \pi k}{\omega^m_{31}},
    \qquad k = 1,\, 2,\, \ldots
\end{equation}
For energies which are not too close to the 1-3 MSW resonance, it reduces to
\begin{equation}
    \label{eq-05:cond2a}
    E_\nu \simeq \frac{\Delta m^2_{31} L(\Theta_\nu)}
    {|4\pi k \pm 2 V L(\Theta_\nu) |} \,,
\end{equation}
which corresponds to the bent curves in the $(\Theta_\nu, E_\nu)$
plane. For very large energies, where $\Delta m^2_{31}/2E \ll V$, the
atmospheric lines approach the same vertical lines as the solar magic
lines (\ref{eq-05:cond1a}).\\

3.  The  condition (\ref{eq-05:C}) determines  the {\it interference phase 
lines} in  the ($\Theta_\nu, E_\nu$) plane. 
In the constant density approximation
$\phi \approx - \phi_{31}^m$. Consequently in the energy range
between the two resonances we have
\begin{equation}
    \label{eq-05:betw}
    \phi_{31}^m \approx \frac{\Delta m^2_{31} L}{4E_\nu} = \phi_A^0 \,,
\end{equation}
{\it i.e.}, in the first approximation $\phi$ does not depend on the matter
density. From \eq~(\ref{eq-05:C}) we then obtain
\begin{equation}
    \label{eq-05:intphase}
    E_\nu = \frac{\Delta m^2_{31} L(\Theta_\nu)}
    {4\pi l - 2(\delta + \delta_0)} \,.
\end{equation}

Thus, in the factorization approximation, the conditions 
(\ref{eq-05:abc}) and (\ref{eq-05:C}) define three sets of lines 
(grid of magic lines) in the oscillograms (see
\Fig~\ref{fig-05:dcp-me}), which play crucial roles in understanding
the CP violation effects. 
Along the lines, the probabilities $P_{\mu e}$, $P_{e\mu}$ $P_{\tau e}$ 
and $P_{e\tau}$ do not depend on the CP-phase in the first order approximation. 
The other probabilities  depend  on the phase weakly.

From \Fig~\ref{fig-05:dcp-me}, we can see that the magic lines described 
above do not coincide exactly with the lines of $\Delta P_{\mu e}^{\rm 
CP} = 0$ which bound the CP-domains.  
Furthermore, interconnections of the latter occur. 
This is due to the breakdown of the factorization approximation.

\subsection{Determination of hierarchy with accelerator experiments}
\label{sec-05:accelerator} 

\begin{figure}[h!!]
\begin{center}
\vspace{1cm}
\includegraphics[height=8cm,angle=0]{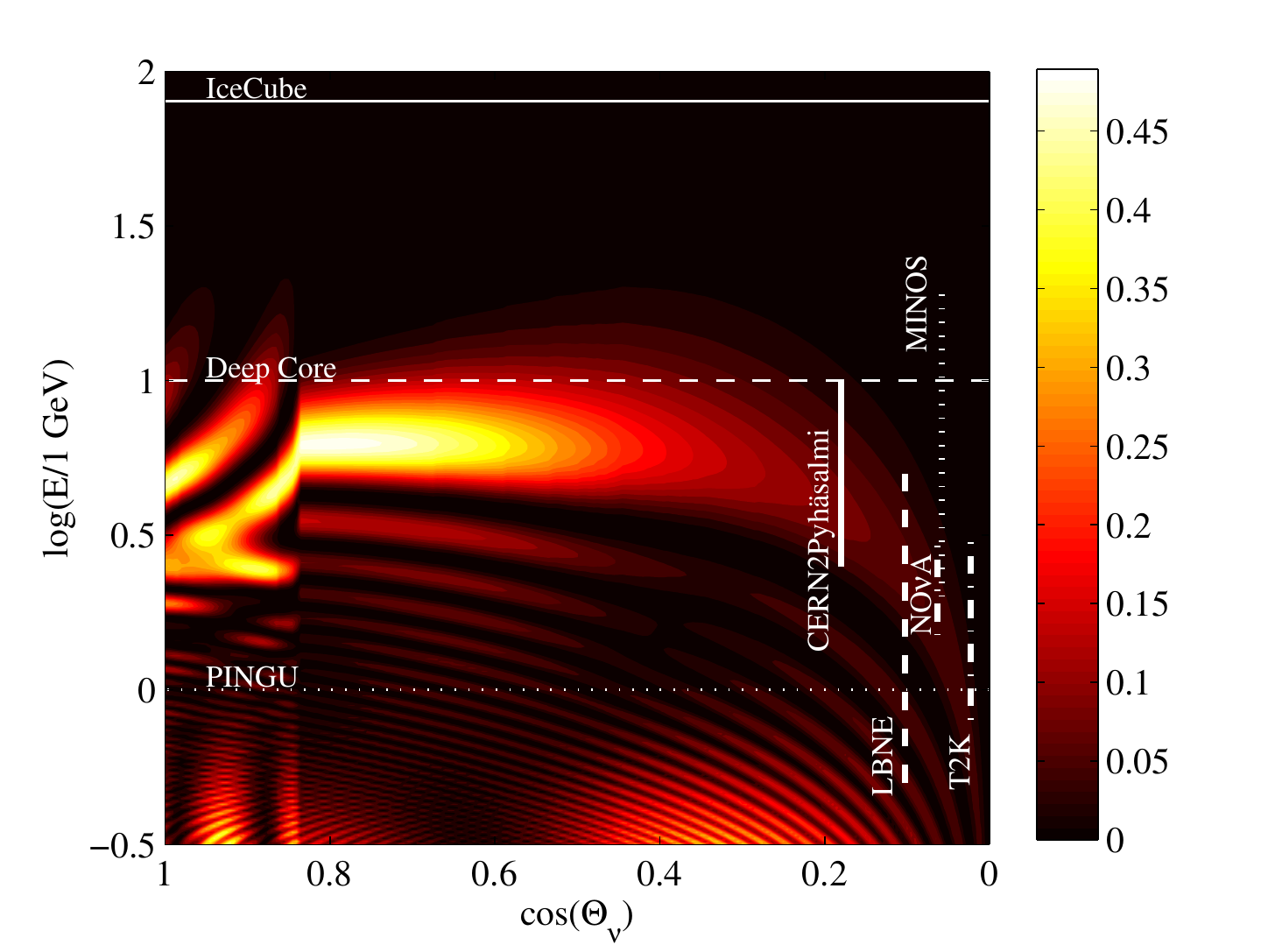}
\caption{Physics reach of the accelerator and 
atmospheric neutrino experiments. Shown are areas in the oscillogram for 
$\nu_e \rightarrow \nu_\mu$ channel which can be covered by 
different accelerator (vertical sections) and 
atmospheric neutrino experiments (sensitive to the area above the corresponding horizontal lines).    
\label{fig-05:oscillogram+exp}}
\end{center}
\end{figure}

An accelerator neutrino experiment has a fixed baseline which corresponds to a vertical line with the length determined by the available energy spectrum. In the oscillogram of \Fig~\ref{fig-05:oscillogram+exp} we have included such lines for a handful of accelerator experiments. Furthermore, this energy spectrum is usually peaked at certain energy
(or narrow energy range) resulting in the experiment being most sensitive to the oscillation probability at that specific energy. An accelerator neutrino experiment would typically 
run for several years in neutrinos or antineutrinos before switching polarity and therefore getting information both on $P_{\alpha\beta}$ and $\bar P_{\alpha\beta}$. The goal of such a search is to observe in which channel the oscillation probability is suppressed and in which it is enhanced. If a neutrino experiment could run at energy similar to the resonant one and at a baseline of several thousand kilometers, then this determination would be quite simple. However, as can be seen from the oscillogram, accelerator neutrino experiments are confined to relatively shallow trajectories with rather poor oscillatory pattern, 
and this severely limits their capabilities leading to various degeneracies. 
In particular, lack of knowledge of the mass hierarchy is part of the famous eightfold degeneracy, which arises as follows. Assume we have access to the values of oscillation 
probabilities 
$P_{\mu e}$ and $\bar P_{\mu e}$ at a given baseline $L$ and energy $E$ only. 
Then there exists three types of ambiguities that give rise to the same values 
of the  probabilities in different parts of the parameter space (mixing angles, 
CP phase, signs of mass differences).

\begin{enumerate}
 \item \emph{Sign (hierarchy) degeneracy:} This is the degeneracy due to the unknown neutrino mass hierarchy. Changing the mass hierarchy, it is often possible to find a point in parameter space that predicts the same oscillation probabilities.
 
 \item \emph{Intrinsic ($\theta_{13}$,$\delta$) degeneracy:} For any combination of ($\theta_{13}$,$\delta$), there exists a different combination ($\hat\theta_{13}$,$\hat\delta$) that also predicts the same oscillation probabilities.
 
 \item \emph{Octant ($\theta_{23}$) degeneracy:} Changing the octant of $\theta_{23}$ also leads to a degeneracy due to $\mu$-$\tau$ symmetry. If $\theta_{23}$ is close to maximal, the effects of this degeneracy are less pronounced.
\end{enumerate}

\begin{figure}[h!!]
\begin{center}
\vspace{1cm}
 \includegraphics[width=0.45\textwidth,angle=0]{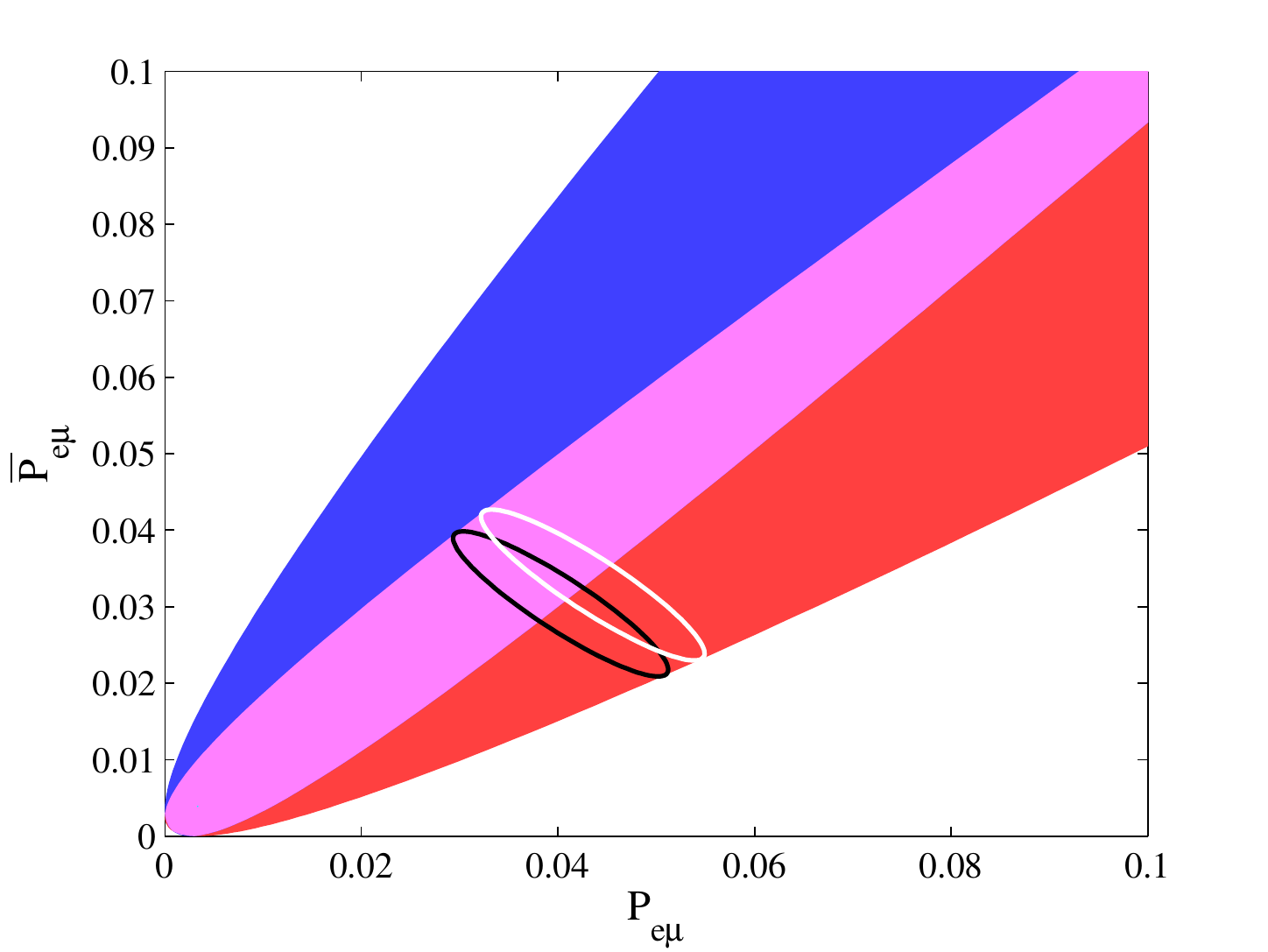} %
 \includegraphics[width=0.45\textwidth,angle=0]{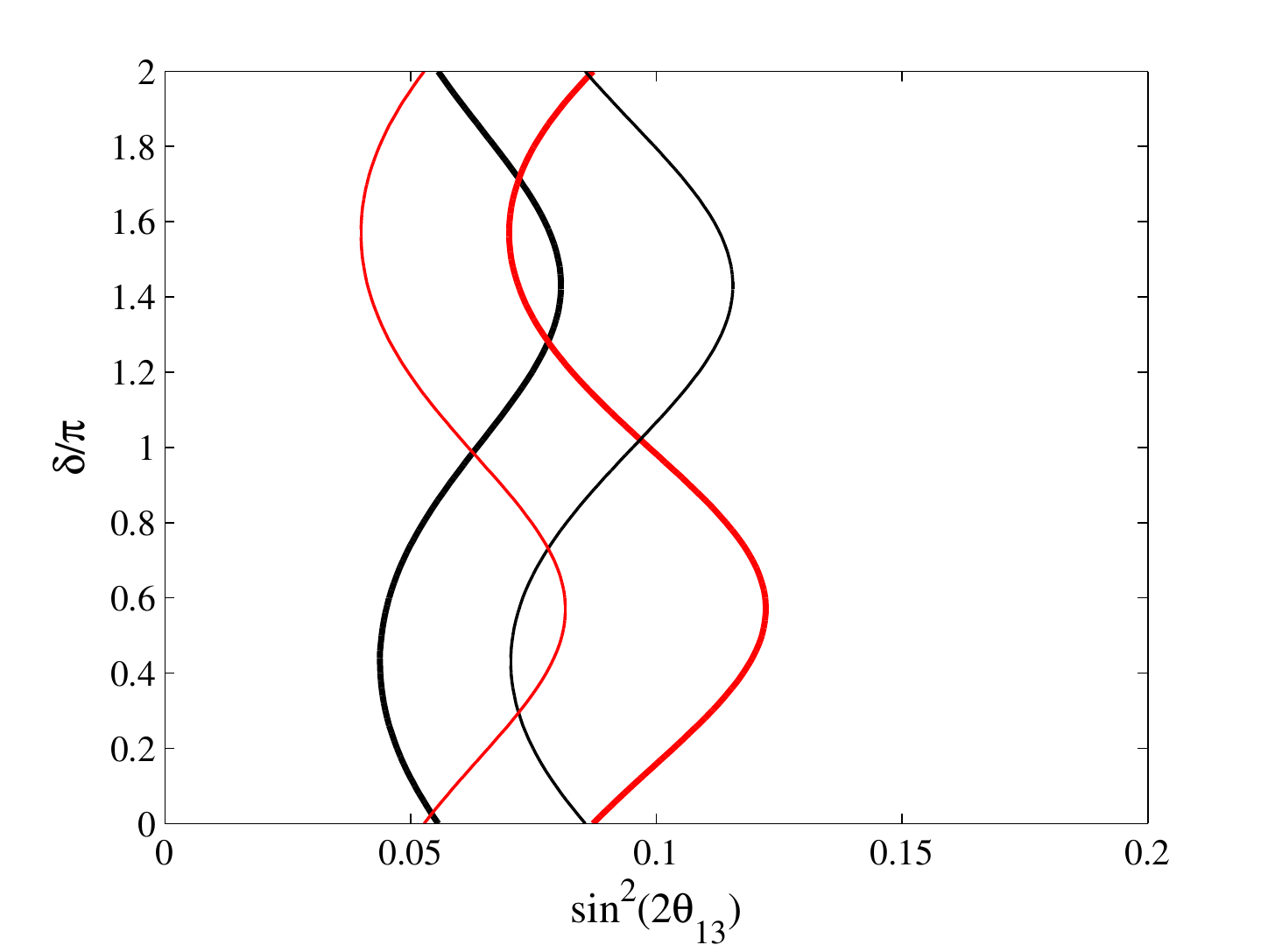}
\caption{Two different illustrations of parameter space degeneracies. \emph{Left panel:} Biprobability plot for $L = 295$~km and $E = 0.65$~GeV. The red band indicates the possible values of the probabilities for normal hierarchy, the blue for inverted, and the pink for the intersection of the two. The black and white ellipses represent the possible values of the probabilities for two different fixed values of $\theta_{13}$. \emph{Right panel:} Probability isocontours of $P_{\mu e}$ (black) and $\bar P_{\mu e}$ (red). The values of the probabilities correspond to those of the intersections between the black and white ellipses in the left panel (with the thick lines representing the upper left intersection). The intersections are where the parameter values reproduce the oscillation probabilities for both neutrinos and anti-neutrinos.
  \label{fig-05:biprobability}}
\end{center}
\end{figure}

Since each of these degeneracies is twofold, an overall degeneracy is eightfold: $2^3 = 8$. 
The first two of these degeneracies can be illustrated in  a bi-probability plot of  \Fig~\ref{fig-05:biprobability}.
As follows from this figure, even if both the probabilities 
(for a given neutrino energy) are known with infinite accuracy, we can not 
identify the hierarchy within the pink region. 

For known mass hierarchy (e.g. normal one) 
a given value of $\theta_{13}$ fixes ellipse in the plot along which the CP phase varies. 
Increasing $\theta_{13}$ moves the  ellipse up and to the right in the plot. Therefore 
for every point on an ellipse, there will be another ellipse corresponding  
another value $\theta_{13}^{prime}$, which crosses this point and therefore 
$\theta_{13}^{prime}$ reproduces the same oscillation probabilities. For example, in the left intersection of the black and white ellipse (\Fig.~\ref{fig-05:biprobability}) both combinations of $\theta_{13}$ and $\delta$ correspond to those precise oscillation probabilities and there are also values of $\theta_{13}$ and $\delta$ that will reproduce them in the inverted hierarchy. For the right intersection, the intrinsic degeneracy is still present, while the sign degeneracy is resolved. It should be remembered that this type of figure is just an illustration. In real experiment the neutrino energy spans over wide range, 
the oscillation probabilities would not be exactly known and strictly this type of consideration becomes invalid.  

\begin{figure}[h!!]
\begin{center}
\vspace{1cm}
 \includegraphics[width=0.99\textwidth,angle=0]{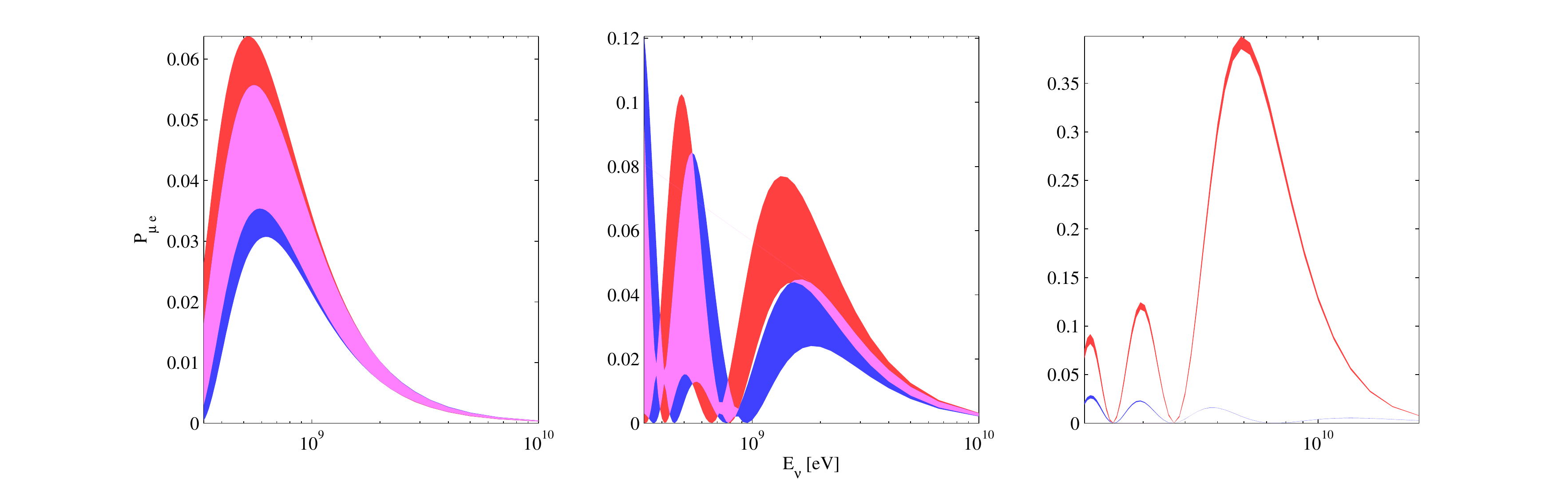}
\caption{ The neutrino oscillation probability $P_{\mu e}$ at baselines of 295 (left), 810 (middle), and 7500~km (right) as a function of the neutrino energy. The red (blue) band corresponds to the normal (inverted) mass hierarchy and the band width is obtained by varying the value of $\delta$. The probabilities for $\bar P_{\mu e}$ look similar with the hierarchies interchanged. Note the different scales of the axes.
  \label{fig-05:probabilities}}
\end{center}
\end{figure}

In order to see how these degeneracies  manifest themselves in an experimental setup, we show the oscillation probability $P_{\mu e}$ as a function of the baseline length in \Fig~\ref{fig-05:probabilities}.
While the 295~km baseline is too short for matter effects to be very significant, as the baseline increases matter effects start being more and more important. In particular, when the oscillation phase maximum occurs at an energy similar to that of the matter resonance, as is the case of 7500~km baseline, we can see the enhancement of the transition probability  in the neutrino channel for the normal hierarchy and the suppression in the inverted. In a simple two-flavor scenario, the amplitude of $P_{\mu e}$ at the resonance is one by definition in the normal mass hierarchy case. At the same time, the oscillation amplitude in the inverted hierarchy at the same energy is given by
\begin{equation}
 \sin^2 2\tilde\theta  = \frac{\sin^2 2\theta}{1+3\cos^2 2\theta} \simeq 
 \frac{1}{4}\sin^2 2\theta,
\end{equation}
where the last equality holds for small $\theta$. On the other hand, 
if the neutrino energy is far below the resonance in order to accumulate a significant oscillation phase, such as in the left and middle panels, then the oscillation amplitude will be effectively given by
\begin{equation}
 \sin^2 2\tilde\theta \simeq \sin^2 2\theta \left[ 1 +\frac{4VE}{\Delta m^2}\cos 2\theta  \right].
\end{equation}
The reason that the 810~km baseline separates the hierarchies better than the 295~km one is based mainly on the fact that the oscillation maximum can be reached for higher energies due to the longer baseline, and thus, the relative difference between probabilities for the two hierarchies increases. Also note that the oscillation probabilities for the 7500~km baseline is not very dependent on the CP-violating phase $\delta$. This is due to the so-called magic baseline effect, which has been discussed before.

In order to successfully determine the neutrino mass hierarchy in a single accelerator experiment, two conditions are of major importance: 1) The baseline must be long enough to allow for a significant value of $VE$ in order to separate the neutrino and anti-neutrino oscillation probabilities. To separate the mass hierarchy determination from the effects of the CP-phase, this separation must be large enough to avoid overlap of the probabilities within the experimental uncertainties. 2) The statistics must be high enough and the systematics low enough in order to make the split statistically significant. The literature contains several proposals for long baseline experiments with baselines of several thousands of kilometers in order to satisfy these conditions. However, as we will discuss later, the large value of $\theta_{13}$ also provides us with an opportunity to pin down the value of $\delta$. Such measurements require the presence of interference terms which will be small at the very long baselines, and instead medium long baselines around 1000~km, such as the 810~km baseline shown in \Fig~\ref{fig-05:probabilities}, may be preferable due to the significant $\delta$ dependence 
of probabilities.

\subsubsection{CP-violation effects and the mass hierarchy}


The figure \Fig~\ref{fig-05:probabilities} shows a significant dependence 
of the probabilities on the CP-violating phase $\delta$, especially at small baselines. 
We are mainly interested in the oscillation probability at the first or second oscillation maximum, where an experiment would typically be placed. 
In these baselines $L$ the $\nu_\mu - \nu_e$ oscillation probability  (the ``golden channel'') 
can be expanded in the small quantity $\Delta m_{21}^2 L/2E$ which gives~\cite{05-Cervera:2000kp}
\begin{equation}
\label{eq-05:cp}
 P_{e\mu} \approx s_{23}^2 P^{2\rm f} + c_{13} \sin 2\theta_{13}  \sin 2\theta_{12} \sin 2\theta_{23} \frac{\Delta m_{21}^2}{2EV} \sin\left(\frac{VL}{2}\right) \sin\left(\frac{\Delta m_{31}^2 L}{4E}\right) \cos\left(\delta - \frac{\Delta m_{31}^2 L}{4E}\right),
\end{equation}
where $P^{2\rm f}$ is the two-flavor oscillation probability discussed earlier. 
In Eq.~\ref{eq-05:cp} we have neglected terms of the second (and higher) order 
in $\Delta m_{21}^2L/2E$\footnote{While the first neglected term is not suppressed by $\theta_{13}$, for the value of $\theta_{13}$ measured by reactor experiments the suppression by the solar mass square splitting is about 6 times stronger.}. as well as the matter effect on $\Delta m_{31}^2$. It is the the second term that is responsible for creating the band of different oscillation probabilities displayed in \Fig~\ref{fig-05:probabilities}, and hence, for creating the sign degeneracy in accelerator neutrino experiments. The appearance of the $\sin(VL/2)$ term is an inheritance from the magic baseline oscillations and will vanish the  $\delta$-dependent term when $VL = 2\pi$. Furthermore, we can observe that this term contains all of the mixing angles in the same way as the Jarlskog invariant, which is expected due to the CP-dependence of the term.

\subsection{Determination of hierarchy with atmospheric neutrinos}
\label{sec-05:atmospheric} 

\subsubsection{Neutrino fluxes}

The original flux of atmospheric neutrinos 
contains incoherent components of $\nu_e$,  $\nu_\mu$ 
and the corresponding antineutrinos, while the original $\nu_\tau$ flux 
is negligible. We introduce $\Phi_e^0$ and $\Phi_{\mu}^0$,
the electron and muon neutrino fluxes, as well as 
$\bar{\Phi}_e^0$ and $\bar{\Phi}_{\mu}^0$,
the electron and muon antineutrino fluxes, at the detector in the absence of 
oscillations. The flavor ratios
\be
r \equiv \frac{\Phi_\mu^0}{\Phi_e^0},  \quad
\bar{r} \equiv \frac{\bar{\Phi}_\mu^0}{\bar{\Phi}_e^0},  
\nonumber
\ee
increase with energy. 

There is a mild neutrino-antineutrino
asymmetry: 
the neutrino flux $\bar{\Phi}_{\mu}^0 /\Phi_{\mu}^0 \approx 0.8 - 0.9$. 
All the fluxes (at $E > 1$ GeV) decrease rapidly with energy 
$\Phi_\alpha^0 \propto E^{-k}$, $k = k(E) = 3 - 5$,  
and an azimuthal dependence shows up at low energies. 

The flux of neutrinos of flavor $\nu_\alpha$ at a detector, with oscillations 
taken into account, is given by 
\be
\Phi_\alpha = \Phi_e^0 P_{e\alpha} + \Phi_\mu^0 
P_{\mu \alpha} =  \Phi_e^0 [ P_{e \alpha} + r(E, \Theta_\nu) 
P_{\mu \alpha}], \quad \alpha = e,~ \mu,~ \tau . 
\label{eq-05:twoterms}
\ee
Similar expressions hold for the antineutrino fluxes. 
Inserting the analytic expressions for the probabilities
\ref{eq-05:Pee} - \ref{eq-05:Pmutau}. 
one finds  
\bea
\frac{\Phi_e}{\Phi_e^0} & = & 1 + (r s_{23}^2 - 1)P_{e\tilde{3}} +  
(r c_{23}^2 - 1) P_{e\tilde{2}} + r P_{\mu e}^\delta ~, 
\label{eq-05:fluxe}\\
\frac{\Phi_{\mu}}{\Phi_{\mu}^0} & \approx &  
1 - 2s_{23}^2 c_{23}^2 \left[1 -  {\rm Re} (A_{\tilde{2}\tilde{2}}^*A_{\tilde{3}\tilde{3}}) 
\right] 
- \frac{s_{23}^2}{r}(rs_{23}^2 -1) P_{e\tilde{3}} - \frac{c_{23}^2}{r} (r c_{23}^2 -1)P_{e\tilde{2}} +  
P_{\mu \mu}^\delta  +\frac{1}{r}  P_{e \mu}^\delta~, 
\label{eq-05:fluxmu} \\
\frac{\Phi_{\tau}}{\Phi_{\mu}^0}  & \approx &  
2s_{23}^2 c_{23}^2 \left[1 -  {\rm Re} (A_{\tilde{2}\tilde{2}}^*A_{\tilde{3}\tilde{3}})   
\right] 
- \frac{c_{23}^2}{r}(rs_{23}^2 -1) P_{e\tilde{3}}   
- \frac{s_{23}^2}{r} (r c_{23}^2 -1)P_{e\tilde{2}} 
 + P_{\mu \tau}^\delta  +\frac{1}{r} P_{e \tau}^\delta~, 
\label{eq-05:fluxtau}
\eea
where $P_{e\tilde{3}} \equiv |A_{e\tilde{3}}|^2$ and $P_{e\tilde{2}} \equiv |A_{e\tilde{2}}|^2$, 
are defined in \Sec~\ref{sec-05:oscprobproperties}. In the factorization approximation they correspond to the atmospheric and solar oscillation modes. 
The $\delta$ dependent terms have been introduced in 
Eqs. (\ref{eq-05:Pme-2}). 

Using unitarity relations  
\be
|A_{\tilde{2}\tilde{2}}|^2 = 1 - |A_{\tilde{2}e}|^2 - |A_{\tilde{2}\tilde{3}}|^2 
\approx 1 - |A_{\tilde{2}e}|^2 = 1 - P_{e\tilde{2}}, 
\ee
where term proportional to  $|A_{\tilde{3}\tilde{2}}|^2$ have been neglected 
we can approximate 
\be 
{\rm Re} (A_{\tilde{2}\tilde{2}}^*A_{\tilde{3}\tilde{3}}) \approx 
\sqrt{(1 - P_{e\tilde{3}})(1 - P_{e\tilde{2}})} \cos \psi . 
\label{eq-05:real}
\ee
Here  $\psi \equiv {\rm arg} A_{33} A_{22}^*$ is the relative phase 
between the two amplitudes.  
For the $\nu_e-$ flux, we the obtain  
\be
\frac{\Phi_e}{\Phi_e^0}  \approx  1  + (r s_{23}^2 - 1)P_{e\tilde{3}} 
+ (r c_{23}^2 - 1) P_{e\tilde{2}}
 +  r \sin 2\theta_{23} \sqrt{P_{e\tilde{3}} P_{e\tilde{2}}} \cos(\phi -\delta) ~.
\label{eq-05:fluxe1}
\ee
The oscillated fluxes satisfy the sum rule 
\be 
\Phi_e  + \Phi_\mu +  \Phi_\tau =  \Phi_e^0 + \Phi_\mu^0,  
\ee
which simply reflects the unitarity of transitions 
and, consequently, conservation of the total flux in oscillations. 

The formulas (\ref{eq-05:fluxe}) - (\ref{eq-05:fluxtau}) 
 also show the screening effect. 
Terms with oscillation probabilities driven by the 
1-2 and 1-3 mixings appear with the ``screening'' factors 
\cite{05-Peres:1999yi,05-Akhmedov:1998xq}: 
$P_{e\tilde{3}}$ with $(rs_{23}^2 -1)$ and $P_{e\tilde{2}}$ with $(r c_{23}^2 -1)$. 
The contribution of the ``atmospheric mode'' vanishes along the line 
$ r(E, \Theta_\nu) = 1/ s_{23}^2 
$, 
whereas the contribution of the ``solar mode'' vanishes along 
$ r(E, \Theta_\nu) = 1/c_{23}^2$. 
For maximal mixing both contributions vanish along the same line, 
$r(E, \Theta_\nu) = 2$. 
For the neutrino energies above 0.1~GeV, $r > 1.8 - 1.9$ and only one of these
contributions can vanish for 
substantial deviation of the 2-3 mixing from 
maximal: $s_{23}^2$ or  $c_{23}^2 < 0.45$. 
Thus, both the effects of 1-2 and 1-3 mixing turn out to be sub-leading and the
oscillation effects are well described by the first order approximation 
of 2-3 vacuum oscillations.

In the $\nu_\mu$ flux, the contributions of the 1-2 and 1-3 modes 
are suppressed by additional factors $s_{23}^2/r$ and $c_{23}^2/r$, respectively. 
There is no suppression of the interference terms, which depend on the 
CP-violation phase. Furthermore, in the $\nu_e-$ flux the interference term 
is enhanced by the flux ratio $r$. 
There is no suppression  of the interference terms of the 1-2 and 1-3 modes in  the  
$\mu -\tau$ mode.

\subsubsection{Sensitivity to mass hierarchy}

Let us discuss the sensitivity of large water or ice detectors of atmospheric neutrinos to the 
neutrino mass hierarchy. The $\nu_\mu$-like events correspond 
to interactions $\nu_\mu + N \rightarrow \mu + X$,  
$\bar{\nu}_\mu + N \rightarrow \mu^+ + X$ and can be observed as events with muon tracks and hadron cascades. 
There are also some contributions 
from $\nu_\tau$ which produce $\tau$ with subsequent decay into $\mu$. 
The number of $\nu_\mu$-like events in the $ij$-bin  
in the $E_\nu - \cos \theta_z$ plane  equals  
\be
N_{ij, \mu}^{\rm NH}  =  2 \pi N_A \rho T \int_{\Delta_i\cos\theta_z} 
d\cos\theta_z \int_{\Delta_jE_\nu} dE_\nu~ V_{\rm eff} (E_\nu) 
D_\mu (E_\nu, \theta_z), 
\label{eq-05:nev}
\ee
where $T$ is the exposure time, $N_A$ is the Avogadro number, $\rho$ is 
the density of ice, $V_{\rm eff} (E_\nu, \theta_z )$ 
is the effective volume of the  detector, and the number density of events per unit 
time per target nucleon is given by 
\be
D_\mu (E_\nu, \theta_z) = 
\left[\sigma^{CC} \left(\Phi_\mu^0  P_{\mu\mu}  + \Phi_e^0  P_{e\mu}
\right) +
{\bar \sigma}^{CC}
\left({\bar \Phi}_\mu^0 {\bar P}_{\mu\mu} + {\bar \Phi}_e^0  {\bar P}_{e\mu}\right) 
\right].  
\label{eq-05:den}
\ee
It is assumed here that experiments do not distinguish the neutrino and antineutrino events and 
corresponding signals are summed up.  

\begin{figure}\begin{center}
\includegraphics[trim=0.in 0.in 2.8in 0.in, clip, height=3in]{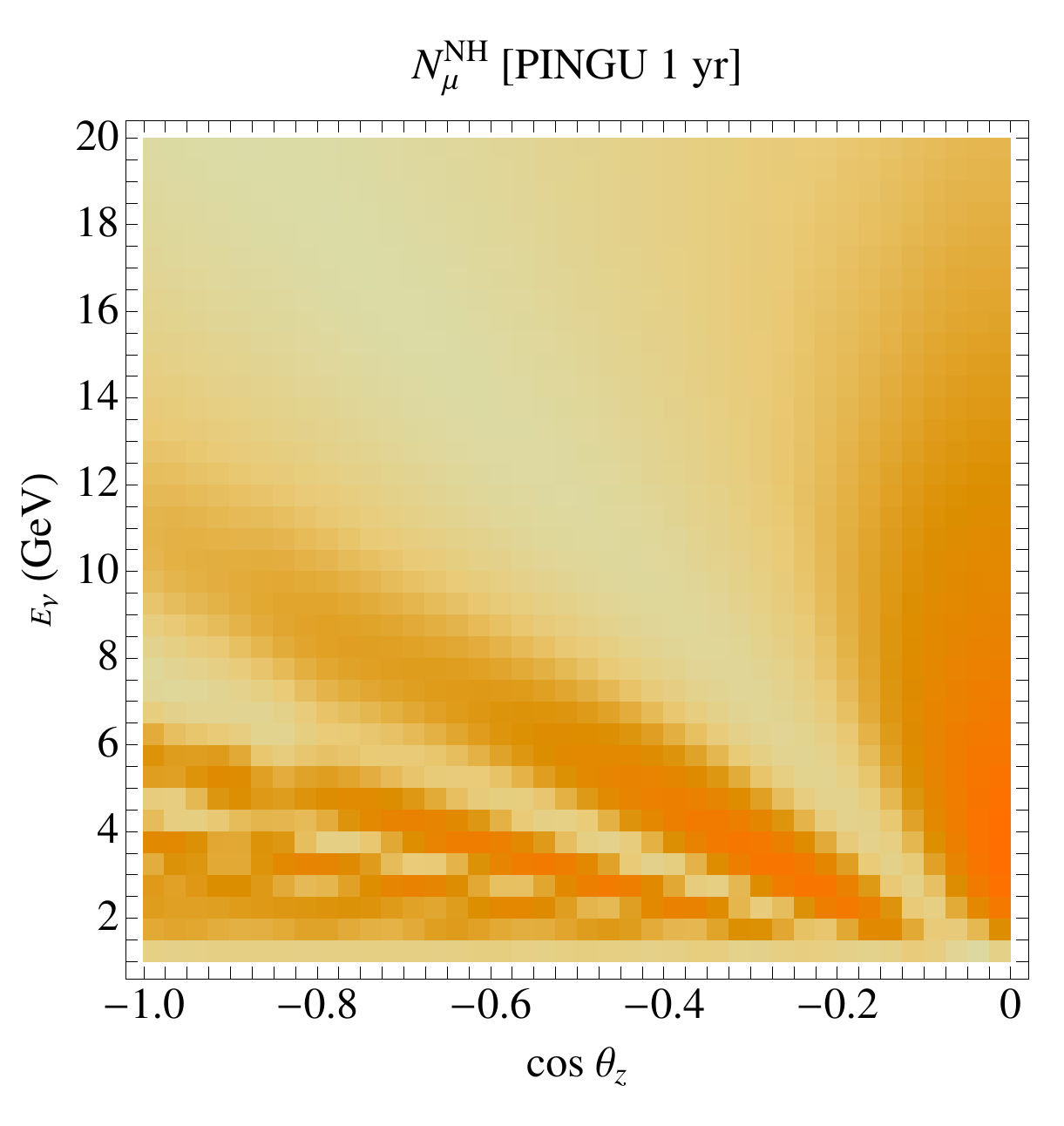}
\includegraphics[trim=0.in 0.in 4.8in 0.in, clip, height=3in]{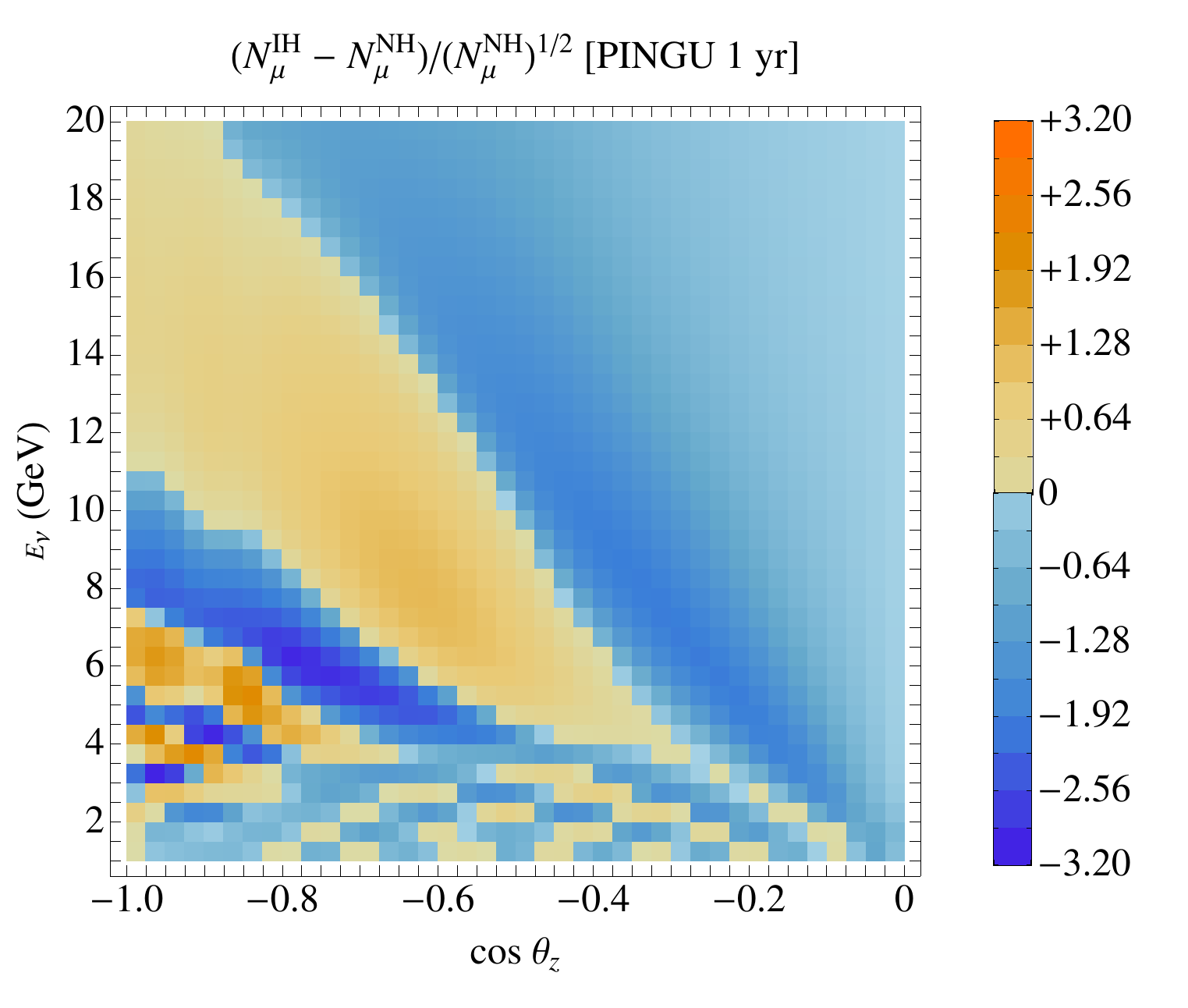}
\caption{\emph{Left:} The binned distribution of the number of $\mu$ events 
in PINGU after 1~year under the assumption that the neutrino hierarchy is normal. \emph{Right:} The N-I
hierarchy asymmetry of $\nu_\mu$ events
in the $E_\nu - \cos \theta_z$ plane.
The absolute value of the asymmetry in a given bin
determines the statistical significance
of the difference of the numbers of events for the inverted and normal
mass hierarchies. Both figures from~\cite{05-Akhmedov:2012ah}.
\label{fig-05:asym-mu}
}
\end{center}
\end{figure}

The fine-binned distribution of events  
(\ref{eq-05:nev}) is shown in \Fig~\ref{fig-05:asym-mu}.   
For illustration we use the effective volume of PINGU with 22 additional strings \cite{05-cowen}. 
which increases from $\sim 2$~Mt at $E_\nu = 2$ GeV to 20~Mt at $E_\nu =  20$~GeV.  
The pattern of the event number distribution follows the oscillatory 
picture due to the $\nu_\mu - \nu_\mu$ mode of oscillations with a certain distortion in the resonance region. The maxima and minima 
are approximately along the lines of equal oscillation phases 
$E_\nu \sim \phi_{32}\Delta m^2_{32}  |\cos \theta_z| R_\oplus$ 
(where $R_\oplus$ is the Earth radius), with distortion in the resonance region $E_\nu = (4 - 10)$ GeV.   
In the high density bins, the number of events reaches 200 and the total 
number of events is about $10^{5}$.

The expression for the density of events (\ref{eq-05:den}) can be written as 
\be
D_{\mu}^{\rm NH} =  \sigma^{CC} (E_\nu) \Phi_\mu^0  
\left[ \left(P_{\mu\mu}  + \frac{1}{r} P_{e\mu}\right) 
 +
\kappa_\mu \left({\bar P}_{\mu\mu} + \frac{1}{\bar r}  {\bar P}_{e\mu}\right) 
\right], 
\label{eq-05:mueventsNH}
\ee
where 
\be
\kappa_\mu \equiv
\frac{{\bar \sigma}^{CC} \bar{ \Phi}_\mu^0}
{\sigma^{CC} \Phi_\mu^0}.
\nonumber
\ee
%
%
Similarly one can determine the number of events for inverted mass hierarchy. 
Let us introduce the N-I hierarchy asymmetry for the $ij$-bin  in the 
$(E_\nu - \cos \theta_z)$ plane as 
\be
A^{N-I}_{\mu, ij}  \equiv \frac{N^{IH}_{\mu, ij} 
- N^{NH}_{\mu,ij}}{\sqrt{N^{NH}_{\mu, ij}}}. 
\label{eq-05:asym}
\ee
The moduli of the asymmetry (\ref{eq-05:asym}) are the measures of statistical 
significance of the difference of the number of events for the  normal 
and inverted mass hierarchies: $S_{ij} = |A_{ij}|$.

The strongest effect of hierarchy change is in the strips along the 
constant phase lines in the energy interval  $E_\nu =  (4 - 12)$ GeV,
where these lines are distorted by the matter effect.  
Here the asymmetry changes sign with the zenith angle 
and  number of intervals with the same sign asymmetry increases with 
decrease of energy. 
The $\nu_\tau \rightarrow \tau \rightarrow \mu$ 
events can be considered as background events and treated within $\sim 5\%$ systematic errors.

\subsubsection{Measurements}

According to \Fig~\ref{fig-05:asym-mu}, the hierarchy asymmetry of the 
$\nu_\mu$ events has opposite signs in different parts of the oscillogram. 
Thus, the integration over $E_\nu$ and $\cos \theta_z$ 
substantially reduces the sensitivity to the hierarchy. Due to this, a
relatively good reconstruction of the neutrino energy and 
direction are required to identify the hierarchy. 
The uncertainties  of the  reconstruction 
of energy $\sigma_E$ and  angle $\sigma_\theta$
should be comparable to or smaller than the sizes of the domains 
with the same sign of the asymmetry.  
The oscillograms for the reconstructed neutrino energy $E_\nu^r$ and 
angle $\theta_z^r$ can be obtained by smearing of the 
$E_\nu -\cos \theta_z$ oscillograms with the reconstruction 
functions of the width $\sigma_E$ and angle $\sigma_\theta$.

Small uncertainties $\sigma_E$ and $\sigma_\theta$  require rather 
precise measurements of the energy
$E_\mu$ and direction $\theta_\mu$ of the muon,
as well as energy of the accompanying hadron cascade $E_h$.  
Then the neutrino energy equals $E_\nu^r  = E_\mu + E_h$. 
The reconstruction of the neutrino direction is more involved.  
In the first approximation, one can use $\theta_\nu \approx \theta_\mu$
with a spread which decreases with energy: 
$\sigma_\theta \sim A \sqrt{m_p/E_\nu}$ (A = O(1)). Knowledge of the 
hadron cascade energy allows to reduce this uncertainty. 
Further improvements could be possible if some information about 
geometry of the cascade is available.  
A possibility to separate (at least partially) the neutrino
and antineutrino samples would significantly improve sensitivity to 
the mass hierarchy, as well as to CP-violation. 

All this imposes conditions on the detector characteristics.
According to \Fig~\ref{fig-05:asym-mu}, the  most sensitive region to the hierarchy is 
around the resonance and above: 
$E = (5 - 15)$ GeV. The number of events in Super-Kamiokande is too small, but (upgraded)
ice and underwater detectors 
of the multi-megaton ($\sim 10$ Mt) scale could collect around the order of $10^5$ 
$\nu_\mu$ events a year in this range so that a high statistics study becomes possible.  

A small enough spacing between the PMTs ($\sim 10 - 20$ m between strings and 3 - 5 m in the vertical direction) 
 will  allow the reduction of the threshold down to a few GeV and perform reasonably good 
measurements of the muon and hadron cascade characteristics. 
Very high statistics will also allow the resolve the problem of parameter degeneracy: 
 effects qualitatively similar to the mass hierarchy effect  can be obtained 
by small (within $1\sigma$ interval) variations of $\Delta m_{32}^2$ 
and $\theta_{23}$. The effect of an unknown CP-phase is small.

High statistics would allow to resolve the degeneracy problem
by selecting specific regions in the $E_\nu -\cos \theta_z$
for the analysis, where effects of $\Delta m^2_{32}$ are suppressed in comparison 
to the hierarchy effects  or averaged out as a result of 
specific integration. High statistics also 
allows to perform an analysis of the data using 
$\Delta m^2_{32}$ and $\theta_{23}$ as fit parameters. 
This will open a possibility to determine the mass hierarchy and measure these parameters simultaneously.  

Note that other experimental techniques using atmospheric neutrinos may also prove valuable for determination of the mass hierarchy. In particular, experiments that can separate neutrinos from anti neutrinos on an event basis need a significantly lower number of events to obtain the same sensitivity. Thus, such detectors can be smaller in size as compared to the neutrino telescopes. In this context, a magnetized iron calorimeter, such as the India based Neutrino Observatory~\cite{05-INO}, could also provide an important 
contribution to the determination of mass  hierarchy. The capabilities of detectors using charge identification were studied in~\cite{05-Petcov:2005rv}.

\subsubsection{Interplay between accelerator and atmospheric neutrinos}

The  atmospheric neutrino data can also be used to compliment the data from accelerator neutrino experiments in order to extract the most information possible. 
As was demonstrated in~\cite{05-Petcov:2005rv}, the atmospheric neutrino determination of the neutrino mass hierarchy can be significantly affected by the addition of external priors and, in particular, may lead to different sensitivity to the neutrino mass hierarchy for 
in the cases of true normal or inverted hierarchy. However, once external input on the neutrino oscillation parameters is included by considering also other experiments, the room to mimic the true oscillation pattern in the wrong hierarchy becomes much more restricted and the sensitivity to the hierarchy increases. Adding the accelerator experiments' own sensitivity to the mass hierarchy, a measurement may be possible even for the current generation of accelerator experiments by the addition of detector capable of lepton charge identification. This has been discussed in~\cite{05-Blennow:2012gj} and the prospects of using a magnetized iron calorimeter detector to augment the current generation of accelerator experiments are a 2--4$\sigma$ determination of the mass hierarchy within 10 years of data taking, depending on the true value of the oscillation parameters and the characteristics of the detector.

\section{Discussion and Conclusions}
\label{sec-05:conclusions}

In this paper, we have described the effects
of neutrino propagation in matter
relevant for experiments with atmospheric
and accelerator neutrinos and aimed at the determination of the
neutrino mass hierarchy and CP-violation.
Thus, to a large extent, we have focused on neutrino propagation
in the Earth matter. 

\noindent
1. At relatively low energies, the dominant effect of neutrino interactions
with matter is the elastic forward scattering,
which is described by an effective potential.
Neutrino evolution in matter is then described
by a Schr\"odinger-like equation including this effective potential.
The potential differences for neutrinos of different types
influence the flavor evolution of the system of mixed neutrinos.\\

In the majority of realistic situations, neutrinos propagate
in normal (unpolarized non-relativistic) matter
with nearly constant or slowly changing density. 

\noindent
2.  Matter modifies the neutrino flavor mixing and changes the
eigenvalues of the Hamiltonian of propagation.
This is equivalent to a modification of
the dispersion relations of neutrinos.
The influence of matter on mixing of neutrinos
has a resonance character.  At energies or densities
for which the eigenfrequency of the neutrino system with mixing
$\omega_{ij} = \Delta_{ij}^2/2E$ equals approximately the
eigenfrequency of the medium $2\pi/l_0$, the mixing
in matter becomes maximal.  Large mixing shifts
the position of the resonance to lower values of the potential. At usual densities, there are
two resonances related to the two mass squared differences $\Delta m_{21}^2$ and
$\Delta m_{31}^2$ between the neutrino mass eigenstates.
The resonances are realized in oscillation channels involving electron neutrinos.\\

\noindent
3.  In many practical situations, knowledge of neutrino
mixing in matter and the eigenstates of the Hamiltonian
in matter allows to immediately find the results
of the neutrino flavor evolution. This includes neutrino oscillations
in matter with constant density and also adiabatic
conversion of neutrinos, where the averaged oscillation
results can be written down immediately. In the non-averaged case,
the problem is reduced to finding the oscillation phase
(integrating the energy splittings over distance).
In this sense the Nature has implemented the most (computationally) simple
setups.

The very convenient presentation of mixing in matter
can be obtained as series expansion in
the ratio of the  two mass squared differences, $r_\Delta$, 
(perturbative diagonalization of the effective Hamiltonian),
which allows to understand  a number of subtle results.

The simplest and physically transparent  description
of dynamics of neutrino flavor evolution can be obtained in
the propagation basis (in the case of the standard parameterization).
In this basis, the CP-violating phase and 2-3 mixing
do not influence the evolution and the amplitudes of transitions do
not depend on $\delta$ or $\theta_{23}$.
The dependence on these parameters appear as a result of projecting the
states of the propagation basis back to the flavor states at
production and detection.

 In many practical cases the $3\nu$ evolution can be reduced
to evolution of two neutrino systems with certain corrections.

\noindent
4. There are two practically important cases:
(i) neutrino propagation in matter with constant
or nearly constant density and (ii) neutrino propagation in
matter with slowly (adiabatically) changing density. 

\noindent
5. In the case of constant density, flavor evolution has a character
of oscillations with parameters determined by mixing
and mass splitting in matter. The oscillations are an effect of a phase
difference increase in the course of neutrino propagation.
The resonance enhancement of oscillations is realized in
an energy region around $E_R$.

If the density is approximately constant, then
the results can be obtained by using perturbation theory
in the deviation of the density distribution from a constant one.
The accuracy improves if the density
profile is symmetric with respect to the middle point of the
neutrino trajectory, as is realized for neutrinos crossing the Earth.

A simple and rather precise semi-analytical description
of neutrino oscillations in matter with varying density
can be obtained in the limits of small density,
$V < \Delta m_{ij}^2/2E$, and high density $V \gg \Delta m_{ij}^2/2E$.
The latter gives a very accurate description of neutrino flavor evolution in the Earth
at $E > (8 - 10)$ GeV. 

\noindent
6. In a medium with slowly changing density, adiabatic conversion
takes place. This effect is related to the change of mixing in matter
due to density change.
Adiabaticity implies that there are no transitions
among the eigenstates of the instantaneous Hamiltonian during propagation.

The strongest flavor transformation is realized when
the initial density is much larger, and the final one is much lower
than the resonance density. In this case, the initial state
(and due to adiabaticity, the state at any other moment of evolution)
practically coincides with one of the eigenstates.
Therefore, oscillation effects are absent and
non-oscillatory flavor conversion takes place.
This is realized for supernova neutrinos and
approximately -- for high energy solar
neutrinos.
In general, if the initial mixing is not strongly suppressed, an
interplay of adiabatic conversion and oscillations occurs.

Adiabatic transformations are also realized for neutrinos with energy
$\leq 1$ GeV propagating in the mantle of the Earth.
In particular, this means that the oscillation depth at the
detector
is determined by mixing at the surface of the Earth
and not by the mixing at average density.

Until now, the mater effects have been observed
in solar neutrinos and, indirectly,
in atmospheric neutrinos and there is good chance that
they will be observed  by new generation of the accelerator and 
atmospheric neutrino experiments. 

\noindent
7. Strong flavor transition can be realized without  enhancement of mixing.
This occurs in matter with periodic or quasi-periodic
density change when the parametric resonance condition is
fulfilled.  For small mixing strong transition requires a large number of periods.

A similar enhancement can take place in matter
with several layers of different densities. Here the enhancement
occurs when a certain correlation between  the oscillation phases
in each layer and amplitudes of oscillations determined by mixing is present.
The case of a medium with 3 layers (1.5 periods) is
of practical interest for neutrinos crossing both the mantle and the core of the Earth.

For a multilayer medium two conditions must be satisfied to
have strong transitions: the amplitude (collinearity)
and the phase conditions. 

\noindent
8. For neutrinos crossing a small amount of matter,
such as accelerator experiments with baselines up to
$(1 - 2)\cdot 10^{3}$~km, the column density of matter is small
and, according to the minimal width condition, the matter effect
on oscillations is small regardless of energy, vacuum mass splitting, and
neutrino mixing. Furthermore, if the oscillation phase is small, then
mimicking of vacuum oscillations occurs.

\noindent
9. A comprehensive description of the neutrino
flavor transitions in the Earth is given in terms
of neutrino oscillograms of the Earth.
After the recent determination of the 1-3 mixing, the
structure of oscillograms is well fixed.
The salient features of oscillograms 
at high energies (due to 1-3 mixing) are
the MSW resonance peak in the mantle domain,
three parametric ridges and the MSW peak in the core domain.
At low energies (due to 1-2 mixing), there are three peaks,
due to the MSW resonance, and the parametric ridge.
The positions of all these and other structures are determined
by the generalized phase and amplitude conditions.

 In the case of normal mass hierarchy,
the  resonance peaks 
induced by the 1-3 mixing are in the neutrino channels.
For inverted mass hierarchy they are in the antineutrino channels.
This is the foundation for determining the neutrino mass hierarchy.
The resonance structures due to the 1-2 mixing
are always in the neutrino channels, since the sign of the small 
mass square difference has been fixed.

\noindent
10. The CP-properties of the oscillograms
(their dependence on CP-phase) are determined by
the CP-domains: areas in which the CP-violation effect
has the same sign. The borders of these domains
are approximately determined by the grids of the  magic lines
(solar and atmospheric magic lines) and the lines where the oscillation phase condition is fulfilled. 

\noindent 
11. Measurements of matter effects in neutrino oscillations
provides a good opportunity to determine the neutrino mass hierarchy.
The 1-2 ordering has been determined 
due to the matter effect of solar neutrinos. The 1-3 ordering can be identified 
by studying the matter effects in accelerator and
atmospheric neutrino experiments.

There is a good chance that future
studies of the atmospheric neutrinos with
multi-megaton underwater (ice) detectors
will be able to establish the mass hierarchy.
With a threshold of a few GeV, these detectors will be sensitive to
the resonance region ($\sim 6 - 10$)~GeV, where the difference of
probabilities for the normal and inverted mass hierarchies is maximal.

The challenges here are the accuracy of reconstruction of the
neutrino energies and directions. Integration over the energy
and angle, as well summation of neutrino and
antineutrino signals, diminish the sensitivity to the hierarchy.
Another problem is the degeneracy of the hierarchy
effects with the effects of other neutrino parameters,
in particular with $\Delta m^2_{32}$ and $\theta_{32}$.

\noindent
12.  In accelerator experiments, many of the problems
mentioned above are absent. However, existing and proposed
accelerator experiments will cover only periferal
regions of oscillograms where enhancement of oscillations
is very weak and oscillatory structures are rather poor.
As a consequence the problem of degeneracy here is even more severe.


\end{document}